\documentclass[fleqn,usenatbib,useAMS]{mnras}
\usepackage{graphicx}
\usepackage{amsmath}
\usepackage{amsfonts}
\usepackage{float}
\usepackage{bm}
\usepackage{perpage} 
\usepackage[T1]{fontenc}
\usepackage{ae,aecompl}

\newcommand{\appropto}{\mathrel{\vcenter{
  \offinterlineskip\halign{\hfil$##$\cr 
    \propto\cr\noalign{\kern2pt}\sim\cr\noalign{\kern-2pt}}}}}

\newcommand{\ssim}{\,{\sim}\,} 
\newcommand{\sga}{\,{\ga}\,} 

\title{Dynamical History of the Local Group in $\Lambda$CDM}{}
\MakePerPage{footnote} 
\author[Indranil Banik \& Hongsheng Zhao]{Indranil Banik$^{1}$\thanks{Email: ib45@st-andrews.ac.uk (Indranil Banik), \newline hz4@st-andrews.ac.uk (Hongsheng Zhao)}, Hongsheng Zhao$^{1}$\\
$^{1}$Scottish Universities Physics Alliance, University of St Andrews, North Haugh, St Andrews, Fife, KY16 9SS, UK}

\pubyear{2016}
\begin{document}
\label{firstpage}
\pagerange{\pageref{firstpage}--\pageref{lastpage}}
\maketitle

\begin{abstract}

The positions and velocities of galaxies in the Local Group (LG) measure the gravitational field within it. This is mostly due to the Milky Way (MW) and Andromeda (M31). We constrain their masses using distance and radial velocity (RV) measurements of 32 LG galaxies. To do this, we follow the trajectories of many simulated particles starting on a pure Hubble flow at redshift 9. For each observed galaxy, we obtain a trajectory which today is at the same position. Its final velocity is the model prediction for the velocity of that galaxy.

Unlike previous simulations based on spherical symmetry, ours are axisymmetric and include gravity from Centaurus A. We find the total LG mass is ${4.33^{+0.37}_{-0.32}\times{10}^{12}M_\odot}$, with $0.14 \pm 0.07$ of this being in the MW. We approximately account for ${\text{IC}~\text{342}}$, M81, the Great Attractor and the Large Magellanic Cloud.

No plausible set of initial conditions yields a good match to the RVs of our sample of LG galaxies. Observed RVs systematically exceed those predicted by the best-fitting $\Lambda$CDM model, with a typical disagreement of ${45.1^{+7.0}_{-5.7}}$ km/s and a maximum of ${110 \pm 13}$ km/s for ${\text{DDO}~\text{99}}$. Interactions between LG dwarf galaxies can't easily explain this.

One possibility is a past close flyby of the MW and M31. This arises in some modified gravity theories but not in $\Lambda$CDM. Gravitational slingshot encounters of material in the LG with either of these massive fast-moving galaxies could plausibly explain why some non-satellite LG galaxies are moving away from us even faster than a pure Hubble flow.

\end{abstract}

\begin{keywords}
galaxies: groups: individual: Local Group -- Galaxy: kinematics and dynamics -- Dark Matter -- methods: numerical -- methods: data analysis -- cosmology: cosmological parameters
\end{keywords}

\section{Introduction}
\label{Introduction}

In a homogeneous universe, particles would follow a pure Hubble flow. This means their velocities would depend on their positions according to
\begin{eqnarray}
	\bm{v} \left( t \right) = H \left( t \right) \bm{r} ~\text{ where } H \equiv \text{Hubble parameter at time }t
	\label{Hubble_flow}
\end{eqnarray}

However, the Universe is inhomogeneous on small scales. The resulting inhomogeneous gravitational field causes motions to deviate from Equation \ref{Hubble_flow}. These deviations $-$ termed `peculiar velocities' $-$ are easily discerned in the Local Group (LG). Thus, the observed positions and velocities of LG galaxies hold important information on the gravitational field in the LG, both now and in the past.\footnote{Due to Hubble drag (paragraph below Equation \ref{Equation_26}), peculiar velocities are mostly sensitive to forces acting at late times.} Therefore, by investigating a range of physically motivated models for the gravitational field of the LG, we can hope to see which ones $-$ if any $-$ plausibly explain these observations. This technique is known as the timing argument.

The timing argument was first applied to the Milky Way (MW) and Andromeda (M31) galaxies over 50 years ago \citep{Kahn_Woltjer_1959}. This pioneering work attempted to match the present relative velocity of the MW and M31, assuming no other major nearby sources of gravity. As M31 must initially have been receding from the MW but is currently approaching it at $\sim$110 km/s \citep{Slipher_1912, Schmidt_1958}, it was clear that models with very little mass in the MW \& M31 could not work.\footnote{In the limit of no mass, M31 would be receding at $\sim$50 km/s.} In fact, their combined mass had to be $\sim$10 times the observed baryonic mass in these galaxies. This provided one of the earliest indications that most of the mass in typical disc galaxies might be dark.

This conclusion has withstood the test of time, at least in the context of Newtonian gravity. More recent works find a total LG mass of $M \sim \left( 4 - 5 \right) \times {10}^{12} M_\odot$ \citep{Li_White_2008, Van_der_Marel_2012, Partridge_2013}. This is roughly consistent with the combined dynamical masses of the MW and M31. For example, analysis of the giant southern stream around Andromeda (a tidally disrupted satellite galaxy) yielded $M_{_{M31}} \approx 2.5 \times {10}^{12} M_\odot$ \citep{Fardal_2013}.\footnote{This is an estimate of $M_{200}$.} Combining a wide variety of observations of our own Galaxy, \citet{McMillan_2011} found that $M_{_{MW}} \approx 1.5 \times {10}^{12} M_\odot$.\footnote{This is an estimate of the virial mass.} However, careful analysis of the Sagittarius tidal stream \citep{Newberg_2002, Majewski_2003} found a mass of about half this \citep{Vasily_2014}, though this depends on the uncertain distance to the progenitor.

The timing argument seems to suggest a higher mass than the sum of the MW \& M31 dynamical masses. The tension would be further exacerbated if the LG mass was smaller in the past, forcing up the present mass inferred by the timing argument. This is quite likely as galaxies accrete mass from their surroundings. 

One possible explanation may be that, in the context of a cosmological simulation, the timing argument overestimates the LG mass \citep{Gonzalez_2014}. However, this trend is not seen in the work of \citet{Partridge_2013}, whose timing argument calculations included the effect of dark energy. In any case, the tension does not appear to be significant. 

The present Galactocentric radial velocity (GRV) of Andromeda provides just one data point. Therefore, it can only be used to constrain one model parameter: the total LG mass. The mass ratio between the MW and M31 can't be constrained in this way, although it is likely on other grounds that $M_{_{MW}} < M_{_{M31}}$ as M31 is larger \citep{Bovy_2013, Courteau_2011} and rotates faster \citep{Carignan_2006, Kafle_2012}.

More importantly, we can't determine if the model itself works with just one data point. As a result, it has been suggested to include more distant LG galaxies in a timing argument analysis \citep{Lynden_Bell_1981}. Such an analysis was attempted a few years later \citep{Sandage_1986}. This work suggested that it was difficult to simultaneously explain all the data then available.

The quality of observational data has improved substantially since that time. More galaxies have also been discovered, providing additional constraints on any model of the LG. This is partly due to wide field surveys such as the Sloan Digital Sky Survey \citep[SDSS,][]{SDSS} and the Pan-Andromeda Archaeological Survey \citep{PANDAS}.

Such surveys have shown that satellite galaxies of the MW are preferentially located in a thin (rms thickness $\sim$25 kpc) co-rotating planar structure \citep{Kroupa_2013}. Known MW satellites were mostly discovered using the SDSS, which has only limited sky coverage. Even when this is taken into account, it is extremely unlikely that the MW satellite system is isotropic \citep{Pawlowski_2016}. In fact, this hypothesis is now ruled out at $>5\sigma$.

A similar pattern is also evident with the satellite galaxies of Andromeda \citep{Ibata_2013}. Roughly half of its satellites are consistent with an isotropic distribution but the other half appears to form a co-rotating planar structure even thinner than that around the MW. However, co-rotation can't be definitively confirmed until proper motions become available.

The observed degree of anisotropy appears very difficult to reconcile with a quiescent origin in a Lambda-Cold Dark Matter ($\Lambda$CDM) universe \citep[][and references therein]{Pawlowski_2014}. This result seems to hold up with more recent higher resolution simulations \citep{Gillet_2015, Pawlowski_2015}. One reason is that filamentary infall is unlikely to work because it leads to radial orbits, inconsistent with observed proper motions of several MW satellites \citep{Angus_2011}.

This result has recently been challenged by \citet{Sawala_2014} and \citet{Sawala_2016} based on the \textsc{eagle} simulations, which include baryonic physics \citep{Schaye_2015, Crain_2015}. When comparing with the observed satellite systems of the MW and M31, these investigations did not take into account all of the available information, in particular the observed distances to the MW satellites. Once this is considered, it becomes clear that the observed distribution of satellites around the MW is very anisotropic, making a quiescent scenario for their origin much less likely \citep{Pawlowski_2015}. Moreover, the inclusion of baryonic physics had very little impact on the extent to which satellite systems are anisotropic. This is what one would expect given the large distances to the MW satellites.

In this context, it seems surprising that a recent investigation found that the observed satellite systems of the MW and M31 are consistent with a quiescent $\Lambda$CDM origin at the 5\% and 9\% levels, respectively \citep{Cautun_2015}. However, this analysis suffers from several problems, in particular not considering several objects orbiting the MW (only its 11 classical satellites are considered). The result for the MW is based on assuming that $\frac{1}{3}$ of the sky is not observable due to the Galactic disc. The actual obscured region is likely smaller, making the observed distribution of MW satellites harder to explain. Some of the more important deficiencies with this investigation have been explained by \citet[][last paragraph of page 2]{Corredoira_2016}.


The MW and M31 are $\sim$ 0.8 Mpc apart now \citep{McConnachie_2012} and have never interacted in $\Lambda$CDM (see Figure \ref{MW_M31_separation_history}). Thus, one might expect their satellite systems to be almost independent in this model. Indeed, it has recently been demonstrated in simulations that the degree of anisotropy of the MW satellite system is not enhanced by the presence of an analogue of M31 \citep{Pawlowski_2014_paired_halos}.

It must be borne in mind that all these authors focused on Local Group satellites merely because they happen to be nearby, allowing for much more accurate measurements of 3D positions and velocities. It is very difficult to conduct similarly detailed investigations further away. Thus, while it may be dangerous to conclude too much about the Universe based on just $\sim$ 50 satellite galaxies, one should at least concede that these are located in two essentially independent systems which were not selected because of their anisotropy.

Although a quiescent origin for these highly anisotropic satellite systems appears unlikely, it is possible that an ancient interaction created them by forming tidal dwarf galaxies \citep[TDGs,][]{Kroupa_2005}. After all, there are several known cases of galaxies forming from material pulled out of interacting progenitor galaxies \citep[e.g. in the Antennae,][]{Mirabel_1992}.

Such TDGs tend to be more metal-rich than primordial galaxies of the same mass \citep[e.g.][]{Croxall_2009}. M31 satellites in the planar system around it seem not to have different chemical abundances to M31 satellites outside this plane \citep{Collins_2015}. This might be a problem for the scenario, had it involved a \emph{recent} interaction. But with a more ancient interaction, the problem seems to be much less severe \citep{Kroupa_2015}. Essentially, this is because gas in the outer parts of the MW/M31 would have been very metal-poor when the interaction occurred. This would lead to TDGs that were initially metal-poor, similar to primordial objects of the same age.

TDGs should be free of dark matter as their escape velocity is much below the virial velocity of their progenitor galaxies \citep{Barnes_1992, Wetzstein_2007}. Thus, a surprising aspect of LG satellite galaxies is their high mass-to-light ($M$/$L$) ratios \citep[e.g.][]{McGaugh_2013}. These ratios are calculated assuming dynamical equilibrium. Tides from the host galaxy are probably not strong enough to invalidate this assumption \citep{McGaugh_2010}. With dark matter unlikely to be present in these systems, the high inferred $M$/$L$ ratios would need to be explained by modified gravity.

One possibility is to use Modified Newtonian Dynamics \citep[MOND,][]{Milgrom_1983}. This imposes an acceleration-dependent modification to the usual Poisson equation of Newtonian gravity \citep{Bekenstein_Milgrom_1984}. Despite having only one free parameter, MOND fares well at explaining rotation curves of disc galaxies \citep[][and references therein]{Famaey_McGaugh_2012}. It also seems to work for LG satellites \citep{McGaugh_2010, McGaugh_2013}, although the relevant observations are challenging.

Applying this theory to the MW and M31, \citet{Zhao_2013} found that they would have undergone an ancient close flyby $\sim$9 billion years (Gyr) ago. The thick disc of the MW would then be a natural outcome of this interaction. Indeed, recent work suggests a tidal origin for the thick disc \citep{Banik_2014}. Moreover, its age seems to be consistent with this scenario \citep{Quillen_2001}.

An ancient flyby of M31 past our Galaxy might have affected the rest of the LG as well. Infalling dwarf galaxies might have been flung out at high speeds by gravitational slingshot encounters with the MW/M31. Material might also have been tidally expelled from within them, perhaps forming a dwarf galaxy later on. As a result, the velocity field of the LG would likely have been dynamically heated. We hope to investigate whether there is any evidence for such a scenario.

To this end, the use of more distant LG galaxies can be particularly useful. Within the context of $\Lambda$CDM, \citet{Jorge_2014} used non-satellite galaxies within $\sim$ 3 Mpc for a timing argument analysis. Satellite galaxies can't easily be used in this way because the velocity field becomes complicated close to the MW or M31 (Figure \ref{LG_Hubble_Diagram}). Intersecting trajectories make it difficult to predict the velocity of a satellite galaxy based solely on its position.

We perform a similar analysis of the same `target' galaxies as in that work. The basic idea is the same: we construct a test particle trajectory that today is at the same position as a target galaxy.\footnote{Our model is effectively two-dimensional, so we used a 2D version of the Newton-Raphson algorithm to achieve this.} The final velocity relative to the MW is then projected onto our line of sight (Equation \ref{Model_GRV}). This model-predicted GRV is corrected for the motion of the Sun with respect to the MW, yielding a heliocentric radial velocity (HRV) prediction which can be compared with observations. When proper motion measurements become available, it will be very interesting to compare the full 3D velocities of LG galaxies with our models.

For simplicity, we assume that the only massive objects in the LG are the MW and M31, which we take to be on a radial orbit. Recent proper motion measurements of M31 indicate only a small tangential motion relative to the MW \citep{M31_motion}. This makes the true orbit almost radial.

The recent work of \citet{Salomon_2016} argues for a high M31 proper motion ($\sim$100 km/s) based on redshift gradients in the M31 satellite system. This measurement is consistent with the more direct measurement of \citet{M31_motion}, though there is some tension. This might be explained by intrinsic rotation of the M31 satellite system. With a field of view of perhaps 5$^\circ$, rotation at only a $\sim$10 km/s level can masquerade as a proper motion of $\sim$100 km/s. In fact, there is strong evidence that nearly half of the M31 satellites rotate coherently around it \citep{Ibata_2013}. Although \citet{Salomon_2016} take this into account to some extent, other rotating satellite planes might also exist around M31. This is suggested by recent investigations into the kinematics of its globular cluster system \citep{Veljanoski_2014}. Moreover, a large tangential velocity between the MW and M31 would show up as a dipole-like feature in the radial velocities of distant LG galaxies. This has been searched for but not found \citep{Jorge_2016}. Thus, we assume the \citet{M31_motion} proper motion measurement is more accurate, making the MW$-$M31 orbit nearly radial.

Starting at some early initial time $t_i$, we evolved forwards a large number of test particles in the gravitational field of the MW and M31. We took the barycentre of the LG at $t = t_i$ as the centre of the expansion. The initial velocities followed a pure Hubble flow (Equation \ref{Initial_conditions}). This is because the Universe was nearly homogeneous at early times $-$ peculiar velocities on the last scattering surface are only $\sim$1 km/s \citep{Planck_2015}, much less than typical values today ($\sim$50 km/s, see Figure \ref{GRV_No_Gravity}).

As both the initial conditions and the gravitational field are axisymmetric, test particles move within meridional planes (i.e. those containing the symmetry axis). This allowed us to use an axisymmetric model. We briefly mention that the gravitational field in our model varies with time, because the MW and M31 move.

A major improvement of our analysis is that the LG is not treated as spherically symmetric. This assumption is not a very good one as the targets considered are at distances of $\sim$1$-$3 Mpc. Meanwhile, the MW$-$M31 separation is $\sim$0.8 Mpc \citep{McConnachie_2012}. This means that the gravitational potential $-$ and thus velocities $-$ are likely to deviate substantially from spherical symmetry in the region of interest. However, we expect only small deviations from axisymmetry for reasons just stated.

Objects outside the LG can have some influence on our results because they can raise tides on the LG. The most important perturbers were identified by \citet{Jorge_2014} as M81, IC 342 and Centaurus A. Their properties are given in Table \ref{Perturbers}. We directly included the gravity of the most massive of these objects, Cen A (Section \ref{Tides_Cen_A}). We took advantage of its location on the sky being almost exactly opposite that of Andromeda. Due to the large distance of Cen A from the LG ($\sim$ 4 Mpc), its velocity is dominated by the Hubble expansion \citep{Karachentsev_2007}. This makes the LG$-$Cen A trajectory almost radial, allowing us to continue using our axisymmetric model.

Our paper is structured as follows: The governing equations and methods are described in Section \ref{Method}. This section also shows some results, to give a rough idea of what happens in our simulations. Comparison of simulation outputs with observations is done in Section \ref{Results}. The posterior probability density functions of all variables and pairs of variables are shown in Figure \ref{Primary_result}. Our results indicate that no model comes close to reproducing all the observations simultaneously.

In Section \ref{Discussion}, we discuss several shortcomings of our model and whether accounting for some of them might help to explain the observations. In Table \ref{Cen_A_effect}, we show how Cen A affects our results. We also estimate carefully the effects of M81 \& IC 342 (Table \ref{IC342_M81_effect}), the Great Attractor (Figure \ref{Great_Attractor_effect}) and the Large Magellanic Cloud (Figure \ref{LMC_trend_s}). These objects seem to little affect GRVs and often worsen the discrepancy with the best-fitting model. We suggest a possible explanation for our results in Section \ref{MW_M31_interaction}. Differences between our approach and the similar study of \citet{Jorge_2014} are described in Section \ref{Comparison_with_Jorge}. Our conclusions are summarized in Section \ref{Conclusions}.

\section{Method}
\label{Method}

The method we follow is to ensure a simulated test particle ends up at the same position as each LG galaxy in our sample (a `target'). At present, only the radial velocities of our targets are available. Thus, the velocity of this particle relative to that of the MW is projected onto the direction towards the particle (Equation \ref{Model_GRV}). This model-predicted GRV is then corrected for solar motion in the MW and compared with observations. The procedure is repeated for different model parameters, which are systematically varied across a grid. Therefore, within the priors we set (Table \ref{Priors}), all model parameter combinations were investigated.

\subsection{Equations of motion}
\label{Equations of motion}

We begin with the metric in the weak field limit
\begin{eqnarray}
	ds^2 = c^2 d\tau^2 ~~~~~~~~~~~~~~~~~~~~~~~~~~~~~~~~~~~~~~~~~~~~~~~~~~~~~~~&&\\
	= \left( 1 + \frac{2 \Phi}{c^2}\right)c^2 dt^2 - \left( 1 - \frac{2 \Phi}{c^2}\right)a^2 \left(d \chi^2 + S^2 \left( \chi \right) d\Omega^2 \right) \\
	S \left( \chi \right) \equiv  
	\left\{
	\begin{array}{ll}
		\sinh \left(  \chi \right)  & \mbox{in an open universe } \\
		\chi & \mbox{in a flat universe } \\
	\end{array} \right. \\
	C \left( \chi \right) \equiv  
	\left\{
	\begin{array}{ll}
		\cosh \left(  \chi \right)  & \mbox{in an open universe } \\
		1 & \mbox{in a flat universe } \\
	\end{array}
\right.\end{eqnarray}

Here, $c$ is the speed of light and $\tau$ is proper time. The scale-factor of the universe is $a$. The spatial part of the metric has been written in spherical polar co-ordinates, with $d\Omega$ representing a change in angle. Using the co-ordinates ${x^0 \equiv t}$, ${x^1 \equiv \chi}$, ${x^2 \equiv \theta}$ and assuming spherical symmetry, we get a diagonal metric where
\begin{eqnarray}
	g_{_{00}} &=& c^2 \left( 1 + \frac{2 \Phi}{c^2} \right) \\
	g_{_{11}} &=& -a^2 \left(1 - \frac{2 \Phi}{c^2} \right) \\
	g_{_{22}} &=& -a^2 \left(1 - \frac{2 \Phi}{c^2} \right) S^2 \left( \chi \right) ~=~ g_{_{11}} S^2 \left( \chi \right)
\end{eqnarray}

As the metric coefficients are independent of $x^2 \equiv \theta$, the geodesic equation tells us that
\begin{eqnarray}
	\overset{.}{x_{_2}} &=& \sum_{b = 0}^3 g_{_{2b}} \overset{.}{x}^b ~~\text{(only non-zero term is } b = 2 \text{)}\\
	&=& a^2 \left(1 - \frac{2 \Phi}{c^2} \right) S^2 \left( \chi \right) {\overset{.}{\theta}} \nonumber \\
	&=& constant
	\label{Angular_momentum_conservation}
\end{eqnarray}

We use $\overset{.}{q}$ to denote the derivative of any quantity $q$ with respect to proper time. In weak gravitational fields ($\Phi \ll c^2$), proper and co-ordinate time are almost equal, making $\tau \approx t$. Bearing this in mind, Equation \ref{Angular_momentum_conservation} tells us that the specific angular momentum of a test particle is conserved. This is due to the spherical symmetry of the situation.

For the radial component of the motion, we use the geodesic equation in the form
\begin{eqnarray}
	\label{Geodesic_equation}
	\overset{..}{x}^1 + \sum_{b = 0}^3 \sum_{c = 0}^3 {\Gamma^1}_{bc} \overset{.}{x}^b \overset{.}{x}^c ~=~ 0 ~\text{~where}\\
	{\Gamma^a}_{bc} ~=~ \frac{1}{2} \sum_{d = 0}^3 g^{ad} \left( \partial_{_b} g_{_{dc}}  +  \partial_{_c} g_{_{bd}}  +  \partial_{_d} g_{_{bc}}\right) 
\end{eqnarray}

Here, we use the notation $\partial_{_b} q \equiv \frac{\partial q}{\partial x^b}$ for any quantity $q$. The non-zero Christoffel symbols relevant to a non-relativistic test particle in this situation are
\begin{eqnarray}
	{\Gamma^1}_{00} &\approx & \frac{\Phi'}{a^2} \\
	{\Gamma^1}_{01} &\approx & \frac{\overset{.}{a}}{a} \equiv H \\
	{\Gamma^1}_{22} &\approx & -S \left( \chi \right) C \left( \chi \right) \\
	{\Gamma^1}_{11} &\approx & -\frac{\Phi'}{c^2}
\end{eqnarray}

Here, $q'$ implies a partial derivative with respect to the co-moving co-ordinate $\chi$ rather than physical distance $a \chi$. Putting in the non-negligible Christoffel symbols\footnote{The ${\Gamma^1}_{11}$ term effectively causes $\Phi' \to \Phi' \left( 1 - \frac{v^2}{c^2} \right)$ where $v$ is the speed of the particle with respect to a co-moving observer at the same place. This leads to a special relativistic correction which makes it difficult for a potential gradient to accelerate a particle if its speed is close to that of light. For non-relativistic particles, the effect of this term is negligible because $v \ll c$.} into Equation \ref{Geodesic_equation}, we get that
\begin{eqnarray}
	\overset{..}{\chi} + \frac{\Phi'}{a^2} + 2H  \overset{.}{\chi} - S \left( \chi \right) C \left( \chi \right) {\overset{.}{\theta}}^2 ~=~ 0
\end{eqnarray}


In terms of physical co-ordinates $r \equiv a \chi$, this becomes
\begin{eqnarray}
	\overset{..}{r} &=& \left(\frac{\overset{..}{a}}{a}\chi + 2H\overset{.}{\chi} + \overset{..}{\chi}\right) a\\ 
	&=& \left(\frac{\overset{..}{a}}{a}\chi - \frac{1}{a^2}\frac{\partial \Phi}{\partial \chi} + S \left( \chi \right) C \left( \chi \right) {\overset{.}{\theta}}^2 \right) a \\
	&=& \frac{\overset{..}{a}}{a} r - \frac{\partial \Phi}{\partial r} + S \left( \chi \right) C \left( \chi \right) a {\overset{.}{\theta}}^2
	\label{Equation_25}
\end{eqnarray}

The real Universe is close to spatially flat \citep{Planck_2015}. Thus, from now on, we will only consider the case of a flat universe. This is defined as one having a density $\rho$ equal to the critical density $\rho_{crit}$, if we count both matter and dark energy towards the total density.
\begin{eqnarray}
	\rho = \rho_{crit} \equiv \frac{3 H^2}{8 \rm{\pi} G}
	\label{Critical_density}
\end{eqnarray}

Equation \ref{Critical_density} is valid at all times, although both $\rho$ and $H$ vary with time. In such a universe, Equation \ref{Equation_25} becomes
\begin{eqnarray}
	\label{Equation_26}	
	\overset{..}{r} &=& \frac{\overset{..}{a}}{a}r - \frac{\partial \Phi}{\partial r} + r{\overset{.}{\theta}}^2
\end{eqnarray}

This looks very similar to the Newtonian equation of motion. The last term corresponds to the centrifugal force while the $\frac{\partial \Phi}{\partial r}$ term corresponds to the potential gradient. The only novel aspect is the term $\frac{\overset{..}{a}}{a}r$. The importance of this term becomes clear if we consider a homogeneous universe, meaning that $\Phi = 0$ everywhere and at all times. In this case, we expect the distance between two non-interacting test particles to behave as $r \propto a \left( t \right)$ (i.e. their co-moving distance is constant). This implies that $\overset{..}{r} = \frac{\overset{..}{a}}{a}r$. This term has also been called Hubble drag because it tends to reduce the magnitude of peculiar velocities.\footnote{If the Universe were contracting, then this term would be a forcing to peculiar velocities rather than a drag upon them.} In the absence of potential gradients, we would get $v_{pec} \propto \frac{1}{a \left( t \right)}$, where the peculiar velocity is defined by
\begin{eqnarray}
	\bm{v_{pec}} ~\equiv~ \overset{.}{\bm{r}} - H \bm{r}
\end{eqnarray}

In general, the Universe is neither homogeneous nor spherically symmetric. For such circumstances, we suppose that the generalization of Equation \ref{Equation_26} is given by
\begin{eqnarray}
	\overset{..}{\bm{r}} &=& \frac{\overset{..}{a}}{a}\bm{r} - \bm{\nabla} \Phi
	\label{Vector_equation_of_motion}
\end{eqnarray}

With the equations of motion in hand, we now need to relate the potential $\Phi$ to the density perturbations that act as its source. To do this, we use the 00 component of the field equation of General Relativity.
\begin{eqnarray}
	\label{Equation_30}
	R_{_{ab}} &=& -\frac{8 \rm{\pi} G}{c^4} \left( T_{_{ab}} - \frac{1}{2}T g_{_{ab}} \right) ~~\text{ where}\\
	T &\equiv & \sum_{d = 0}^3 {T^d}_d
\end{eqnarray}

Here, $T_{_{ab}}$ is the energy-momentum tensor while $R_{_{ab}}$ is the Ricci tensor, related to the curvature of the metric. Perturbations to the solution for a homogeneous universe must satisfy the equation
\begin{eqnarray}
	\Delta R_{_{00}} &=& -\frac{8 \rm{\pi} G}{c^4} \Delta \left( T_{_{00}} - \frac{1}{2}T g_{_{00}} \right)
\end{eqnarray}

In this case, for non-relativistic sources which are almost pressureless (like baryons and cold dark matter), we get that
\begin{eqnarray}
	R_{_{00}} \approx - \sum_{i = 1}^3 \frac{\partial_{_i}  \left( \partial_{_i} \Phi \right)}{a^2 c^2} ~~\text{(sum over spatial indices only)}
	\label{Equation_34}
\end{eqnarray}

The stress-energy tensor takes on a particularly simple form: its only non-zero element is $T_{_{00}} \approx \rho c^2$. Thus,
\begin{eqnarray}
	g_{_{00}} T ~\approx ~ T_{_{00}}  ~\approx ~ \rho c^2
	\label{Equation_35}
\end{eqnarray}

Using Equations \ref{Equation_34} and \ref{Equation_35} in Equation \ref{Equation_30}, we get that
\begin{eqnarray}
	{\nabla}^2 \Phi &=& 4 \rm{\pi} G \Delta \rho
	\label{Poisson_equation_comoving}
\end{eqnarray}


Here, ${\nabla}^2 \Phi$ is the Laplacian of $\Phi$ with respect to physical co-ordinates. In spherical symmetry, it is
\begin{eqnarray}
	{\nabla}^2 \Phi &=& \frac{1}{r^2} \frac{\partial}{\partial r} \left( r^2 \frac{\partial \Phi}{\partial r} \right)
\end{eqnarray}

Equation \ref{Poisson_equation_comoving} is very similar to the usual Poisson equation of Newtonian gravity. Note, however, that only deviations from the background density act as a source for $\Phi$ (i.e. it is sourced by $\Delta \rho$ rather than $\rho$). 




\subsection{Simulations}
\label{Simulations}

\subsubsection{Including the Milky Way and Andromeda}

The LG is assumed to consist of two point masses (the MW and M31) plus a uniform distribution of matter at the same density as the cosmic mean value $\overline{\rho}$ (see Section \ref{Reduced_LG_Mass} for further discussion of this point). Our simulations start when the scale factor of the Universe $a = a_{_i}$. We used $a_{_i} = 0.1$, though our results change negligibly if $a_{_i} = \frac{1}{15}$ instead (see Section \ref{MW_M31_interaction}).

The initial separation of the MW and M31 $d_i$ is varied to match their presently observed separation $d_0$ using a Newton-Raphson technique. Note that altering $d_i$ alters their initial velocities because the galaxies are assumed to have zero peculiar velocity at the start of the simulation ($t = t_i$). Thus, their final attained separation $d_f$ depends strongly on $d_i$.

The MW$-$M31 orbit is taken to be radial, a reasonable assumption given their small tangential motion \citep[$\sim$17 km/s compared to a radial velocity of $\sim$110 km/s,][]{M31_motion}. This makes the gravitational field in the LG axisymmetric. As the initial conditions are spherically symmetric (Equation \ref{Initial_conditions}), a 2D model is sufficient for this investigation.

Applying Equation \ref{Vector_equation_of_motion} to a radial orbit, the distance $d$ between the galaxies satisfies
\begin{eqnarray}
	\label{Separation_history_equation}
	\overset{..}{d} &=& -\frac{GM}{d^2} + \frac{\overset{..}{a}}{a} d	\\
	\overset{.}{d} &=& H_{_i} d_i ~~~\text{ initially }
\end{eqnarray}

\begin{figure}
	\centering 
		\includegraphics [width = 8.5cm] {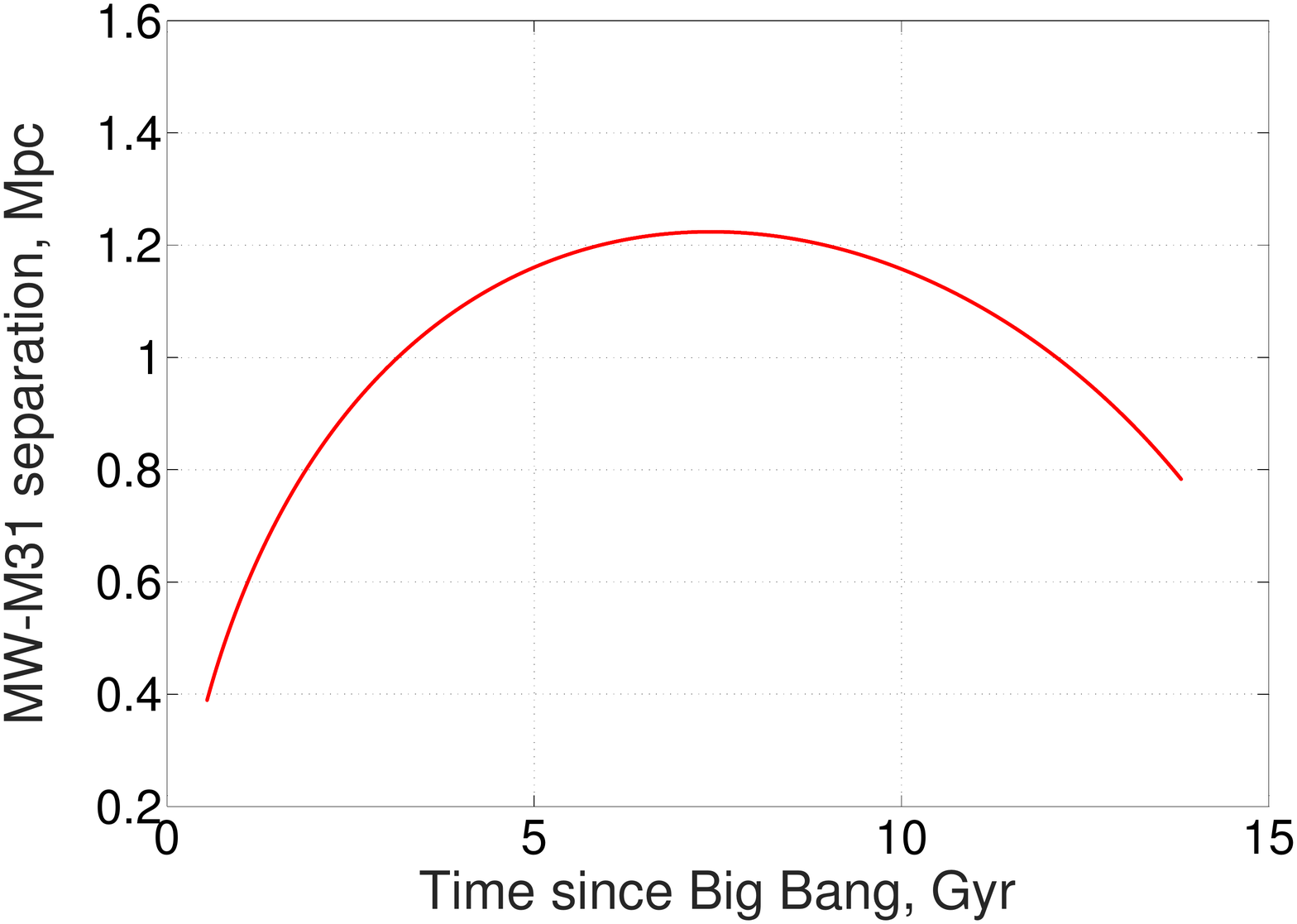}
	\caption{MW$-$M31 separation $d(t)$ for a typical case where $q_{_1} = 0.2$ and $M_i = 3.4 \times {10}^{12} M_\odot$. $d(t)$ always looks broadly similar $-$ in $\Lambda$CDM, the MW and M31 have never approached each other closely for any plausible model parameters.}
	\label{MW_M31_separation_history}
\end{figure}

$H_{_i}$ is the value of the Hubble constant $\frac{\overset{.}{a}}{a}$ when $t = t_i$, while $d_i$ is the MW$-$M31 separation at that time. $M$ is the combined mass of the MW \& M31. It can be verified straightforwardly that when $M = 0$ (i.e. non-interacting test particles), we recover $d \propto a$. In this case, the galaxies trace the cosmic expansion but don't influence each other. 

Equation \ref{Separation_history_equation} implicitly assumes that the MW and M31 are surrounded by a distribution of matter with the same density as the cosmic mean value $\overline{\rho}$. This point is discussed more thoroughly in Section \ref{Reduced_LG_Mass} where we also redo our entire analysis assuming instead that the surroundings of the MW and M31 are empty.

We use a standard flat\footnote{$\Omega_{m,0} + \Omega_{\Lambda, 0} = 1$} dark energy-dominated cosmology with parameters given in Table \ref{Priors}. Therefore,
\begin{eqnarray}
	\frac{\overset{..}{a}}{a} &=& -\frac{4 \rm{\pi} G}{3} \left( \rho_m - 2 \rho_\Lambda \right) \\
	&=& {H_{_0}}^2 \left(- \frac{1}{2} \Omega_{m,0}~a^{-3} + \Omega_{\Lambda, 0}  \right)
	\label{Friedmann_equation}
\end{eqnarray}

Defining time $t$ to start when ${a = 0}$ and requiring that ${\overset{.}{a} = H_{_0}}$ when ${a = 1}$ (the present time), we get that
\begin{eqnarray}
	a(t) &=& {{\left( \frac{{{\Omega }_{m,0}}}{{{\Omega }_{\Lambda ,0}}} \right)}^{\frac{1}{3}}}{{\sinh }^{\frac{2}{3}}}\left( \frac{3}{2}\sqrt{{{\Omega }_{\Lambda ,0}}}{{H}_{0}}t \right)
\label{Expansion_history}
\end{eqnarray}

The present values of $H_{_0}$ and $\Omega_{m,0}$ uniquely determine the present age of the Universe $t_f$ via inversion of Equation \ref{Expansion_history} to solve for when $a = 1$. We also use it to determine when $a = a_{_i}$, thereby fixing the start time of our simulations.

The timing argument is particularly sensitive to late times (Figure \ref{Impulsed_trajectories}). This makes it important to correctly account for the late-time effect of dark energy. Because this tends to increase radial velocities of LG galaxies, one is forced to increase the mass of the LG to bring their predicted radial velocities back down to the observed values. As a result, the inclusion of dark energy in timing argument analyses of the LG increases its inferred mass by a non-negligible amount \citep{Partridge_2013}.


Once we obtained a trajectory that (very nearly) satisfied $d_f \equiv d \left( t_f \right) = d_0$, we had the ability to find the gravitational field everywhere in the Local Group at all times. A large number of test particles were then evolved forwards, all starting on a pure Hubble flow with the centre of expansion at the barycentre of the Local Group.
\begin{eqnarray}
	\bm{v_{_i}} = H_{_i} \bm{r}_{_i}
	\label{Initial_conditions}
\end{eqnarray}

Note that Equation \ref{Initial_conditions} also applies to the MW and M31, which we model as point masses. A point mass approximation should work for determining $d(t)$ as Andromeda never gets very close to the Milky Way (Figure \ref{MW_M31_separation_history}). However, it is not good for handling close encounters of test particles with either galaxy. Thus, we adjust the forces they exert on test particles to be $\propto \frac{1}{r}$ at low $r$ (i.e. close to the attracting body). This is for consistency with the observed flat rotation curves of the MW and M31. To recover $g \propto \frac{1}{r^2}$ at large $r$, we set the gravity towards each galaxy to be
\begin{eqnarray}
	g &=& \frac{GM}{r^2} \frac{b}{\sqrt{1 + b^2}} \label{Force_balance} ~~\text{ where}\\
	b &\equiv& \frac{r}{r_{_S}}
\end{eqnarray}

$r_{_S}$ is chosen so that the force at $r \ll r_{_S}$ leads to the correct flatline level of rotation curve for each galaxy, i.e. $r_{_S} = \frac{GM}{{v_{_f}}^2}$. For the MW, we take $v_{_f} = 180$ km/s \citep{Kafle_2012} while for Andromeda, we take $v_{_f} = 225$ km/s \citep{Carignan_2006}.

Combining Equation \ref{Force_balance} with the cosmological acceleration term, the equation of motion for our test particles is
\begin{eqnarray}
	{\overset{..}{\bm r} } ~~= ~~~{\frac{\overset{..}{a}}{a}}  {\bm r} ~~- \sum_{j=\text{MW, M31}}  { \frac{G M_j \left( \bm r - \bm r_{_j} \right)} {{\left( |\bm r - \bm r_{_j} |^2 + {r_{_{S,j}}}^2 \right)}^{\frac{1}{2}} |{\bm r}- \bm r_{_j} |^2}}
	\label{Equation_of_motion}
\end{eqnarray}

Some trajectories go very close to the MW or M31. Approaches within a distance of $r_{acc}$ (given in Table \ref{Priors}) are handled by terminating the trajectory and assuming the particle was accreted by the nearby galaxy. This causes the mass of that galaxy to increase.

\begin{figure}
	\centering
		\includegraphics [width = 8.5cm] {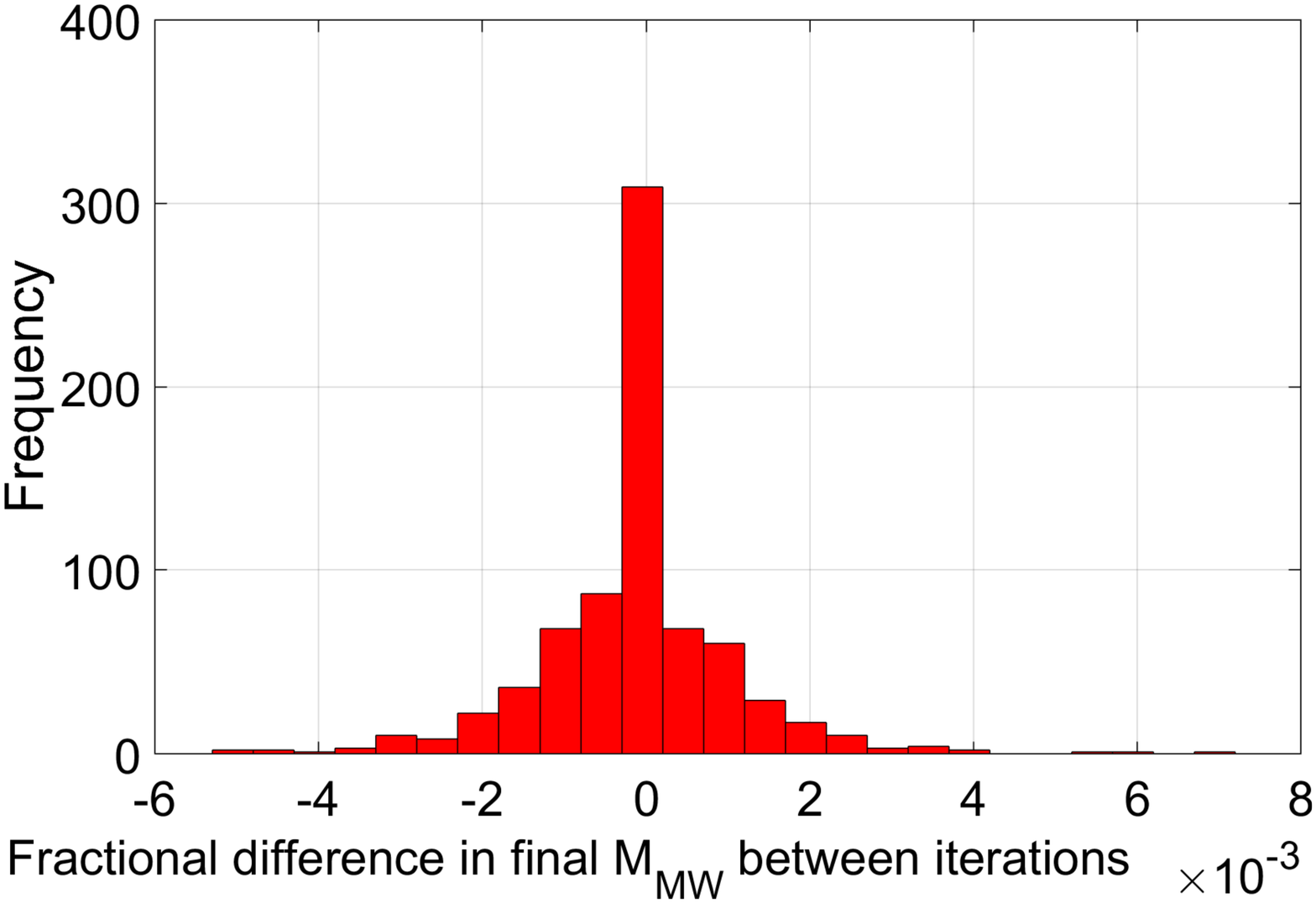}
	\caption{Fractional difference between the final attained MW mass on the first and second runs of each simulation. The very small values show that our solution for $M_{_{MW}}(t)$ converged well. Simulations without Centaurus A give similar results.}
	\label{Mass_convergence_Cen_A}
\end{figure}

As we solved the test particle trajectories sequentially, it wasn't possible until the very end to have the mass histories $M_{_{MW}}(t)$ and $M_{_{M31}}(t)$. Thus, we assumed constant masses for the force calculations. We then repeated the process, using the previously stored mass histories for each galaxy. This meant that the initial MW$-$M31 separation $d_i$ also had to be adjusted. In this way, we found that the final mass had converged fairly well with just two iterations (Figure \ref{Mass_convergence_Cen_A}).


The changing mass of the MW and M31 meant that one could not trivially convert the separation history $d(t)$ into MW and M31 positions ($y_{_{MW}}(t)$ and $y_{_{M31}}(t)$, respectively). However, the instantaneous acceleration of the MW must be due to the gravity of Andromeda.\footnote{This is not strictly true at early times due to the $\frac{\overset{..}{a}}{a}y$ term, but we do not expect either galaxy to have accreted much mass at that stage because no test particle starts very close to the MW or M31. Without mass accretion, the ratio of this term between the two galaxies is also inverse to that of their masses.} Thus, the magnitude of this acceleration must be a fraction $\frac{M_{_{M31}}}{M_{_{MW}} + M_{_{M31}}}$ of the total mutual acceleration. This means that
\begin{eqnarray}
	{\overset{..}{y}}_{_{MW}}(t) &=& \frac{M_{_{M31}}(t)}{M_{_{MW}}(t) + M_{_{M31}}(t)} ~\overset{..}{d} \\
	{\overset{..}{y}}_{_{M31}}(t) &=& -~\frac{M_{_{MW}}(t)}{M_{_{MW}}(t) + M_{_{M31}}(t)} ~\overset{..}{d}
	\label{MW_M31_Motion}
\end{eqnarray}

In practice, we solved Equation \ref{Separation_history_equation} to determine $d(t)$. We found the change in separation over each time timestep $d(t + \Delta t) - d(t)$ and apportioned this to the MW and M31 in inverse proportion to their masses at $t + \frac{1}{2} \Delta t$.




Our equations are referred to the frame of reference in which the origin corresponds to the initial centre of mass position (considering only the MW \& M31). This makes our reference frame inertial. We do not keep track of how the centre of mass moves after our simulations start.


The initial masses of the MW and M31 imply that they must have accreted material in some region prior to the start of our simulation. Thus, we do not allow test particles to start within a certain excluded region. This is defined by an equipotential $U_{exc}$, chosen so as to enclose the correct total volume (i.e. $V_{exc}~\rho_{_{M,i}} = M_i$, the initial LG mass). The density of matter $\rho_{_{M,i}}$ at the initial time $t_i$ includes contributions from both baryonic and dark matter. For most parameters, the resulting excluded region is a single region encompassing both the MW and M31 rather than distinct regions around each galaxy.

The potential resulting from integrating Equation \ref{Force_balance} is
\begin{eqnarray}
	~~~~~~~~~~U ~~=~ \sum_{j = MW,M31}{{\frac{GM}{r_{_{S,j}}} Ln\left( \frac{\sqrt{1 + {b_j}^2} - 1}{b_j} \right)}}
	\label{U}
\end{eqnarray}

\begin{figure}
 \centering 
  \includegraphics [width = 8.5cm] {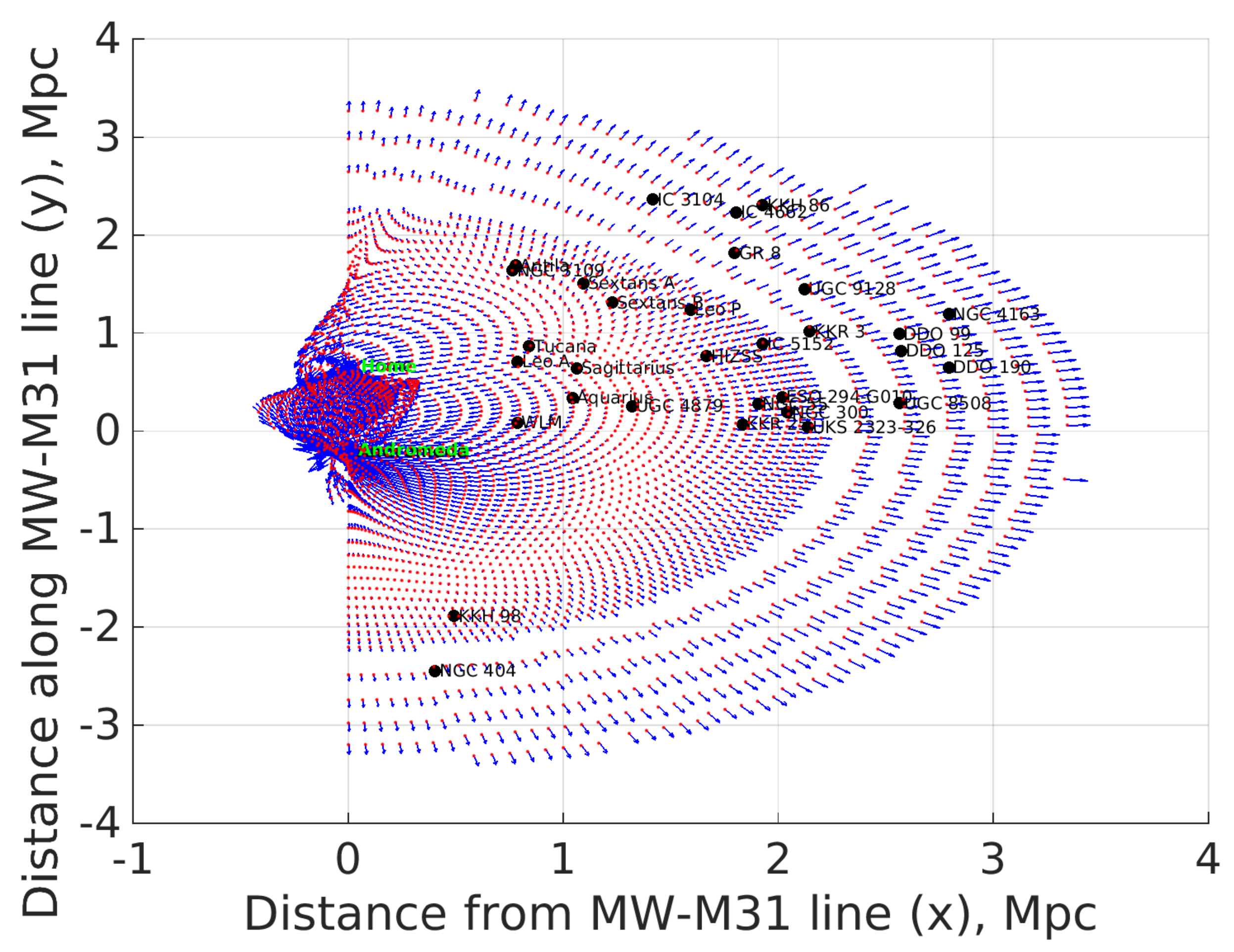}
  \includegraphics [width = 8.5cm] {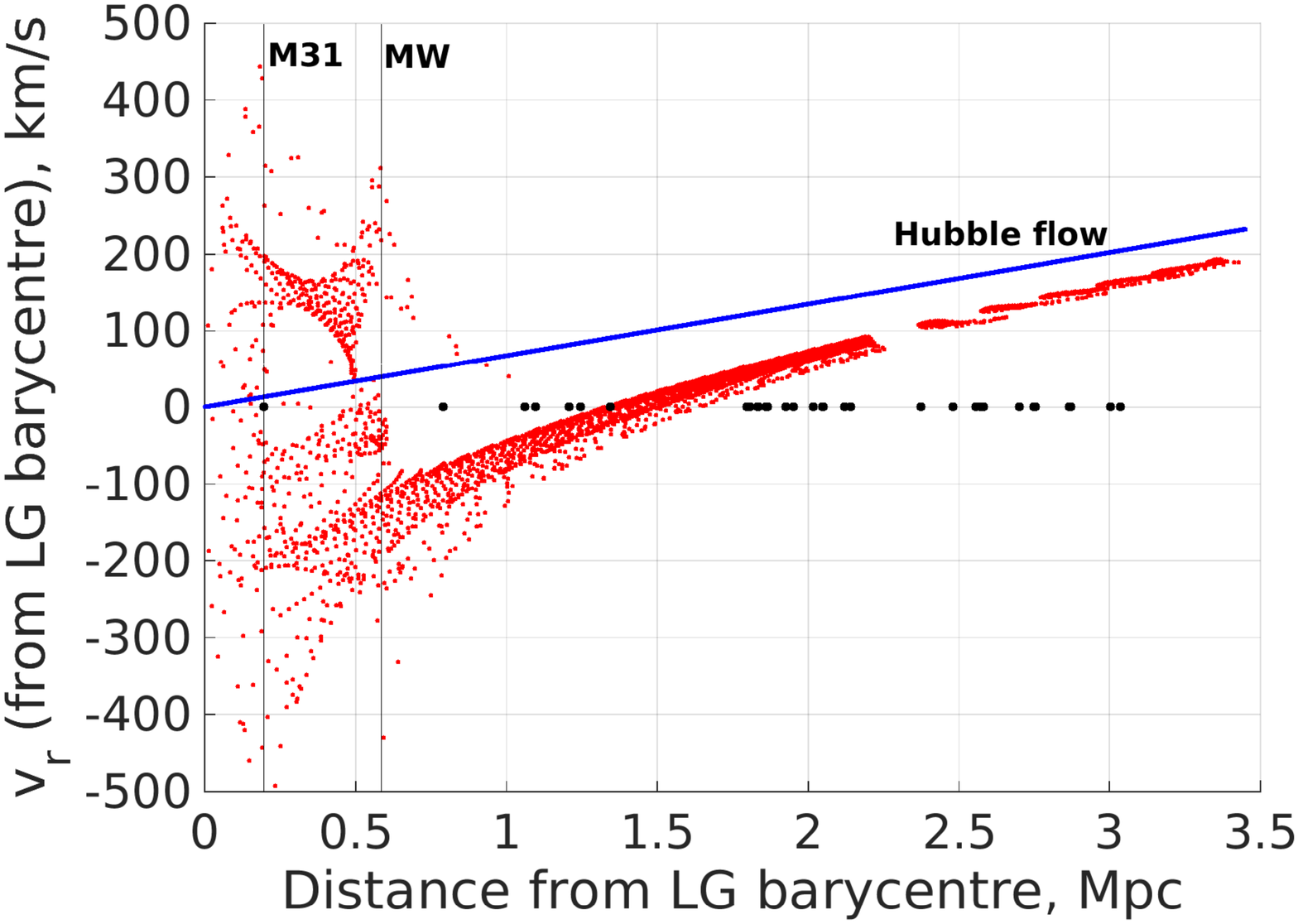}
 \caption{\emph{Top:} Local Group velocity field for the case $q_{_1} = 0.3$, $M_i = 4 \times {10}^{12} M_\odot$ and no Centaurus A. Locations of indicated galaxies are shown. The MW is just above the origin. Only particles starting at $x \geq 0$ (and thus $v_x \geq 0$) were considered. Thus, the presence of particles at $x < 0$ indicates intersecting trajectories and a disturbed velocity field. \emph{Bottom:} Radial velocities of test particles with respect to the LG barycentre. Vertical lines represent the distances of M31 and the MW from there. Increased velocity dispersion near these galaxies is apparent. Black dots on the $x$-axis show the distances of target galaxies from the LG barycentre. Without proper motions, observations can't be put on such a Hubble diagram as the MW is not at the LG barycentre.}
 \label{LG_Hubble_Diagram}
\end{figure}

We start our test particles on a grid of plane polar co-ordinates. At some particular angle $\theta$, we consider a sequence of trajectories which start further and further out. Trajectories are skipped if they start within the `exclusion zone' ($U < U_{exc}$ at $t = t_i$). Once we obtain a trajectory that finishes further than 2.15 Mpc from the LG barycentre, we skip 3 out of every 4 steps as the velocity field is fairly smooth at such large distances (Figure \ref{LG_Hubble_Diagram}). Once we reach beyond 3.2 Mpc, we move on to the next value of $\theta$. This is because we do not need the velocity field further than ${\sim 3}$ Mpc from the LG as there are no target galaxies further away.\footnote{This requires trajectories starting out to distances of $\sim$0.5 Mpc.}

We use a fourth-order Runge-Kutta algorithm with an adaptive timestep designed to be 30$-$70 times shorter than the instantaneous dynamical time $t_{dyn}$. This is estimated by dividing the distance to each galaxy by the speed of a test particle with zero total energy, ignoring the presence of the other galaxy. Faced with two estimates of $t_{dyn}$, we use the shorter one in order to maximize the resolution.

The worst time resolution we use is $\frac{1}{1000}$ of the total duration ($\sim$13.5 Gyr). This was sufficient for distances $\sga r_{_S}$ from each galaxy. At smaller distances, we found that $t_{dyn} \propto r^{\frac{5}{4}}$ is a good approximation. If required, we improve the time resolution in powers of 2 up to a maximum of 5 times (for a $2^5 = 32\times$ reduction in $\Delta t$). This should provide adequate resolution for distances from the MW and M31 greater than their respective `accretion radii' $r_{acc}$, which we chose to be a few disc scale-lengths (Table \ref{Priors}).

\subsubsection{Including Centaurus A}
\label{Tides_Cen_A}

Although none of our target galaxies are too close to any of the perturbers listed in Table \ref{Perturbers} \citep[due to pre-selection by][]{Jorge_2014}, we were still concerned that their gravity might have noticeably affected our target galaxies. To test this scenario, we decided to directly include the most massive perturber, Centaurus A.

Due to the large distance of Cen A from the LG (${\sim 4 \text{ Mpc}}$), any peculiar velocity it has is likely to be much smaller than its radial velocity. Indeed, this is borne out observationally for motion along our line of sight \citep{Karachentsev_2007}. As a result, Cen A is probably on an almost radial orbit with respect to the LG barycentre. Fortunately, Cen A is currently located almost directly opposite M31 on our sky ($\cos \theta = -0.99$, where $\theta$ is the angle on the sky between M31 and Cen A). This allowed us to continue using our axisymmetric model.


To initialize each simulation, we need trajectories for the MW, M31 and Cen A that match the presently observed distances to M31 and Cen A. This is done using a 2D Newton-Raphson algorithm\footnote{For stability, we under-relaxed the algorithm, meaning that in each iteration, we altered the parameters by 80\% of what the algorithm would normally have altered them by} on the initial relative positions of all three galaxies along a line. As before, initial velocities were found using Equation \ref{Initial_conditions}.

Test particle trajectories were then solved in the usual way, with the grid of initial positions centred on the initial barycentre of the MW \& M31 as before. Including Cen A, Equation \ref{Equation_of_motion} becomes
\begin{eqnarray}
	{\overset{..}{\bm r} } ~=~ {\frac{\overset{..}{a}}{a}}  {\bm r} - \sum_{\begin{array}{c} \text{j = MW,}\\ \text{M31, Cen A}\end{array}}{ \frac{G M_j \left( \bm r - \bm r_{_j} \right)} {{\left( |\bm r - \bm r_{_j} |^2 + {r_{_{S,j}}}^2 \right)}^{\frac{1}{2}} |{\bm r}- \bm r_{_j} |^2}}
	\label{Equation_of_motion_Cen_A}
\end{eqnarray}

For simplicity, we keep the mass of Cen A fixed at ${4 \times 10^{12} M_\odot}$ \citep{Karachentsev_2005} but still allow the MW and M31 to accrete mass.

\begin{table}
 \begin{tabular}{lllll}
	\hline
  Name & $b$ & $l$ & $d_{MW}$ & $M$ \\
  \hline
  Centaurus A & $19.4173^\circ$ & $309.5159^\circ$ & 3.8 & 4 \\
  M81 & $40.9001^\circ$ & $142.0918^\circ$ & 3.6 & 1.03 \\
  IC 342 & $10.5799^\circ$ & $138.1726^\circ$ & 3.45 & 1.76 \\
  \hline
 \end{tabular} 
 \caption{Properties of mass concentrations outside the Local Group which we considered. Distances are in Mpc and sky positions are in the Galactic system. The estimate for Cen A is from \citet{Harris_2010}, that of M81 is from \citet{Gerke_2011} and that of IC 342 is from \citet{Wu_2014}. Masses are in units of $10^{12} M_\odot$ and were obtained from \citet{Karachentsev_2005} for Cen A and IC 342. For M81, we used \citet{Karachentsev_2006}.}
 \label{Perturbers}
\end{table}

\subsection{Observations \& Sample Selection}
\label{Observations}

Our dataset comes mostly from the catalogue of LG galaxies compiled by \citet{McConnachie_2012}. We used the subset of these that were used for a timing argument analysis by \citet{Jorge_2014}. This implicitly applies a number of criteria. The basic idea behind them was to ensure that gravity from the MW and M31 dominates over gravity from anything else. For this reason, targets $\ga$ 3 Mpc from the LG were not considered. The authors also avoided galaxies too close to any major mass concentrations outside the LG. The perturbers they considered are listed in Table \ref{Perturbers}.


Very close to the MW or M31, there are crossing trajectories and so the model-predicted velocity in such regions is not well-defined (top panel of Figure \ref{LG_Hubble_Diagram}). Further away, this issue does not arise. Thus, \citet{Jorge_2014} restricted their sample to \emph{non-satellite} galaxies. We further restricted their sample by excluding Andromeda XVIII as it is in the disturbed region around M31, even if it is unbound.

We treated HIZSS 3A \& B as one object as they are almost certainly a binary system. Naturally, we used the velocity of its centre of mass, assuming a mass ratio of 13:1 \citep{Begum_2005}. To allow for uncertainty in this ratio, we inflated the error on the heliocentric radial velocity (HRV) to 3.5 km/s. This decision turns out not to matter very much because the uncertainty in its distance has a much larger effect than uncertainty in its HRV (this is true for most of our targets $-$ distances are harder to measure accurately).





In regions close to the MW and M31, the presence of crossing trajectories makes it impossible to uniquely predict the velocity of a target galaxy based solely on its position (Figure \ref{LG_Hubble_Diagram}). In such cases, we should reject the target (i.e. not use it in our analysis). In practice, we accepted all of our targets in all cases. We checked the velocity field to ensure none of our target galaxies fell in regions with crossing trajectories. Although none of them did so, NGC 3109 and Antila came close. We tried raising $H_{_0}$ and altering the distances to these galaxies within their uncertainties, but we still could not get them in a region of crossing trajectories. In any case, excluding them would not much affect our conclusions, as will become apparent later (Figure \ref{Tide_correlation}). As a final check, IB looked at all the $\sigma_{pos}$ and GRV maps (like those in Figure \ref{GRV_Sextans_A}) and confirmed that they were smooth.

If we had been less fortunate regarding the locations of our target galaxies, then we might have rejected some of them in some simulations using criteria designed to search for intersecting trajectories. The best options seem to be a high density of test particles near the present position of the target and a high velocity dispersion at that position. In this case, we might have to alter Equation \ref{Chi_sq_total} by multiplying the first term on the right by $\frac{32}{N}$, where $N$ is the number of target galaxies `accepted' by the algorithm. Additional care would have to be taken to ensure the analysis remained valid despite $N$ varying with the model parameters (i.e. some models might be constrained using fewer observations than others).\footnote{If a target galaxy is problematic in only some parts of parameter space, then one can simply avoid including it in the analysis altogether, thereby avoiding issues due to $N$ being model-dependent. However, this makes poorer use of the available information.}


To convert observations into the same co-ordinates as our simulations, we first defined Cartesian $xy$ co-ordinates centred on the LG barycentre, with $\widehat{\bm y}$ towards the MW. The positions of observed galaxies were converted into this system using the equations
\begin{eqnarray}
 ~~x &=& d_{_{MW}} ~|\hat{\bm d}_{_{MW}} \times \hat{\bm r}_{_{MW}}| \\
 y_{rel} ~\equiv ~ y - y_{_{MW}} &=&  d_{_{MW}} \left( \hat{\bm d}_{_{MW}} \cdot \hat{\bm r}_{_{MW}} \right)
 \label{y_rel}
\end{eqnarray}

$y_{_{MW}}$ is the present distance of the MW from the initial position of the LG barycentre. $\widehat{\bm{y}} \equiv \hat{\bm r}_{_{MW}}$ is the direction from M31 towards the MW. This is just the opposite of the direction in which we observe Andromeda. $d_{_{MW}}$ is the distance from the MW to the target galaxy. This is essentially equivalent to its heliocentric distance. We neglected the difference that arises because the Sun is not at the centre of the Milky Way.\footnote{Target galaxies are $\ga$800 kpc away while the Sun is only $\sim$8kpc from the Galactic Centre, well below typical distance errors.} For this reason, we can approximate the direction between the MW centre and the target galaxy $\hat{\bm d}_{_{MW}}$ as the direction in which we observe it.

Although the position of the Sun with respect to the Galactic Centre is unimportant for this work, its \emph{velocity} relative to the MW is very important because this velocity is $\sim$250 km/s \citep{McMillan_2011}. For observational reasons, we split this velocity into two components. The MW is a disc galaxy, so most of the Sun's velocity is just ordered circular motion within the disc plane. In the absence of non-circular motions, its speed would be $v_{c, \odot}$. This is known as the Local Standard of Rest (LSR) because particles moving tangentially at this speed would be at rest in a rotating reference frame.

We temporarily define a 3D Cartesian co-ordinate system with $\hat{\bm x}$ pointing from the Sun towards the Galactic Centre, $\hat{\bm z}$ pointing towards the North Galactic Pole and $\hat{\bm y}$ chosen so as to make the system right-handed. Fortunately, $\hat{\bm y}$ points along the direction of rotation. In this system, the velocity of the Sun with respect to the Milky Way (including its non-circular motion) is
\begin{eqnarray}
 \bm v_\odot = \begin{bmatrix}
 U_\odot \\
 V_\odot + v_{c, \odot} \\
 W_\odot
 \end{bmatrix}
 \label{Solar_velocity}
\end{eqnarray}

The direction towards another galaxy can be determined from its Galactic co-ordinates using
\begin{eqnarray}
 \hat{\bm d}_{_{MW}} = \begin{bmatrix}
 \cos b ~ \cos l \\
 \cos b ~ \sin l \\
 \sin b
 \end{bmatrix}
\end{eqnarray}

where $b$ is the Galactic latitude and $l$ is the Galactic longitude, whose zero point is the direction towards the Galactic Centre. Galactic co-ordinates are actually heliocentric, though the distinction is unimportant for very distant objects.

Without proper motions of LG galaxies, only their GRV can be constrained. Thus, we project the velocity of the Sun with respect to the MW onto the direction towards the desired galaxy. This is then added onto its observed heliocentric radial velocity.
\begin{eqnarray}
 GRV_{obs} ~=~ HRV_{obs} + \bm v_\odot \cdot \hat{\bm d}_{_{MW}}
 \label{GRV_obs}
\end{eqnarray}

This estimate of $GRV_{obs}$ is dependent on the model used for $\bm v_\odot$, in particular the adopted LSR speed. Thus, for a range of plausible values of $v_{c, \odot}$, we stored the resulting values of $\bm v_\odot \cdot \hat{\bm d}_{_{MW}}$ for each target galaxy. This quantity is the difference between its GRV and its HRV.


\subsection{Comparing simulations with observations}
\label{Simulation_observation_comparison}

Our simulations yield a velocity field for the LG. To determine the model-predicted GRV of an observed galaxy, we need a test particle landing at exactly the same position. To achieve this, we started with whichever test particle landed closest to the targeted final position. We then used a 2D Newton-Raphson algorithm on the initial position of this particle. The dependence of its final position on its initial one was found using finite differencing. For this, we used trajectories starting at $\left(x_{_0}, y_{_0} \right)$, $\left(x_{_0} + dx_{_0}, y_{_0} \right)$ and $\left(x_{_0}, y_{_0} + dy_{_0} \right)$, with $dx_{_0} = dy_{_0} = $ 307 pc. Note that we have reverted to the usual $xy$ co-ordinates, with $\hat{\bm y}$ pointing from M31 towards the MW and $\hat{\bm x}$ orthogonal to this direction.

We considered the Newton-Raphson algorithm to have converged once the error in the final position $\left( x, y \right)$ was below 0.001\% of the distance between the target and the LG barycentre. The final velocity of this trajectory $\left( v_x ,v_y \right)$ was used to determine the model-predicted GRV of the target galaxy. We then corrected this for the motion of the Sun with respect to the MW to obtain its model-predicted HRV.
\begin{eqnarray}
	\label{Model_GRV}
	GRV_{model} &=& \frac{v_x x + (v_y - \overset{.}{y}_{_{MW}})(y - y_{_{MW}})}{\sqrt{x^2 + {(y - y_{_{MW}})}^2}} \\
	HRV_{model} &=& GRV_{model} - \bm v_\odot \cdot \hat{\bm d}_{_{MW}}
\end{eqnarray}

If the MW or M31 mass is altered, then another simulation is required. However, if we only wish to alter the adopted $v_{c,\odot}$, then this is not necessary. We simply use the same $GRV_{model}$ but a different $\bm v_\odot$ (Equation \ref{Solar_velocity}). In general, this alters $HRV_{model}$.

To account for distance uncertainties, the target was moved to the $1\sigma$ upper limit of its observed distance $d_{_{MW}}$ (using the 1$\sigma$ lower limit instead had a negligible impact on our analysis). The Newton-Raphson procedure was then repeated targeting the revised position. Once this converged, we extracted the GRV from the final trajectory. We took the difference between these GRV estimates and called this $\sigma_{pos}$. This is the uncertainty in the model-predicted GRV of a target galaxy due to uncertainty in its position.
\begin{eqnarray}
	\sigma_{pos} \equiv \left| GRV_{model} \left(d_{_{MW}} + \sigma_{d_{MW}} \right) - GRV_{model} \left(d_{_{MW}} \right) \right|
\end{eqnarray}

Here, $\sigma_{d_{MW}}$ is the uncertainty in the distance to a target galaxy. We assume negligible uncertainty in the direction towards it, constraining its position to be along a line. The velocity field is treated as linear over the part of this line where the target galaxy is likely be. Thus, assuming distance errors to be Gaussian, ${GRV}_{model}$ would also have a Gaussian distribution.


To determine $\sigma_{pos}$ for M31, we use a slightly different procedure because it is not massless. Once we have the time history of the MW$-$Cen A separation, we keep this fixed and vary the initial MW$-$M31 separation to target a revised final value. For consistency, we also do not change $M(t)$. The effect on the final GRV of M31 is used for $\sigma_{pos}$.

We expect this procedure to be approximately correct because Cen A only affects the GRV of M31 by $\sim$10 km/s, making it not crucial to handle tides from Cen A very accurately. It would be possible to do so by recalculating trajectories for all three galaxies with revised target positions, but due to numerical difficulties this would probably have been less precise. It will become clear later that our results are not much affected by the value of $\sigma_{pos}$ for M31.


Altering the MW$-$M31 separation changes the gravitational field in the rest of the LG, affecting GRVs of objects within it. We expect this to be a very small effect and so we neglected it.

We conducted simulations across a wide range of total initial masses $M_i$ and mass fractions in the MW $q_{_1}$ (see Table \ref{Priors}). For situations with $q_{_1} > \frac{1}{2}$, we took advantage of a symmetry that arises between situations with ${q_{_1} \leftrightarrow 1 - q_{_1}}$. Essentially, the behaviour of M31 in the low-$q_{_1}$ case is equivalent to the behaviour of the MW in the high-$q_{_1}$ case. Thus, we did not repeat all our calculations for the latter.

The positions of observed galaxies were altered in the following way:
\begin{eqnarray}
 x & \to & x ~~~\text{(unaltered)}\\
 y &=& y_{_{M31}} - y_{rel} \text{  ~~instead of } \left( y_{_{MW}} + y_{rel} \right)
 \label{Equation_54}
\end{eqnarray}

where $y_{rel}$ is still obtained using Equation \ref{y_rel} and is therefore unchanged. Equation \ref{Equation_54} also applies to Cen A, so its final position is now different. This meant we had to find a new solution for the trajectories of the MW, M31 and Cen A respecting the revised constraint on Cen A. Once this was done, we had to deal with altered positions $\left(x, y\right)$ for our target galaxies by finding new test particle trajectories with the right final positions. The final GRVs of these trajectories were obtained using Equation \ref{Model_GRV}, but referred to M31 rather than to the MW.

The step we did not repeat was the calculation of the LG velocity field. This meant we had a much poorer guess for the initial position of each target galaxy. For this, we simply re-used the values of $x$ and $y_{rel}$ at the initial time in the low-$q_{_1}$ case. Despite this, our algorithm still converged.

This procedure implicitly assumes that the accretion radii of the MW and M31 are swapped (i.e. that the galaxy with the higher mass always has the larger accretion radius). However, with the very low amounts of mass accreted by these galaxies (Figure \ref{Mass_accretion}), this should hardly affect our results. This is especially true when considering that our analysis tends to disfavour $q_{_1} > \frac{1}{2}$ (Figure \ref{Primary_result}).

As well as uncertainties due to position ($\sigma_{pos}$) and measurement error on the radial velocity ($\sigma_{v_h}$), we also included an extra variance term, $\sigma_{extra}$. This was to account for effects not handled in our algorithm, for instance interactions between LG dwarf galaxies and tides raised by large-scale structures (LSS). $\sigma_{extra}$ is a measure of how much model-predicted and actual radial velocities disagree. Including it, the contribution to $\chi^2$ of any particular galaxy $i$ is
\begin{eqnarray}
	{\chi_{_i}}^2 &\equiv & \left( \frac{HRV_{model} - HRV_{obs}}{\sigma} \right)^2 ~~\text{ where} \\
	\sigma &=& \sqrt{{\sigma_{pos}}^2 + {\sigma_{v_h}}^2 + {\sigma_{extra}}^2}
	\label{sigma}
\end{eqnarray}



The uncertainty on the motion of the Sun can introduce systematic errors into our analysis. Thus, we treated $v_{c, \odot}$ as another model parameter. However, it is independently constrained \citep[$239 \pm 5$ km/s,][]{McMillan_2011}. This was accounted for using a Gaussian prior, or equivalently by adding an extra contribution to $\chi^2$. Therefore, the total $\chi^2$ for any particular model ($\equiv$ combination of model parameters) is
\begin{eqnarray}
	\chi^2 = \sum_{\begin{array}{c} \text{Target}\\ \text{galaxies}\end{array}} {\chi_{_i}}^2 ~+~ \left( \frac{v_{c, \odot} - v_{c, \odot \text{,nominal}}}{\sigma_{v_{c, \odot}}} \right)^2
	\label{Chi_sq_total}
\end{eqnarray}

Models with higher $\sigma_{extra}$ will necessarily achieve a lower $\chi^2$. Thus, we can't use $\chi^2$ alone to decide which models are best. We made use of the fact that the probability of a model matching an individual observation
\begin{eqnarray}
	P(\text{Observation of galaxy } i|\text{ Model}) \propto \frac{1}{\sigma} {\rm e}^{^{-\frac{{\chi_{_i}}^2}{2}}}
\end{eqnarray}

Thus, we recorded both ${\chi_{_i}}^2$ and $\sigma_{_i}$ for each observed galaxy. The relative model likelihoods were then found using
\begin{eqnarray}
	P(\text{Model } | \text{ Observations}) \propto \left( \prod_{i} \frac{1}{\sigma_{_i}} \right) {\rm e}^{^{-\frac{\chi^2}{2}}}
	\label{P_final}
\end{eqnarray}



If model-predicted and observed HRVs often disagree by much more than observational errors, then non-zero values of $\sigma_{extra}$ will be preferred. Once $\chi^2$ becomes comparable to the number of target galaxies (32), increasing $\sigma_{extra}$ further will not much reduce $\chi^2$. As a result, instead of $P$ increasing with $\sigma_{extra}$, it will actually start to \emph{decrease} because of the factors of $\frac{1}{\sigma_{_i}}$ in Equation \ref{P_final}. One can imagine this as penalizing models where $\chi^2$ is so small that such good agreement with observations is `too good to be true'.

\begin{figure}
 \centering
  \includegraphics [width = 8.5cm] {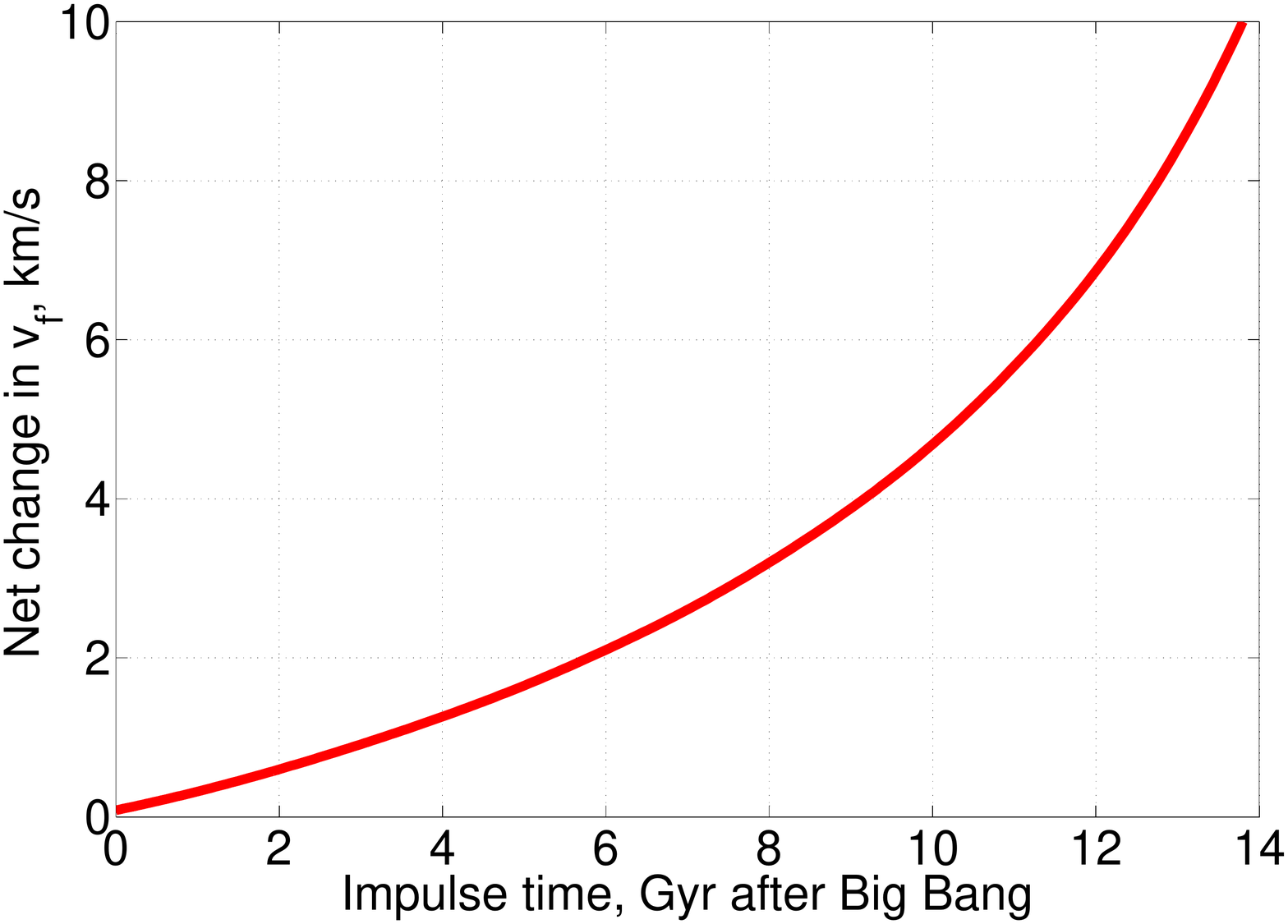}
 \caption{The overall effect on the present GRV of Andromeda due to a 10 km/s impulse to its GRV in the past, with the present distance to Andromeda constrained. This constraint is maintained by altering the initial MW$-$M31 separation (see text). Because of this, impulses applied longer ago have a smaller net effect on present motions.}
 \label{Impulsed_trajectories}
\end{figure}

In this way, we hoped to constrain $\sigma_{extra}$. If model-predicted and observed HRVs agree well given observational uncertainties, then the posterior distribution of $\sigma_{extra}$ would peak at or near 0. If that does not occur, then this might indicate underestimated observational errors or a failure of the model.

Physically, we expect the main source of astrophysical noise contributing to $\sigma_{extra}$ to be interactions between LG dwarf galaxies. However, Andromeda is much heavier than them, suggesting that it should be treated somewhat differently. This is because a minor merger would affect its velocity very little. Thus, whatever the adopted value of $\sigma_{extra}$ for other LG galaxies, a smaller value of $\sigma_{extra,~M31}$ should be adopted for M31. We used $\frac{\sigma_{extra,~M31}}{\sigma_{extra}} = 0.1$. This alters Equation \ref{sigma} for M31 and thus its contribution to $\chi^2$.

We considered the effect of a minor merger with Andromeda or the Milky Way in the past. This was modelled as an impulse, meaning that we instantaneously altered the GRV of M31 at some time in the past. The effect on its present GRV was then determined. For simplicity, Centaurus A was omitted and the total LG mass was held constant at $4 \times 10^{12} M_\odot$. This roughly reproduces the present GRV and distance of M31.

One might think that the longer ago the impulse was, the bigger its effect on the present GRV of M31, $\Delta v_{_f}$. After all, pushing the galaxies towards each other increases the force between them at later times, further reinforcing the original impulse. 

However, this would lead to the constraint on the present distance to M31 being violated (in this example, it would end up too close). Consequently, we had to alter the initial separation of the galaxies $d_i$ compared with a non-impulsed trajectory. This tends to counteract the direct effect of the impulse.

The results we obtained for $\Delta v_{_f}$ as a function of the impulse time are shown in Figure \ref{Impulsed_trajectories}. An impulse applied very recently hardly affects $d_f$ and so $d_i$ doesn't have to be altered much. Thus, $\Delta v_{_f}$ is almost equal to the impulse.

For impulses applied longer ago, $\Delta v_{_f}$ rapidly becomes very small. The dependence on impulse time is even steeper than for Hubble drag ($\Delta v_{_f} \appropto a^{2.4}$, where $a$ was the cosmic scale factor when the impulse was applied). This underlines just how difficult it is to alter the present GRV of M31. Consequently, a realistic model needs to match this constraint very well.

As well as interactions between LG dwarf galaxies, our model does not fully account for the presence of large-scale structures in the Universe beyond the LG. We attempted to include some of these structures in Section \ref{Tides}, but others remain beyond the scope of this investigation. The leading order effect of LSS on the LG is to accelerate it as a whole without altering the relative velocities between objects within it. However, LSS also raise tides on the LG, affecting the GRVs of our target galaxies. Such effects are larger for galaxies further from the MW. As M31 is the closest galaxy in our sample, its GRV should be least affected by tides raised by LSS. This further justifies our decision to use a value for $\frac{\sigma_{extra,~M31}}{\sigma_{extra}}$ that is much smaller than 1.

\section{Results}
\label{Results}

\begin{figure}
 \centering
  \includegraphics [width = 8.5cm] {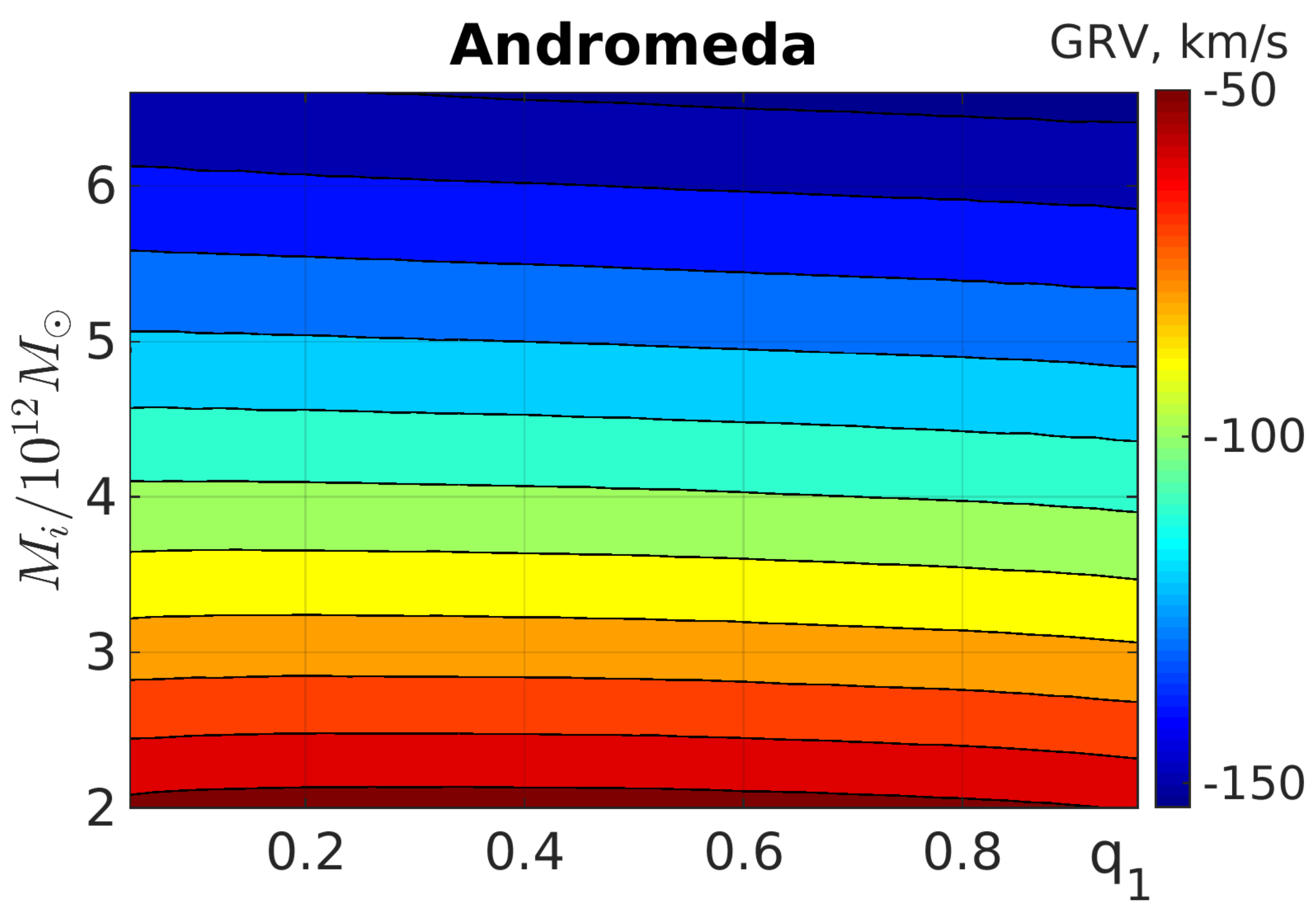}
  \includegraphics [width = 8.5cm] {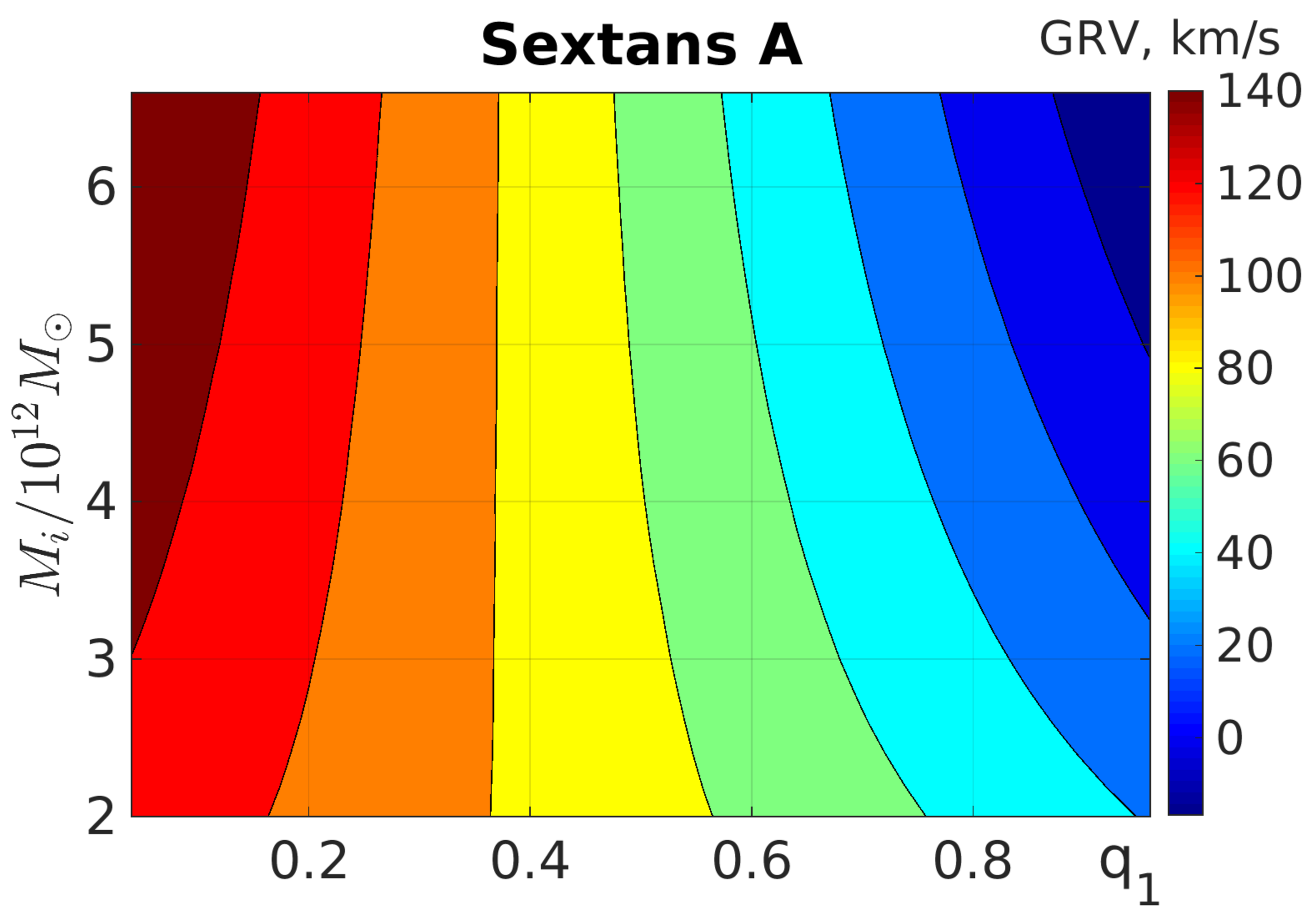}
 \caption{Simulated galactocentric radial velocities (GRVs) as a function of model parameters for two target galaxies. Different galaxies constrain a different combination of model parameters, with Andromeda mostly telling us the total LG mass and Sextans A mostly telling us the fraction of this mass in the MW (because the GRV of Sextans A is much higher at low $q_{_1}$).}
 \label{GRV_Sextans_A}
\end{figure}

Our analysis works by determining the model-predicted GRV of each target galaxy in the LG. A range of models are tried out, with different initial LG masses $M_i$ and mass fractions in the MW $q_{_1}$ (see Table \ref{Priors}). The results for two target galaxies are shown in Figure \ref{GRV_Sextans_A}. Each of these GRV predictions are converted into a range of HRV predictions using $v_{c,\odot}$ within 3$\sigma$ of its most likely value according to the work of \citet{McMillan_2011}. By comparing these with observed HRVs, we obtain complementary constraints on the model parameters. As we have $\gg$ 2 target galaxies, we also test the model itself.


\begin{figure}
 \centering
  \includegraphics [width = 8.5cm] {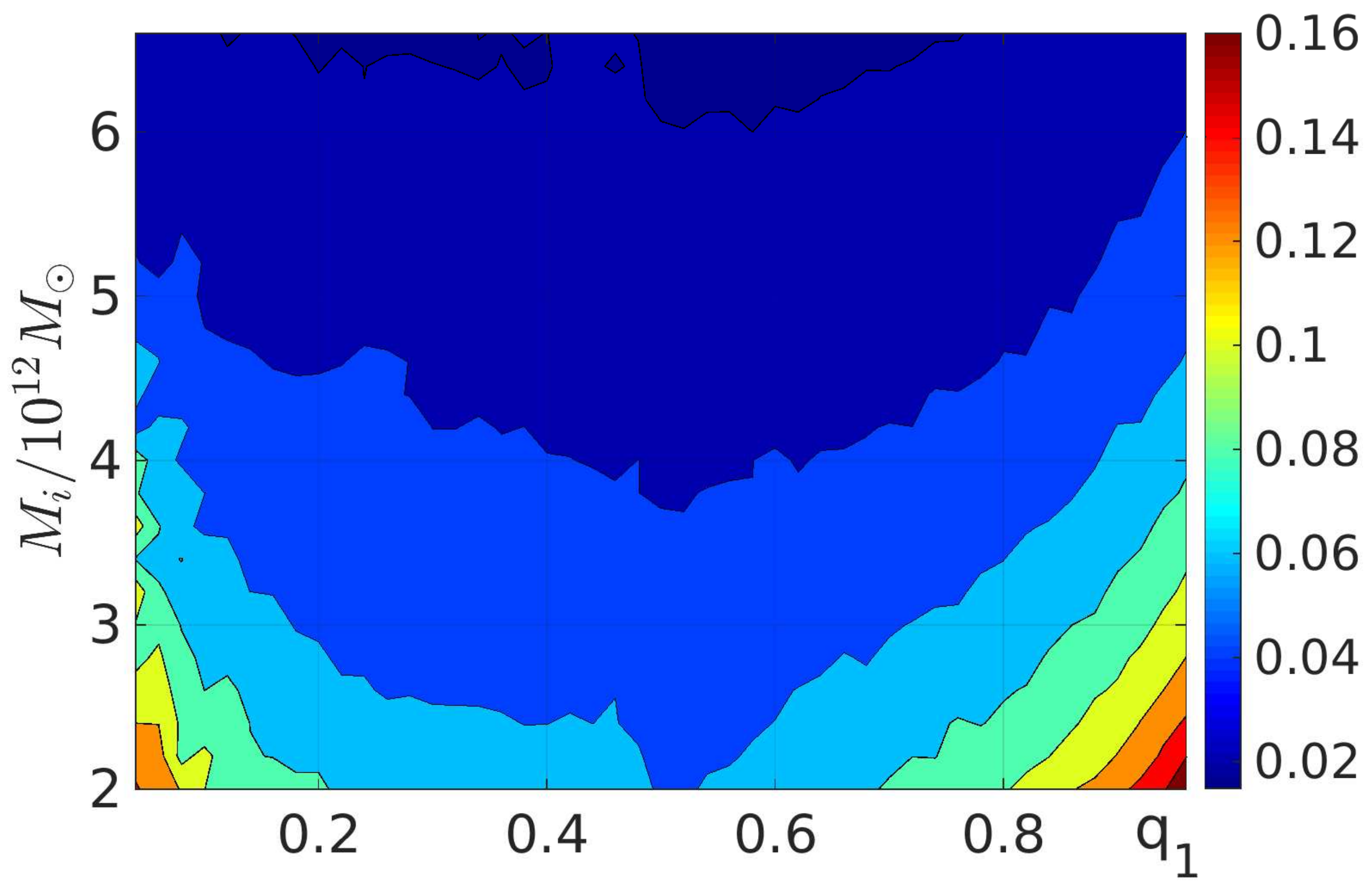}
 \caption{Fraction of the initial Milky Way mass that it accretes by the end of our simulations. Due to symmetry, approximate results for Andromeda can be obtained by setting $q_{_1} \to 1 - q_{_1}$.}
 \label{Mass_accretion}
\end{figure}

Our simulations allowed the MW and M31 to accrete mass. In Figure \ref{Mass_accretion}, we show the fraction by which the original mass of each galaxy increased. The galaxies only increase their mass by a few percent in our simulations. Thus, accretion is unimportant in them. This is mainly because a test particle needs to pass within a few disc scale-lengths of the MW or M31 for us to consider the particle accreted (Table \ref{Priors}).

This aspect of our models is not totally realistic. If more distant approaches were also treated as leading to accretion, then the MW \& M31 would gain more mass. We do not consider this an important effect because we tried a wide range of initial masses for both galaxies.

\begin{figure}
		\includegraphics [width = 16.2cm] {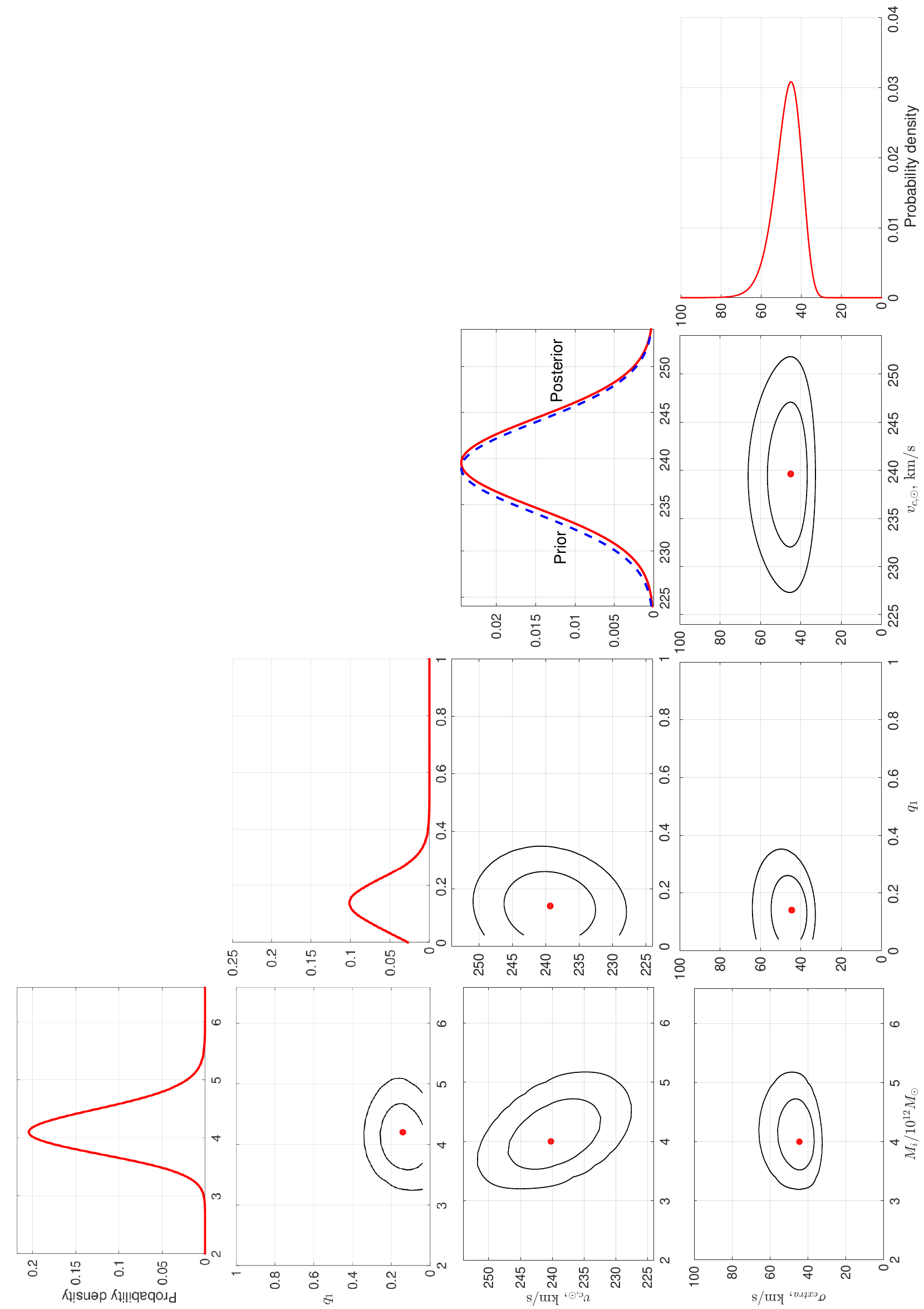}
	\caption{Posterior probability distributions of the parameters considered in this work (defined in Table \ref{Priors}). In each figure, we marginalized over the parameters not shown. For variables plotted against other variables, we show the contours of the probability density corresponding to the $1 \sigma$ (68.3\%) and $2 \sigma$ (95.4\%) confidence levels, as well as the most likely pair of values. Rotate figure $90^\circ$ clockwise for viewing.}
	\label{Primary_result}
\end{figure}

Figure \ref{Primary_result} shows the posterior probability distributions of all our model parameters and pairs of parameters based on a set of 1128 simulations\footnote{spanning a linear grid with 24 steps in $M_i$ and 47 steps in $q_{_1}$, although some shortcuts were taken for $q_{_1} > \frac{1}{2}$} that include Centaurus A with a mass of $4 \times 10^{12} M_\odot$. Each simulation was compared with observations using 101 values of $v_{c, \odot}$ and 201 values of $\sigma_{extra}$ (priors are given in Table \ref{Priors}). Of particular importance is the posterior on $\sigma_{extra}$, which we constrain to be $45_{-6}^{+7}$ km/s. As observational errors are typically $\sim$5$-$10 km/s and are already included in our analysis, this is very surprising.

We checked if varying the start time of our simulations from $a_{_i} = \frac{1}{10} \to \frac{1}{15}$ affected our results. This reduced the most likely value of $\sigma_{extra}$ by $\sim$ 1 km/s. Our results are not much affected by the epoch at which our simulations are started. Some reasons for this are given in Section \ref{MW_M31_interaction}).

We considered a different estimate for the LSR speed \citep[$238 \pm 9$ km/s,][]{Schonrich_2012}. As might be expected, this affected $\sigma_{extra}$ by $\la$1 km/s. This is because we consider $v_{c, \odot}$ to be well constrained independently of our work. It is also apparent that there is very little tension between these independent measurements and our timing argument analysis (Table \ref{Priors}).

Our special treatment of M31 forces up $\sigma_{extra}$ to some extent as it essentially forces our models to match its GRV (given the small uncertainty on $v_{c, \odot}$). As this may be overly restrictive, we redid our analysis using the same value of $\sigma_{extra}$ for M31 as for other LG galaxies (i.e. $\frac{\sigma_{extra, M31}}{\sigma_{extra}} = 1$ instead of 0.1). This lowered $\sigma_{extra}$ by $\sim$ 2 km/s.


Our method of handling distance uncertainties is very similar to that used by \citet{Jorge_2014}. We rely on an assumption that the velocity field is approximately linear over the range of positions where the galaxy could plausibly be. To test this, we repeated our calculations with $\sigma_{pos}$ estimated based on how much the simulated GRV of each target galaxy changed if we altered its distance from its observed value to its 1$\sigma$ lower limit. This gave almost identical results to when we used the 1$\sigma$ upper limit instead ($\sigma_{extra}$ decreased by $\sim$ 0.1 km/s when using the lower limit).


Our analysis favours a very low value for $q_{_1}$, the fraction of the LG mass originally in the MW. This is related to the fact that observed HRVs tend to systematically exceed the predictions of the best-fitting model (Figure \ref{Delta_GRV_histogram_detailed}). Thus, our analysis will prefer those models that generally lead to increased GRVs. Reducing $q_{_1}$ has this effect because it causes particles projected orthogonally to the MW$-$M31 line; to curve towards M31 and away from the MW. It also implies a faster motion of the MW relative to the LG barycentre and a greater distance from there, enhancing projection effects.

Most of our target galaxies are in fact roughly orthogonal to the MW$-$M31 line as perceived from the LG barycentre (Figure \ref{LG_Hubble_Diagram}). This might be why $q_{_1}$ seems to have a strong impact on GRVs (bottom panel of Figure \ref{GRV_Sextans_A}). Thus, one might expect our analysis to prefer very low values of $q_{_1}$, which indeed it does.



Certain correlations are apparent between some of our model parameters. Because we require our models to accurately match the observed HRV of M31, our timing argument estimate of the LG mass is quite sensitive to anything which affects its predicted HRV. M31 is almost directly `ahead' of the Sun in its motion around the MW. Thus, for the same GRV of M31, increasing $v_{c,\odot}$ decreases its HRV (Equation \ref{GRV_obs}). To increase its HRV back up to its observed value, its GRV would have to be increased, which is only possible in a different model where the retarding effect of gravity is smaller (i.e. $M_i$ is lower).

A lot of our target galaxies have HRVs which exceed the predictions of the best-fitting model (Figure \ref{GRV_LCDM_Comparison}). This means that a lower LG mass fits the data slightly better, explaining the correlation between $M_i$ and $\sigma_{extra}$. For the same reason, increasing $v_{c, \odot}$ indirectly improves the fit to the data, reducing $\sigma_{extra}$ slightly.



Some effects are inevitably not considered in our model. If they were included, we might achieve a better fit to the observations. We consider some of these effects in the next section. We pay special attention to tides from objects beyond the Local Group (Section \ref{Tides}) and the Large Magellanic Cloud (Section \ref{Large_Magellanic_Cloud}).


%
%
%
%
%
%
%
%

\begin{table}
  \centering
    \begin{tabular}{llll}
\hline
Name & Meaning and units & Prior & Result\\
\hline
$\sigma_{extra}$ & Extra velocity dispersion & 0 $-$ 100&${45.1}_{-5.7}^{+7.0}$\\
& along line of sight, km/s & & \\ [5pt]
$M_i$ & Initial MW $+$ M31 mass, & 2 $-$ 6.6&4.1$\pm$0.3\\
& trillions of solar masses & & \\ [5pt]
$q_{_1}$ & Fraction of MW $+$ M31 &0.04$-$0.96&0.14$\pm$0.07\\
& mass initially in the MW & & \\ [5pt]
$v_{c, \odot}$ & Circular speed of MW at & $239 \pm 5$&239.5$\pm$4.8\\
& position of Sun, km/s & & \\ [5pt]
\hline
\multicolumn{4}{c}{Fixed parameters} \\
\hline
$d_0$ & Distance to M31, kpc & $783 \pm 25$ &\\ [5pt]
$H_{_0}$ & Hubble constant at the & 67.3&\\
& present time, km/s/Mpc & & \\ [6pt]
$\Omega_{m,0}$ & Present matter density in & 0.315&\\
& the Universe $\div \frac{3{H_{_0}}^2}{8 \rm{\pi} G}$ & & \\ [5pt]
$a_{_i}$ & Scale factor of Universe & 0.1 &  \\
& at start of simulation & & \\ [5pt]
$r_{_{acc, MW}}$ & Accretion radius of MW & \multicolumn{2}{l}{15,337 parsecs} \\
$r_{_{acc, M31}}$ & Accretion radius of M31 & \multicolumn{2}{l}{21,472 parsecs} \\ [5pt]
$U_\odot$ & See Equation \ref{Solar_velocity} & \multicolumn{2}{l}{14.1 km/s}  \\
$V_\odot$ & See Equation \ref{Solar_velocity} & \multicolumn{2}{l}{14.6 km/s}  \\
$W_\odot$ & See Equation \ref{Solar_velocity} & \multicolumn{2}{l}{6.9 km/s}  \\
\hline
 \end{tabular}
  \caption{Priors and $1 \sigma$ confidence levels on model parameters. The latter are far from the boundaries imposed by the former, showing that our results are not strongly affected by our priors. Due to accretion, the present-day masses of the MW and M31 are $\sim$5\% higher than when the simulations start. We use the measurement of $d_0$ by \citet{McConnachie_2012}. Cosmological parameters are from \citet{Planck_2015}. We obtained $v_{c, \odot}$ from \citet{McMillan_2011} and the Sun's non-circular velocity from \citet{Francis_2014}. Uncertainty in the latter is much less than in the former. We assume $v_{c, \odot}$ is within 3$\sigma$ of its most likely value.}
  \label{Priors}
\end{table}

\onecolumn

\twocolumn

\section{Discussion}
\label{Discussion}


Our analysis reveals an astrophysical noise in velocities of LG galaxies that greatly exceeds observational errors. With $N = 32$ targets, the fractional uncertainty in this extra noise $\sigma_{extra}$ should be $\approx \frac{1}{\sqrt{2N}} = \frac{1}{8}$. This agrees closely with the width of the posterior probability distribution of $\sigma_{extra}$ (Figure \ref{Primary_result}).


We considered several factors which could influence our analysis. Perhaps most obviously, the LG contains gravity from objects other than the MW and M31. For example, the non-satellite LG galaxies that we modelled as test particles in reality exert gravity on each other. This would lead to roughly isotropic and random impulses on them. Considering that our analysis is based solely on line of sight velocities, we would need to assume typical impulses of $\sim \sqrt{3} ~ \sigma_{extra} \approx 80$ km/s. 

However, our target galaxies have typical velocity dispersions/rotation speeds of $\la 15$ km/s \citep[e.g.][]{Kirby_2014}. For some impact parameters, these galaxies could perhaps impulse each other by twice this while avoiding a merger. Thus, the high value of $\sigma_{extra}$ inferred by our analysis seems difficult to explain as a result of interactions amongst the galaxies we considered. 

Additional inaccuracies in our model may arise from the effects of large scale structure. Moreover, even distant encounters between LG dwarf galaxies can affect their motion. The likely magnitude of such effects can be estimated based on more detailed cosmological simulations of the $\Lambda$CDM paradigm. Considering analogues of the LG in such simulations, it has been found that the dispersion in radial velocity with respect to the LG barycentre at fixed distance from there should be $\sigma_{_H} \sim 30$ km/s \citep{Aragon_Calvo_2011}.

Looking at the bottom panel of Figure \ref{LG_Hubble_Diagram}, it is clear that our models do not produce such a large velocity dispersion. We seem to get $\sigma_{_H} \sim 10$ km/s, though this rises slightly to $\sim 15$ km/s once we include Centaurus A. Thus, even if $\Lambda$CDM were correct, it would be reasonable for our analysis to infer $\sigma_{extra} \sim 25$ km/s.

In this section, we hope to correct some of our model deficiencies and make it a more accurate representation of $\Lambda$CDM. Table \ref{sigma_H} shows some of the effects we consider and a rough idea of their contributions to $\sigma_{_H}$. Combining everything in quadrature suggests that the objects we consider are sufficient to attain a dispersion in the Hubble flow of $\sim$20 km/s. Thus, a lot of the `scatter' about the Hubble flow found by \citet{Aragon_Calvo_2011} arises because the LG is not spherically symmetric rather than actually being a dispersion in velocities at the same position. Nonetheless, another $\sim$20 km/s must come from factors we do not consider. This means that values of $\sigma_{extra}$ much greater than $\sim$20 km/s would be problematic for $\Lambda$CDM.

\begin{table}
  \centering
    \begin{tabular}{llll}
\hline
Object & Contribution & Comments & Section \\
 & to $\sigma_{_H}$ (km/s) & & \\
\hline
MW, M31 \& Cen A & $\sim$ 15 & 10 at 2 Mpc & \ref{Simulations} \\
IC 342 \& M81 & $\sim$ 5 & Table \ref{IC342_M81_effect} & \ref{Tides_IC342_M81} \\
The Great Attractor & $\sim$ 10 & Equation \ref{Distant_tide_approximation} & \ref{Great_Attractor} \\
\hline
 \end{tabular}
  \caption{Contributions of various sources to $\sigma_{_H}$, the radial velocity dispersion with respect to the Local Group barycentre at fixed distance from there. The Great Attractor leads to a $\sim$40 km/s range in radial velocities at 3 Mpc (Equation \ref{Distant_tide_approximation}). Combining everything in quadrature, we can account for $\sim$20 km/s of the $\sim$30 km/s dispersion in the Hubble flow found by \citet{Aragon_Calvo_2011}. This suggests that our model should represent $\Lambda$CDM to an accuracy of $\sim$20 km/s.}
  \label{sigma_H}
\end{table}


\begin{figure}
	\centering
		\includegraphics [width = 8.4cm] {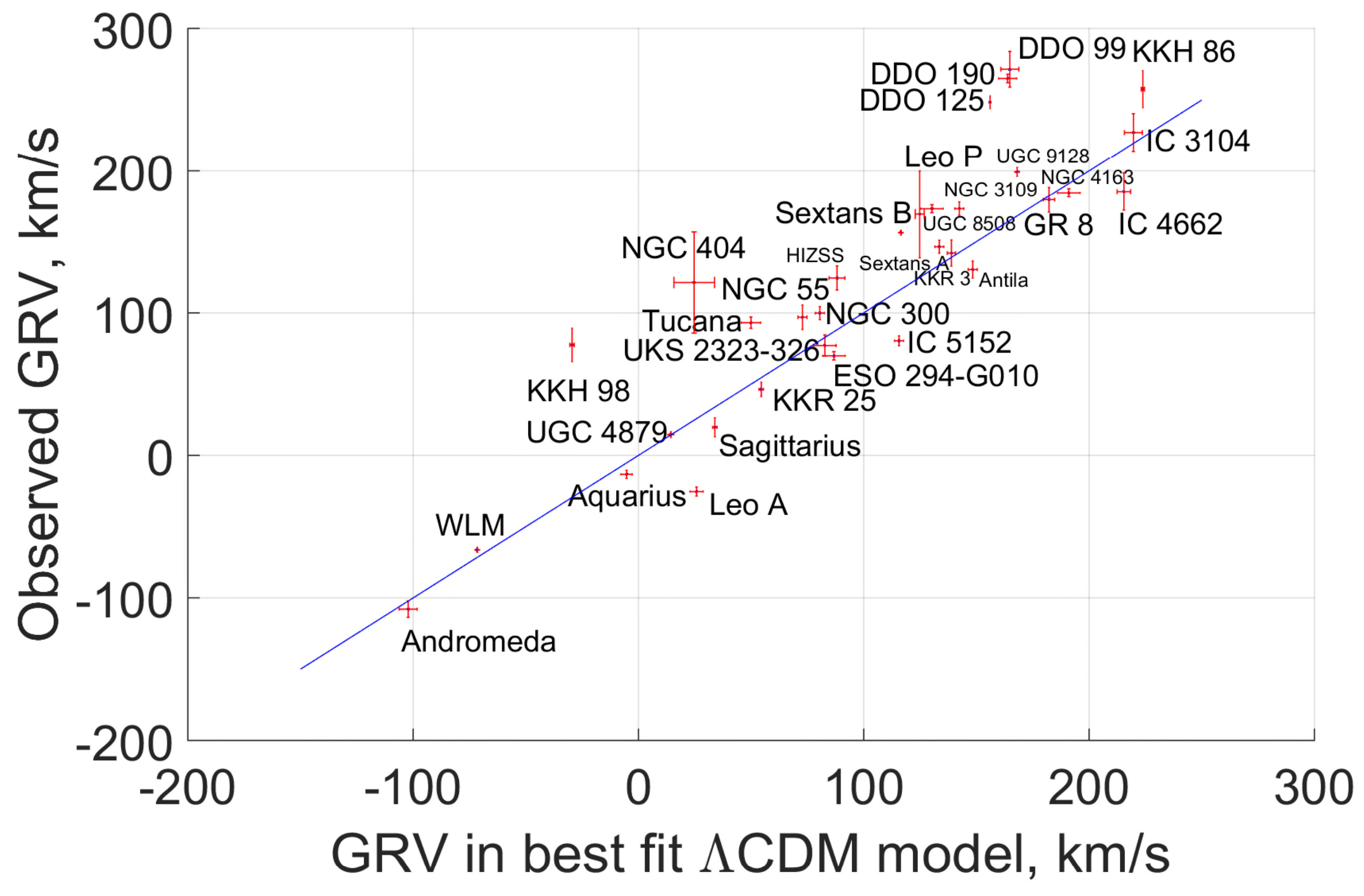}		
	\caption{Comparison between predicted and observed GRVs based on the most likely model parameters ($q_{_1} = 0.14$, $M_i = 4.2 \times 10^{12} M_\odot$, $v_{c,\odot} = 239$ km/s). The line of equality is in blue.}
	\label{GRV_LCDM_Comparison}
\end{figure}

\begin{figure}
	\centering
		\includegraphics [width = 8.5cm] {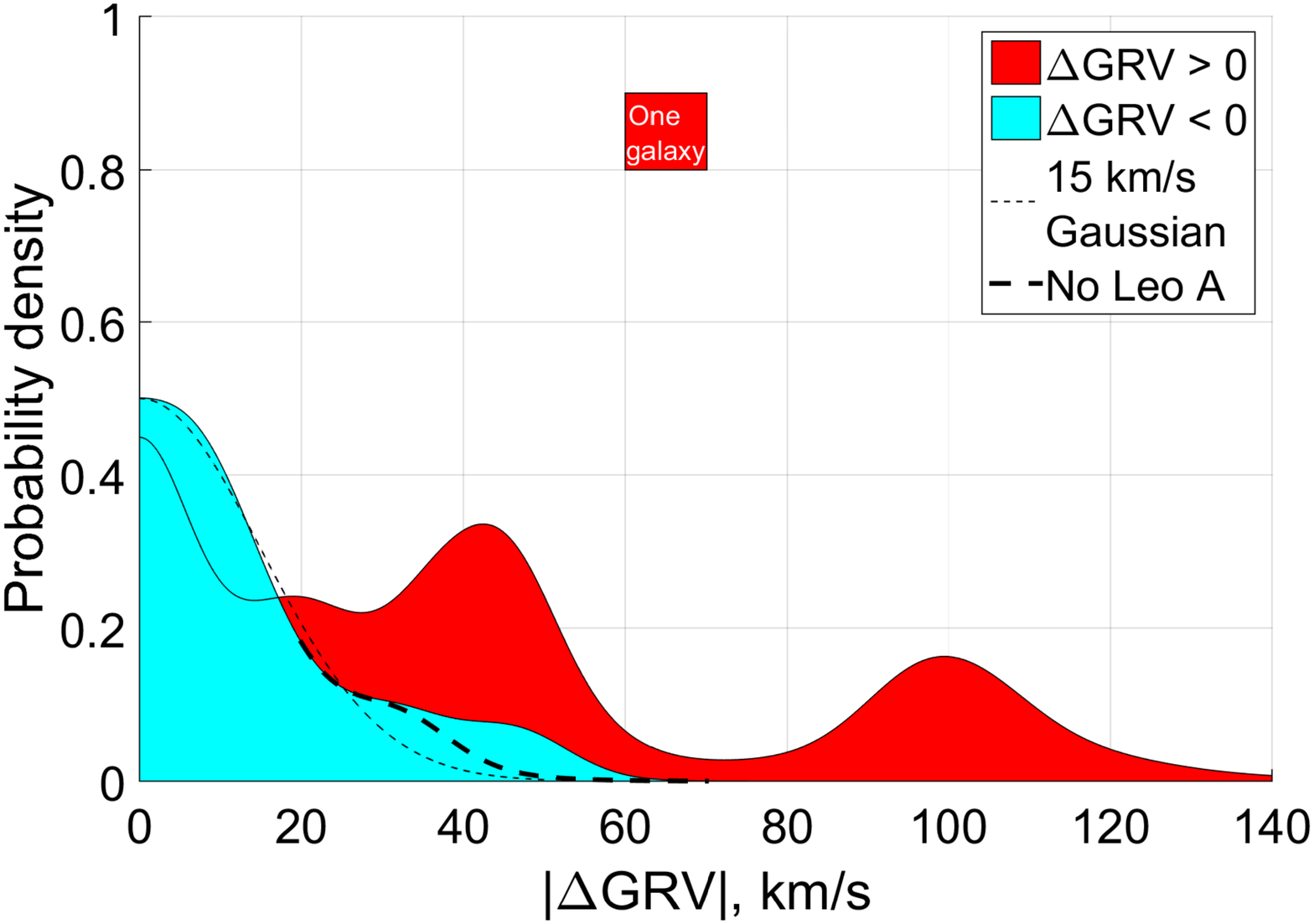}
	\caption{Histogram showing observed $-$ predicted GRVs (i.e. $\Delta GRV$s) of our target galaxies using our most plausible model ($q_{_1} = 0.2$ instead of 0.14, other parameters as in Figure \ref{GRV_LCDM_Comparison}). Each data point was convolved with a Gaussian of width $\sigma = \sqrt{{\sigma_{pos}}^2 + {\sigma_{v_h}}^2 + {\sigma_{v_{c, \odot}}}^2}$. We divided our sample into those with $\Delta GRV < 0$ (blue) and those with $\Delta GRV > 0$ (red). The area corresponding to one galaxy is shown as a red square. A Gaussian of width 15 km/s is overplotted as a short-dashed line. This matches the $\Delta GRV < 0$ subsample quite well, especially when Leo A is excluded (long-dashed line).} 
	\label{Delta_GRV_histogram_detailed}
\end{figure}

Our best-fitting model has $q_{_1} = 0.14$. Results from this model are compared with observations in Figure \ref{GRV_LCDM_Comparison}. However, we consider it unrealistic for the MW to have only $\frac{1}{6}$ as much mass as M31. Assuming that the virial mass of a halo scales as the cube of its velocity dispersion \citep{Evrard_2008} and that the ratio of the latter between the MW and M31 is $\frac{225}{180} = 1.25$ \citep{Carignan_2006, Kafle_2012}, we see that it is unlikely for M31 to have much more than twice as much mass as the MW. We believe the best compromise between this argument and the low value of $q_{_1}$ preferred by our timing argument analysis ($0.14 \pm 0.07$) is found if we set $q_{_1} = 0.2$. Thus, when comparing our model predictions with observations (Figure \ref{Delta_GRV_histogram_detailed} onwards), we use the model parameters which best fit the data but with $q_{_1}$ raised to 0.2. This raises $\sigma_{extra}$ by $\sim$ 1 km/s and has only a small impact on our results, but should make them more realistic.

Observed GRVs seem to systematically exceed model predictions (Figure \ref{GRV_LCDM_Comparison}). We used our most plausible model including Cen A to subtract model-predicted radial velocities from observed ones, yielding $\Delta GRV$ for each target galaxy. We then created a histogram of the resulting $\Delta GRV$s in Figure \ref{Delta_GRV_histogram_detailed}, smoothing each data point over its respective uncertainty. As before, this includes $\sigma_{pos}$ and $\sigma_{v_h}$. Because it is unclear exactly how to convert heliocentric radial velocities into Galactocentric ones, we also added ${\sigma_{v_{c, \odot}}}$ in quadrature to all the uncertainties. 


If one assumes that factors outside our model are just as likely to raise GRVs of target galaxies as to reduce them, then it should be possible to use the population of $\Delta GRV < 0$ galaxies to gain a good idea of how accurately our model represents $\Lambda$CDM. The $\Delta GRV < 0$ galaxies are well described by a Gaussian of width 15 km/s (blue area in Figure \ref{Delta_GRV_histogram_detailed}). Most of the mismatch is due to Leo A. To account for tides raised by IC 342 and M81, its radial velocity prediction should be reduced by $\sim$ 5 km/s, making it more consistent with observations (Table \ref{IC342_M81_effect}). In any case, considering the $\Delta GRV < 0$ galaxies suggests that inaccuracies in our model probably do not exceed 25 km/s, slightly less than the $\sigma_{_H} \sim 30$ km/s found by \citet{Aragon_Calvo_2011} due to our careful modelling. Thus, one might expect a Gaussian of around this width to also describe the distribution of $\Delta GRV$s for galaxies with $\Delta GRV > 0$.

However, unlike galaxies with $\Delta GRV < 0$, those with $\Delta GRV > 0$ are not well described by a 15 km/s Gaussian (red area in Figure \ref{Delta_GRV_histogram_detailed}). There appears to be a population of $\Delta GRV > 0$ galaxies which might be described by such a Gaussian, but in this case we would need perhaps two additional populations to fully account for the observations. A possible mechanism for generating these populations is described in Section \ref{MW_M31_interaction}.

\begin{figure}
	\centering
		\includegraphics [width = 8.5cm] {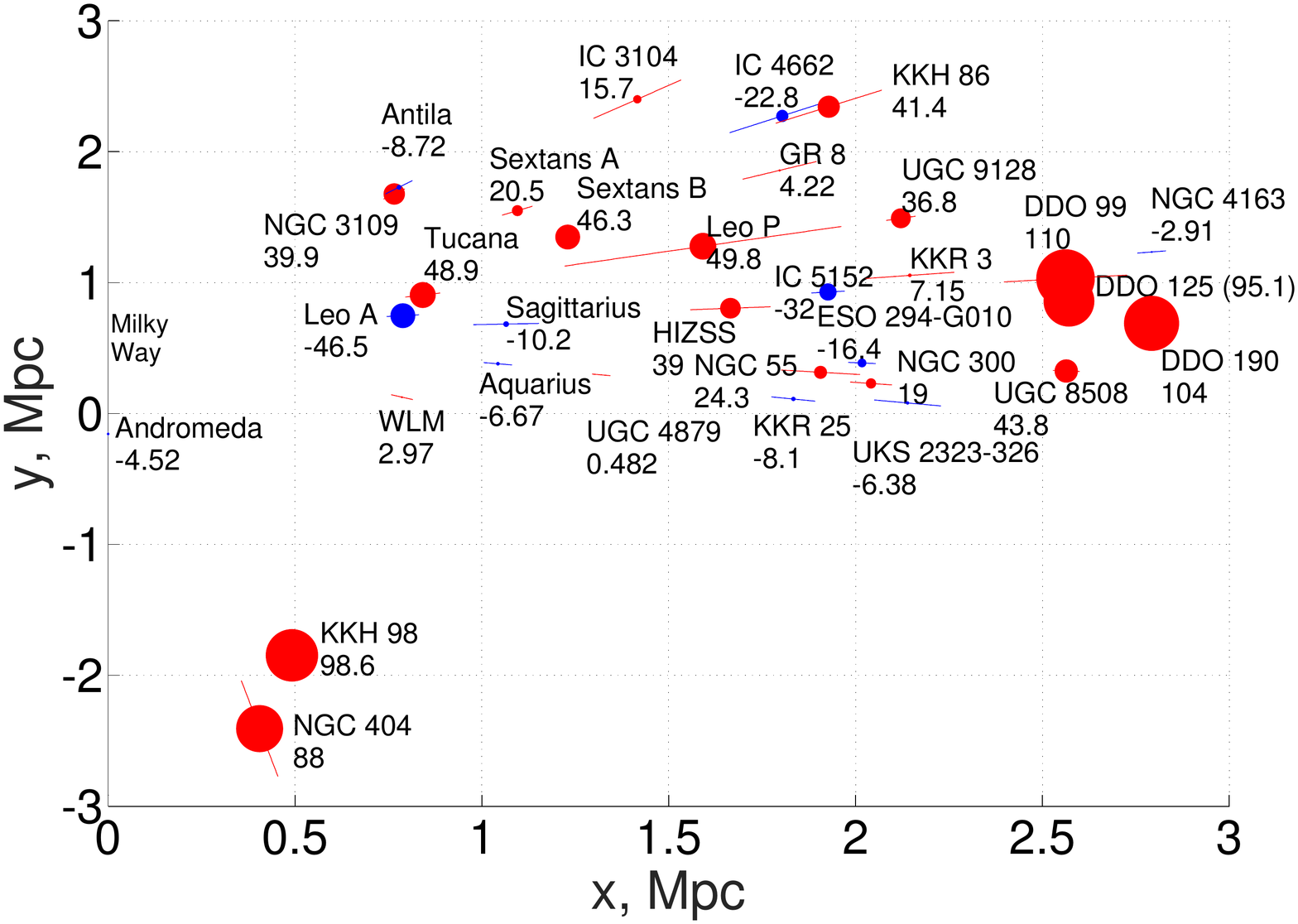}
	\caption{Positions of target galaxies are shown in the same way as the top panel of Figure \ref{LG_Hubble_Diagram}. Distance uncertainties are indicated by a thin line. The size of each marker is directly proportional to $\left| \Delta GRV \right|$ for the associated galaxy (name beside marker) based on the same model as shown in Figure \ref{Delta_GRV_histogram_detailed}, with colour indicating sign (red means positive). Uncertainties in $\Delta GRV$ are roughly proportional to that in distance. For Leo P, this causes a 30 km/s uncertainty, though typical values are much smaller. We expect our model to represent $\Lambda$CDM to an accuracy of $\sim$ 25 km/s (see text), roughly the same as $\Delta GRV$ of NGC 55.}	
	\label{Cylindrical_projection_GRV_map}
\end{figure}

We tried to see if there was any correlation between the position of a target galaxy and it's associated $\Delta GRV$. This is shown in Figure \ref{Cylindrical_projection_GRV_map}, with the size of the marker for each galaxy directly proportional to its $\Delta GRV$. It is assumed that the LG is axisymmetric, so positions are shown using the same co-ordinate system as our simulations. The uncertain distance to each galaxy is indicated by a thin line.

The objects with the highest $\Delta GRV$s tend to be furthest from the MW/M31. This might be a sign that tides from objects outside the LG are responsible for the discrepancies. As we already included Centaurus A in our simulations, we might be seeing the effects of other objects. We will investigate some possibilities in Section \ref{Tides}. In particular, we will show that IC 342 and M81 are unlikely to be responsible for the discrepancies (Section \ref{Tides_IC342_M81}). This is also true of the Great Attractor (Section \ref{Great_Attractor}). An explanation for this trend is suggested in Section \ref{MW_M31_interaction}.

\subsection{Reduced Local Group Mass}
\label{Reduced_LG_Mass}

Comparison with cosmological simulations suggests that the timing argument may overestimate the LG mass \citep{Gonzalez_2014}. Moreover, observed radial velocities tend to be systematically more outwards than in our models. These considerations suggest that a lower LG mass could help to explain the observations. To test this, we removed the effect of gravity altogether and used Equation \ref{Hubble_flow} to predict velocities. As before, we took the barycentre of the LG as the centre of expansion and assumed that $v_{c, \odot} = 239$ km/s.

The MW was assumed to be going towards this point at $\sim$90 km/s, which is reasonable given the observed HRV of M31 and a plausible mass ratio between the galaxies. In theory, the MW should be going away from the LG barycentre in the absence of gravity. However, using the correct MW velocity ensures that velocities with respect to the LG barycentre are correctly converted into velocities with respect to us. One slightly unusual consequence of this is that the model predicts M31 to have a negative GRV.

\begin{figure}
	\centering
		\includegraphics [width = 8.5cm] {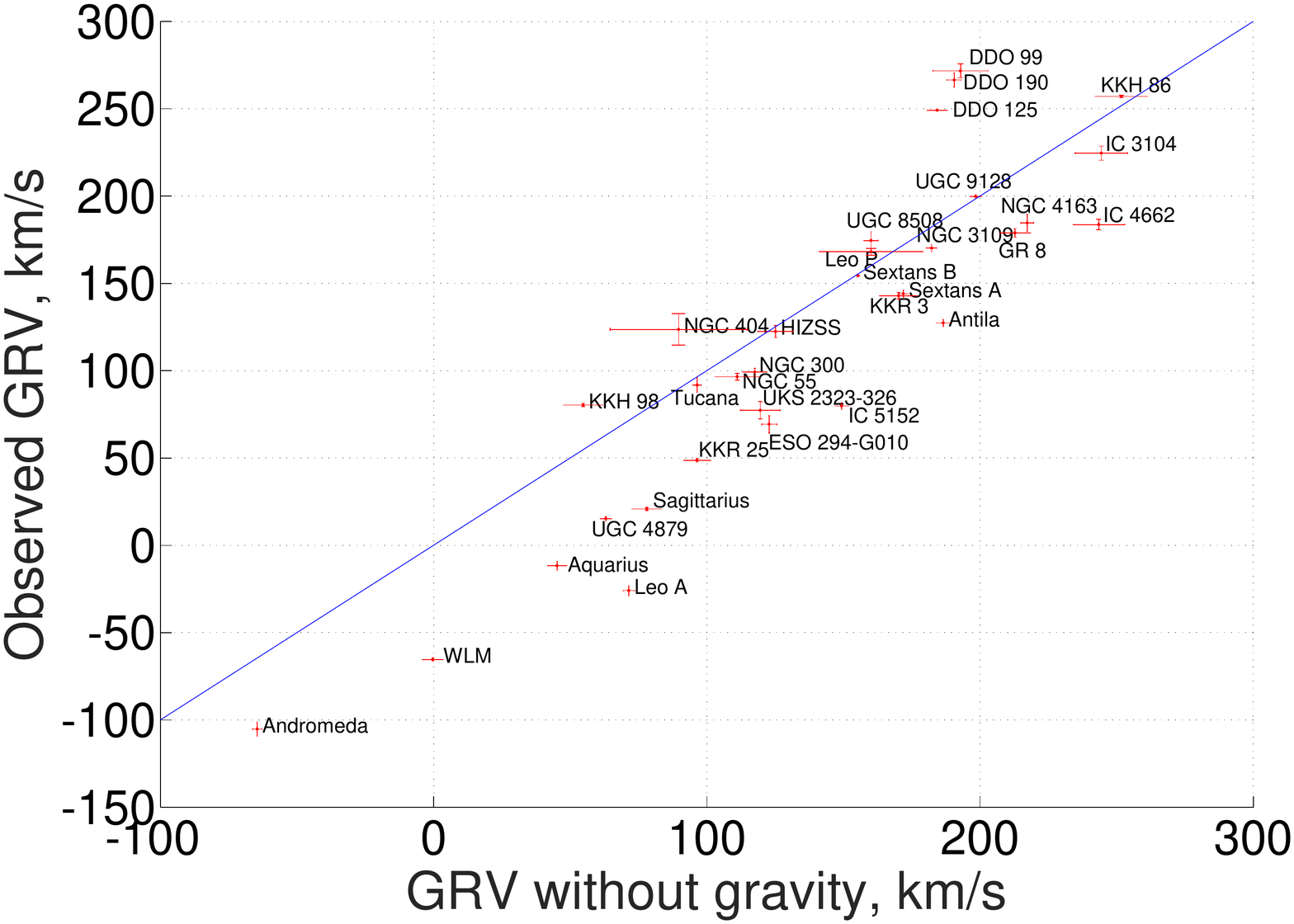}		
	\caption{Comparison between observed GRVs and the predictions of Equation \ref{Hubble_flow}, with the centre of expansion at the barycentre of the LG assuming $q_{_1} = 0.2$. The MW is taken to be moving towards this point at 90 km/s. Notice that some galaxies are moving outwards even faster than a pure Hubble flow (blue line).}
	\label{GRV_No_Gravity}
\end{figure}

GRV predictions obtained in this way are compared with observations in Figure \ref{GRV_No_Gravity}. Due to the effect of gravity, observed GRVs tend to be less than in a pure Hubble flow. Surprisingly, this is not true for some of our target galaxies, especially the DDO objects. Other examples of this behaviour have been identified recently \citep{Pawlowski_McGaugh_2014}. Thus, reducing $M_i$ does not explain the observations, at least if considered on its own.

Moreover, there is limited scope to alter $M_i$ because it is tightly correlated with the present GRV of M31 (Figure \ref{GRV_Sextans_A}). A model needs to match its GRV fairly well because it is unlikely that a minor merger with M31 or the MW could have substantially affected their relative motion. Even if such an event did occur, its net effect on the present GRV of M31 would be greatly diminished unless it occurred recently (Figure \ref{Impulsed_trajectories}).

\begin{figure}
	\centering
		\includegraphics [width = 8.5cm] {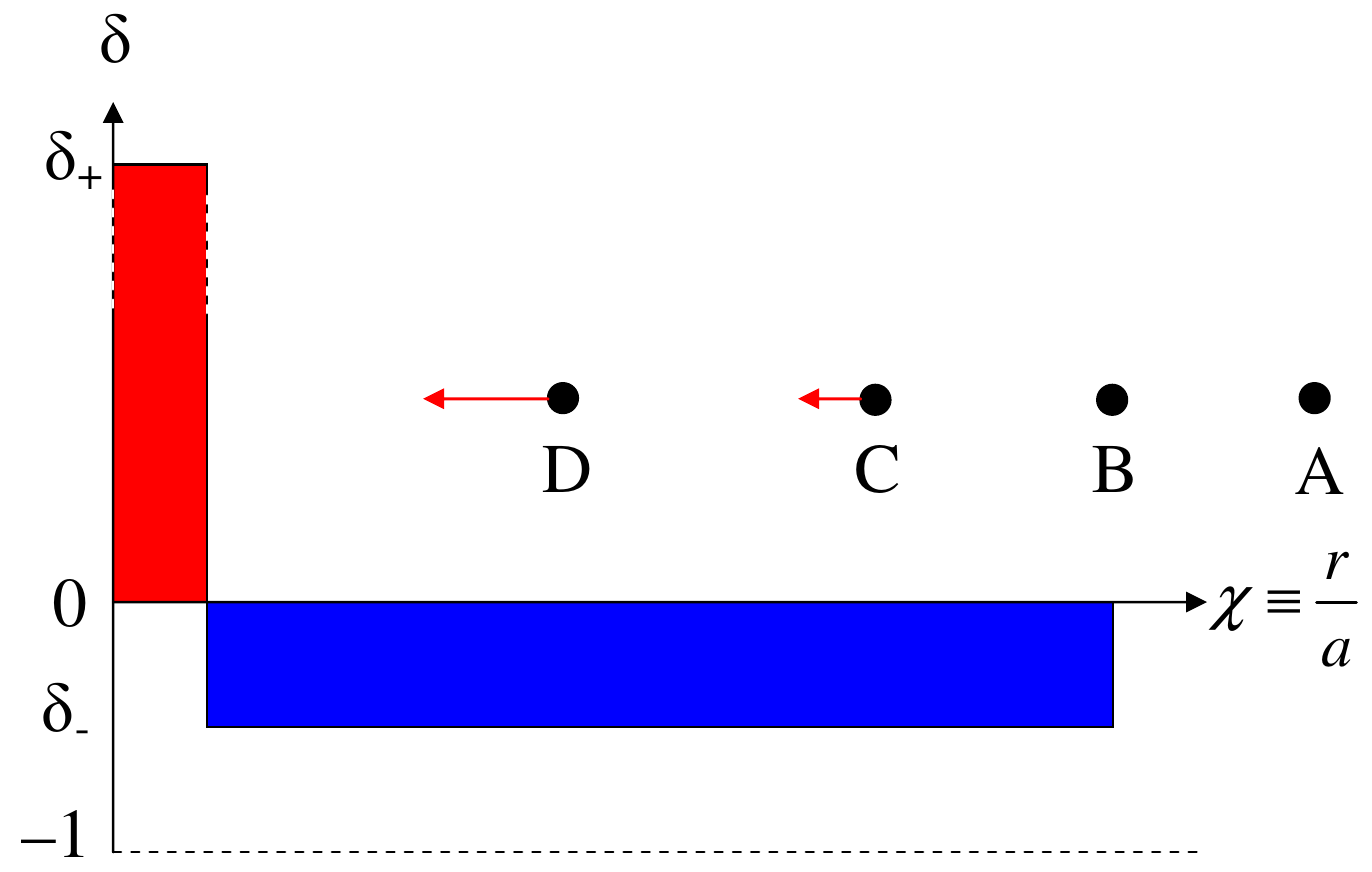}
	\caption{This shows a simple 1D model for the fractional over-density $\delta$ at the start of our simulations, plotted in co-moving co-ordinates $\chi$. The MW and M31 are treated as one object which formed from the over-density at low $\chi$. The over-density must be surrounded by an under-dense region so that the two cancel out on large scales (i.e. $\int \delta \left( \chi \right) \chi^2 d\chi = 0$). To avoid negative density, $\delta_{-} > -1$. We also show some test particle trajectories based on solving Equation \ref{Equation_61}. Particles starting further out than B simply trace the cosmic expansion ($\chi$ is constant) because the mass enclosed interior to their radius is the same as in a homogeneous universe. This is not true for particle C $-$ the region it encloses is over-dense, so $\chi$ of particle C decreases slightly. The effect is larger for particles starting closer to the LG barycentre (e.g. D).}
	\label{Density_profile}
\end{figure}

In our models, the mass in the LG is present not only in the MW and M31. We assume that the rest of the LG (RLG) contains a uniform distribution of matter with the same density as the mean cosmic density of matter $\overline{\rho}$. At this density, a sphere of co-moving radius 2.9 Mpc would have a mass equal to that of the MW and M31, assuming $M_{_{MW + M31}} = 4 \times 10^{12} M_\odot$.

At early times, it is possible that a small region encompassed the material that would later end up in the MW and M31. A surrounding under-dense region would be required so that the mean density in the union of both regions was $\overline{\rho}$. This is depicted schematically in Figure \ref{Density_profile}.

If we assume the under-dense region was completely empty (i.e. $\delta_{-} = -1$), then it would have to extend out to a co-moving radius of 2.9 Mpc. As a result, there would be no mass in the RLG, assuming this was defined to have a radius below 2.9 Mpc. It can be seen from Figure \ref{LG_Hubble_Diagram} that all our target galaxies have distances from the LG barycentre of ${< 2.9}$ Mpc. Thus, they could all be in a void.

However, one must bear in mind that test particles are retarded by the gravity of the MW and M31. This reduces the co-moving volume spanned by a cloud of test particles. Thus, if the RLG is not completely empty, then the LG contains material initially outside its present co-moving volume.

To investigate the interplay between these effects, we now consider the opposite limit in which $\left| \delta_{-} \right| \ll 1$. This corresponds to a much larger under-dense region surrounding the MW and M31 at the start of the simulations ($\sim$500 million years after the Big Bang). To better understand this case, we solved some test particle trajectories assuming a point mass in an otherwise homogeneous universe. We kept fixed the mass enclosed interior to the radius of any given test particle. This makes the equation of motion\footnote{It can be verified that a pure Hubble flow is recovered for the case $M = \delta_{-} = 0$.}
\begin{eqnarray}
	\label{Equation_61}	
	\overset{..}{r} ~&=&~ -\frac{GM_{eff}}{r^2} + {H_{_0}}^2 \Omega_{_{\Lambda, 0}} r \\
	M_{eff} &=& M + \frac{4 \rm{\pi}}{3} {r_{_i}}^3 \overline{\rho}_{_i} \left( 1 + \delta_{-} \right) ~\text{ (note: } \delta_{-} < 0 \text{)}
\end{eqnarray}

$\overline{\rho}_{_i}$ is the mean density of matter in the Universe at the time our simulations are started. $M_{eff}$ includes both the point mass $M$ and any material originally present at radii below the initial radius of the test particle. We assume that $M_{eff}$ \emph{remains constant} because the system avoids crossing of shells. This can be achieved if the massive object accretes any objects that come sufficiently close to it, rather than just letting them escape on the other side.

To obtain a final distance from the LG of 2.9 Mpc, we need an initial distance of 0.46 Mpc for a starting time corresponding to when $a_{_i} = 0.1$. This means that $16 \times 10^{12} M_\odot$ would end up within the RLG at the present time. If the RLG were to contain matter at $\overline{\rho}$, then it would only contain $4 \times 10^{12} M_\odot$. Thus, the RLG might have up to $\left( \frac{0.46}{0.29} \right)^3 = 4$ times as much material as was assumed in our calculations.

It is difficult to know how much mass is actually present in the RLG. There might be a diffuse component of dark matter or concentrations of it that have no detectable stars. Some regions are difficult to survey because of e.g. the disc of our Galaxy. Recently, significant amounts of hot gas have been discovered around the MW \citep{Salem_2015} and around M31 \citep{Lehner_2015}.

For these reasons, we assumed neither of the extreme cases just outlined. Instead, we used an intermediate assumption that the RLG contains matter with a mean density of $\overline{\rho}$ and little density variation. Roughly speaking, this corresponds to $\delta_{-} = -0.43$ and an under-density out to 3.5 Mpc. We think this is reasonable considering the distances to major mass concentrations just outside the LG (Table \ref{Perturbers}).


We investigated whether altering this assumption might affect our conclusions regarding the inferred value of $\sigma_{extra}$. To do this, we assumed the extreme case that the RLG has no mass. This means that the $\frac{\overset{..}{a}}{a}$ term present in the equations of motion (e.g. Equation \ref{Separation_history_equation}) should be replaced with ${H_{_0}}^2 \Omega_{_{\Lambda, 0}}$. We re-ran our entire analysis using equations of motion altered in this way.

If we used the same procedure as before to prevent test particles starting too close to the MW/M31, then we would end up with no test particles within $\sim$2.9 Mpc of the LG barycentre. In this case, there would be no way to obtain HRV predictions for our target galaxies, consistent with the assumption of an empty RLG. Clearly, this assumption is wrong at some level. Thus, we allow test particle trajectories to start anywhere as long as they end up at the correct position.

Altering the setup of our simulations in this way reduced $\sigma_{extra}$ by $\sim$7 km/s. Individual radial velocities are often increased by larger amounts. This tends to reduce the discrepancy with observations. However, the overall effect is small because, for the same MW and M31 mass, the GRV of M31 is increased. To bring it back down to the observed value, the mass of the MW and M31 have to be increased, reducing the predicted GRVs of other LG galaxies.

We checked this explicitly by comparing the marginalised posterior probability distribution of $M_i$. Assuming the RLG has a mean density of $\overline{\rho}$, the total initial LG mass in units of $10^{12} M_\odot$ is $4.2 \pm 0.4$. However, if we assume an empty RLG, this rises to $5.2 \pm 0.4$. This seems rather high, but is in line with similar calculations by other workers \citep{Partridge_2013}.\footnote{Part of the difference arises because Cen A is not usually included in timing argument analyses of the LG.} We suggest that this result points towards a RLG that can't be considered empty for the purposes of the timing argument. However, more reasonable values for $M_i$ are obtained if one includes the kinematic effect of a sufficiently massive Large Magellanic Cloud (Figure \ref{LMC_joint_M_q}).

\subsection{Increased Hubble constant}
\label{High_H_0}

Another way to increase model-predicted HRVs is to increase $H_{_0}$. The cosmological value seems to be fairly well constrained \citep{Planck_2015}. Once certain biases are taken into account, this measurement seems to be consistent with surveys of Type Ia supernovae \citep{Rigault_2015}. However, there is also some cosmic variance: under-dense regions of the Universe expand faster than the average. If we are in such a region, this would lead to the value of $H_{_0}$ appropriate for the local Universe being higher than that for the Universe as a whole \citep{Wojtak_2013}.

To account for this possibility, we performed another simulation with $H_{_0}$ raised by 5 km/s/Mpc. However, we were careful to bear in mind that Planck gives a tight constraint on the age of the Universe. To avoid altering this, we had to further adjust the adopted cosmology. For simplicity, we kept this flat. The parameters used are shown in Table \ref{Revised_cosmology}.

\begin{figure}
	\centering
		\includegraphics [width = 8.5cm] {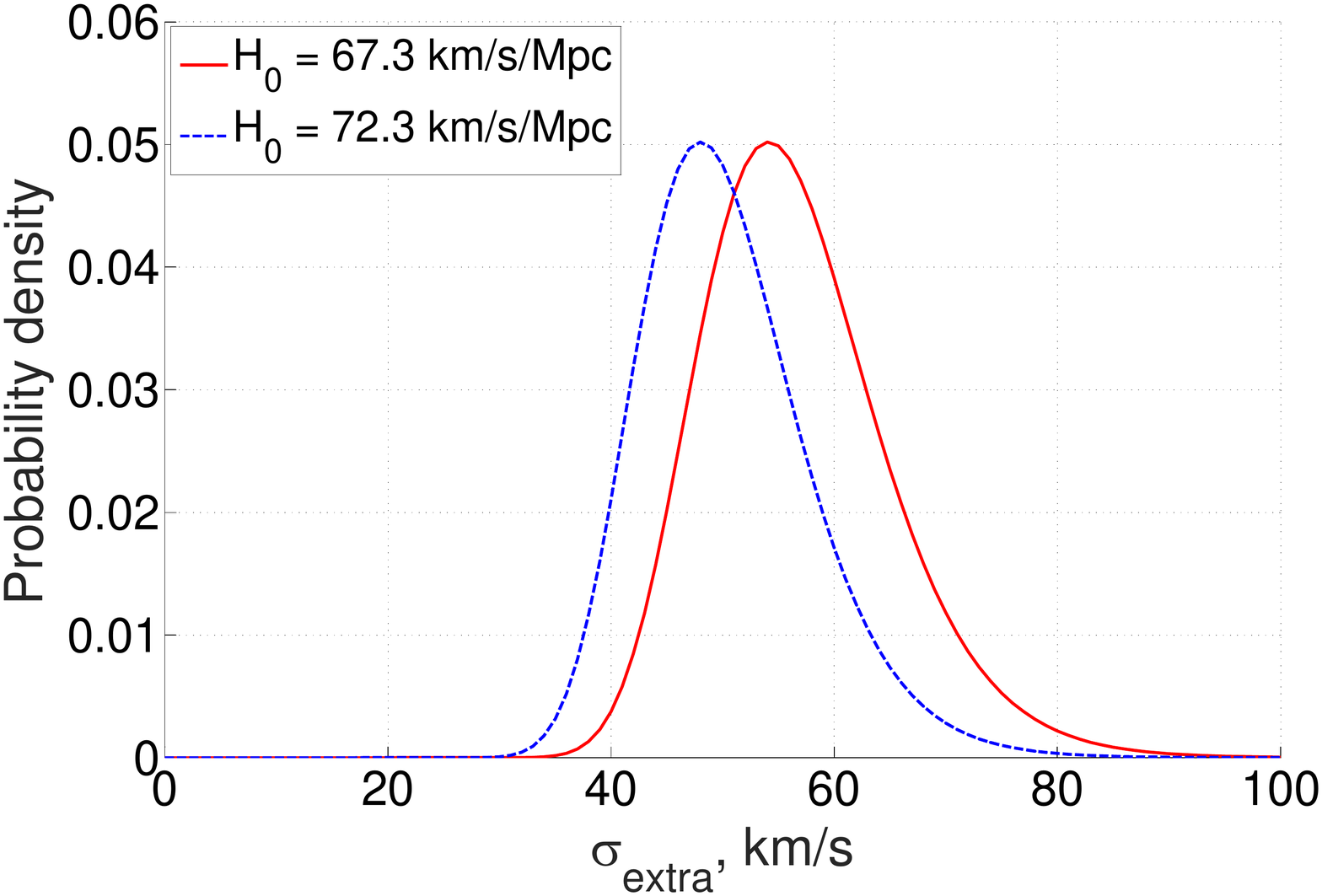}		
	\caption{The posterior on $\sigma_{extra}$ is shown for two plausible values of the Hubble constant $H_{_0}$. The age of the universe is the same in both models, with parameters adjusted accordingly (Table \ref{Revised_cosmology}). Raising $H_{_0}$ by 5 km/s/Mpc reduces $\sigma_{extra}$ by $\sim$5 km/s.} 
	\label{Sigma_extra_high_H}
\end{figure}

The resulting posterior on $\sigma_{extra}$ is shown in Figure \ref{Sigma_extra_high_H}. As might be expected, increasing $H_{_0}$ lowers the inferred value of $\sigma_{extra}$, but only by $\sim$5 km/s. This is similar to the effect of assuming the rest of the LG is empty instead of filled with matter at a density of $\overline{\rho}$ (Section \ref{Reduced_LG_Mass}). This is reassuring as the simulations work in slightly different ways.

Of course, it is only possible to count this reduction in $\sigma_{extra}$ once: to account for the RLG being less dense than in our models, one can either alter the equations of motion to make the RLG empty or one can raise $H_{_0}$ slightly. Whichever method one prefers to use, the effect is not sufficient to explain the observations, although it does help.

\begin{table}
	\begin{tabular}{lll}
	\hline
	Parameter & Old value & New value \\
	\hline
	$H_{_0}$ & 67.3 km/s/Mpc & 72.3 km/s/Mpc \\
	$\Omega_{m, 0}$ & 0.315 & 0.243 \\
	$\Omega_{\Lambda, 0}$ & 0.685	& 0.757 \\
	\hline	
	\end{tabular}
	\caption{Alterations to cosmological parameters in Table \ref{Priors} for the high $H_{_0}$ model shown in Figure \ref{Sigma_extra_high_H}. In both cases, the universe is flat and equally old (13.81 Gyr).}
\label{Revised_cosmology}
\end{table}

One thing that may be in favour of models with an under-dense RLG is the inferred value of $q_{_1}$, the fraction of the LG mass in the MW. When we tried to make the RLG empty by altering the equations of motion (Section \ref{Reduced_LG_Mass}), our analysis preferred ${0.38^{+0.06}_{-0.05}}$ instead of ${0.14 \pm 0.07}$. One might expect a similar effect to occur when we raise $H_{_0} ~-$ after all, the effect on $\sigma_{extra}$ is very similar. However, our calculations show almost no change in the inferred value of $q_{_1}$ due to a higher Hubble constant.

\subsection{Tides from objects outside the Local Group}
\label{Tides}

\subsubsection{Centaurus A}

To better understand the effect of Cen A on our results, we repeated our analysis without including it. The results are shown in Table \ref{Cen_A_effect}. Broadly speaking, the results are similar in both cases, although there are some subtle differences.


\begin{table}
	\begin{tabular}{llll}
	\hline
	Parameter & Prior & Posterior & Posterior with\\
	 \& units & & without Cen A & Centaurus A \\
	\hline
	$\sigma_{extra}$, km/s & 0 $-$ 100 & ${54}^{+8.9}_{-7.0}$ & $46^{+7.4}_{-5.6}$ \\ [5pt]
	$M_i$, ${10}^{12} M_\odot$ & 2 $-$ 6.6 & ${3.42}_{-0.32}^{+0.35}$ & ${4.14}_{-0.31}^{+0.35}$ \\ [5pt]
	$q_{_1}$ & 0.2 $-$ 0.8 & ${0.20}_{-0}^{+0.060}$ & ${0.20}_{-0}^{+0.049}$ \\ [5pt]
	$v_{c, \odot}$, km/s & $239 \pm 5$ & ${242.0}^{+4.9}_{-4.7}$ & ${239.9}^{+4.9}_{-4.7}$ \\
	\hline	
	\end{tabular}
	\caption{Effect of Centaurus A on posteriors of model parameters. Both analyses shown here have a uniform prior on $q_{_1}$ over the range 0.2$-$0.8.}
\label{Cen_A_effect}
\end{table}

\begin{figure}
	\centering
		\includegraphics [width = 8.5cm] {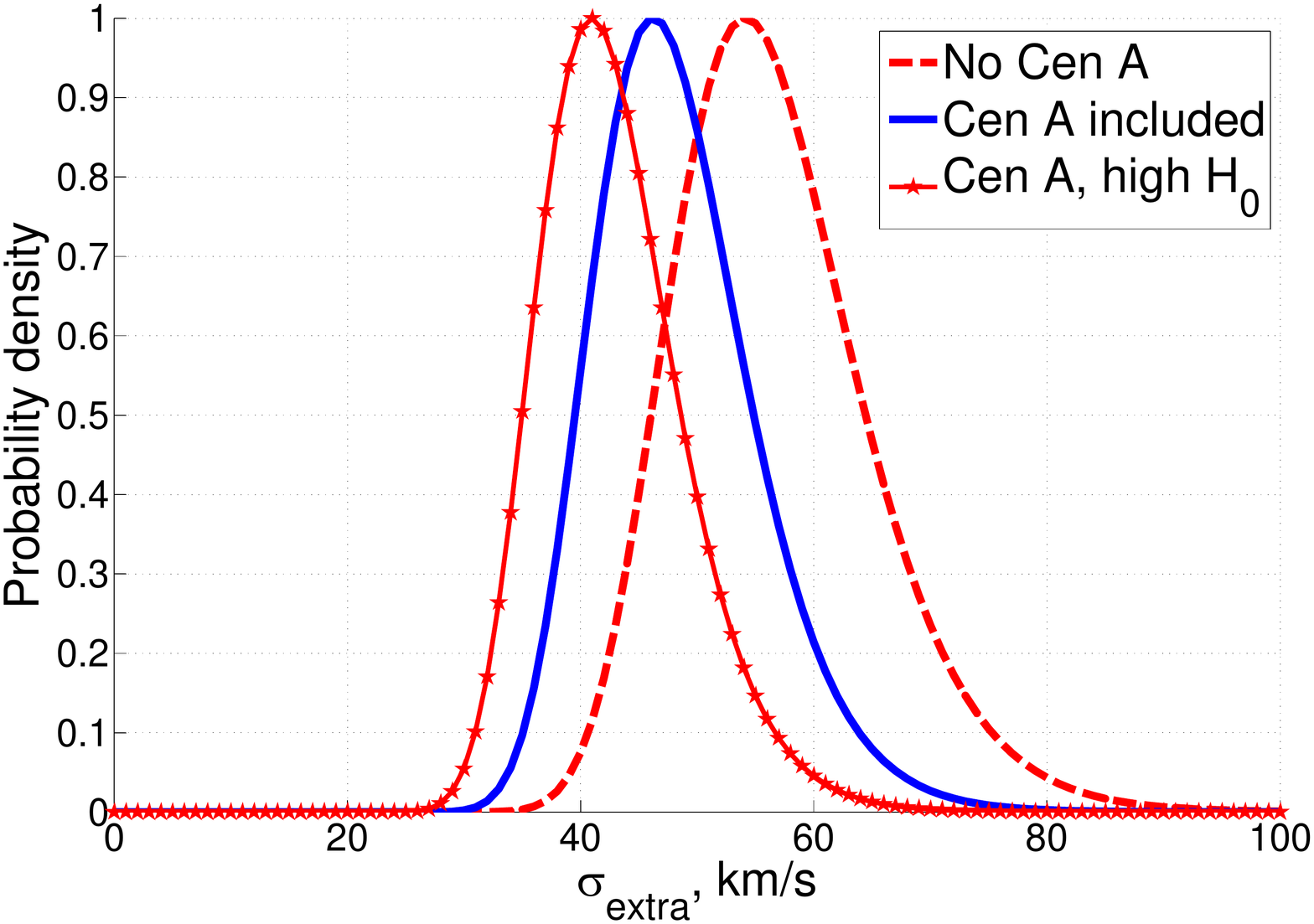}		
	\caption{Posterior probability distribution of $\sigma_{extra}$ under different model assumptions. Including Centaurus A reduces $\sigma_{extra}$ by $\sim$8 km/s. Also raising $H_{_0}$ as described in Table \ref{Revised_cosmology} reduces $\sigma_{extra}$ by a further $\sim$5 km/s.}
	\label{Sigma_extra_comparison}
\end{figure}

Being close to the MW$-$M31 line, Cen A pulls the MW and M31 apart, increasing the GRV of M31 by $\sim$10 km/s. To bring it back down to the observed value, $M_i$ would need to be increased by $\ssim {10}^{12} M_\odot$ (Figure \ref{GRV_Sextans_A}). This is indeed roughly what happens to the posterior on $M_i$. Other effects are harder to understand, such as why including Cen A leads to better agreement with the LSR speed measured by \citet{McMillan_2011}.

The posterior distribution of $\sigma_{extra}$ is shown in Figure \ref{Sigma_extra_comparison}. Including Cen A reduces its most likely value from $54_{-7}^{+9}$ km/s to $46_{-6}^{+7}$ km/s. If $H_{_0}$ is also increased to 72.3 km/s/Mpc, then $\sigma_{extra}$ is further reduced to $41_{-5}^{+6}$ km/s. 

We only tried one possible mass of Cen A ($4 \times 10^{12} M_\odot$). Including it at this mass reduces $\sigma_{extra}$ by $\sim$ 8 km/s. It is possible that adopting a higher mass would reduce it further.\footnote{Though it might not, see Figure \ref{LMC_trend_s}.} However, it is unlikely that Cen A is more massive than $5 \times 10^{12} M_\odot$ (see Figure 1 of \citet{Karachentsev_2005}). This is only 25\% higher than our adopted mass. Thus, using the highest plausible Cen A mass rather than our adopted value might well reduce $\sigma_{extra}$, but only by another $\sim$ 2 km/s.


The inclusion of Centaurus A affects the Hubble diagram for the LG, increasing $\sigma_{_H}$. This is a tidal effect and is therefore larger at greater distances. At 3 Mpc, we found a range in radial velocity from the LG barycentre of $\sim$ 70 km/s, falling to perhaps half that at 2 Mpc. This corresponds to $\sigma_{_H} \sim 10 - 15$ km/s. Although this is below the $\sim$30 km/s found in cosmological simulations \citep{Aragon_Calvo_2011}, it does suggest that some of the `scatter' about the Hubble flow found in such simulations can be accounted for using an axisymmetric model rather than a spherically symmetric one.


\subsubsection{IC 342 and M81}
\label{Tides_IC342_M81}

\begin{figure}
	\centering
		\includegraphics [width = 8.5cm] {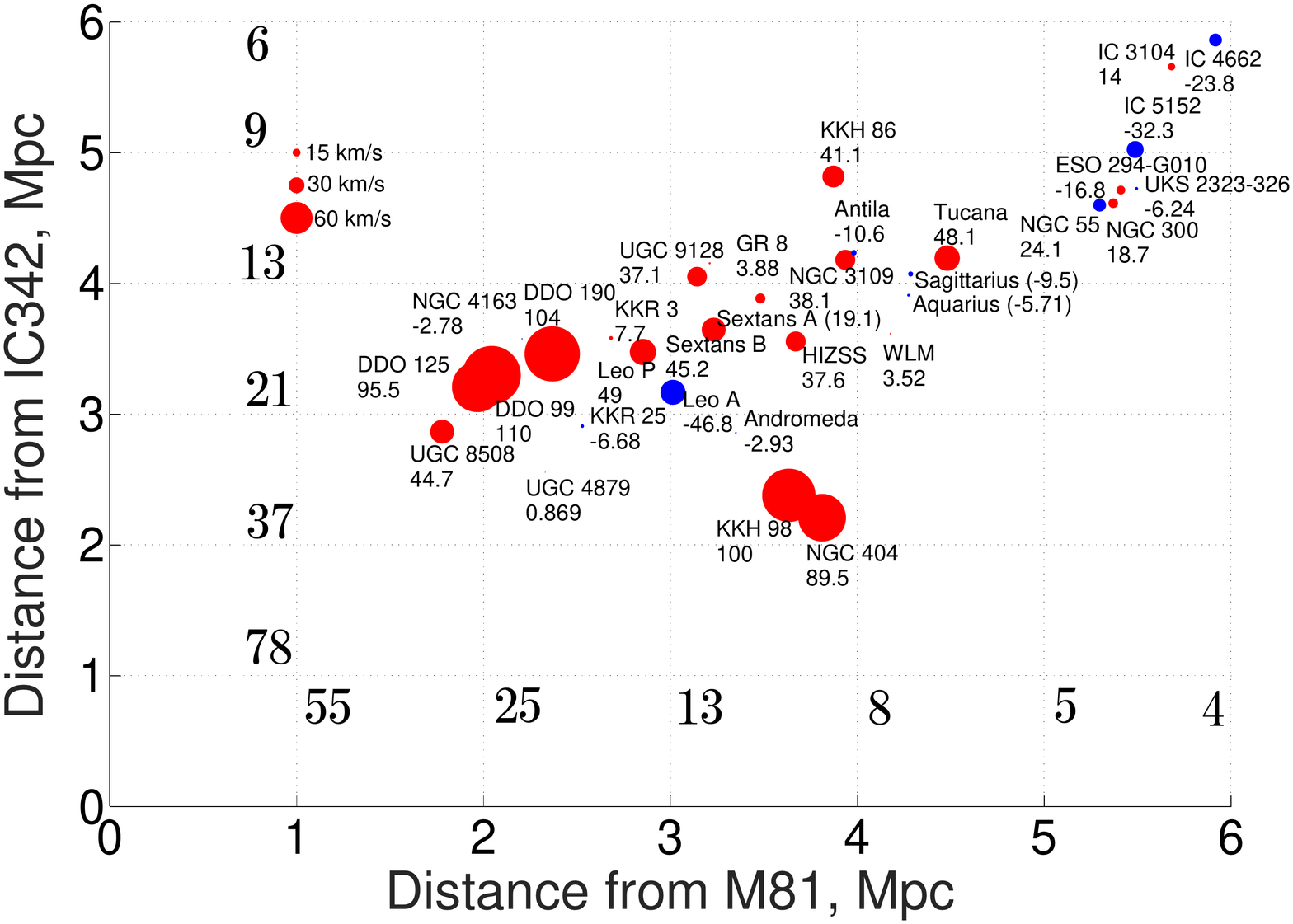}		
	\caption{Observed $-$ predicted HRVs (i.e. $\Delta HRV$s) of indicated galaxies as a function of distance from the perturbers in Table \ref{Perturbers}. The radius of each marker $\propto$ the discrepancy, which we list below the name of each galaxy. Blue indicates a measured HRV below that in the most plausible model ($q_{_1} = 0.2$, $M_i = 4.2 \times {10}^{12} M_\odot$), while red shows the opposite. The numbers on the gridlines show the peculiar velocity in km/s at that distance from each perturber if it was the only object in the universe (see text). A more detailed model for how perturbers affect observations is shown in Table \ref{IC342_M81_effect}.}
	\label{Tide_correlation}
\end{figure}

Including Centaurus A improves the fit to the data slightly but still leaves a very poor fit. As it is the most massive perturber, this suggests that tides can't explain the discrepant HRVs. To check this conclusion, we cross-correlated the discrepancies in the HRVs with the distances between our target galaxies and the remaining perturbers in Table \ref{Perturbers}. This is shown in Figure \ref{Tide_correlation}. The discrepancy seems to be larger for objects closer to IC 342 or to M81. Thus, we tried to see if tides from these objects might help to explain the observations.


We provide two ways of estimating the effects of tides raised by IC 342 and M81 on the Local Group. First, we treat each perturber as the only object in the universe. We solve test particle trajectories in the usual way and target a particular final separation with the perturber. We then record the peculiar velocity of this trajectory. Using the perturber masses in Table \ref{Perturbers}, the results obtained in this way are indicated in km/s on the gridlines of Figure \ref{Tide_correlation}.

In this very simplistic model, the effect of each perturber is just an extra velocity towards it with the calculated magnitude. However, the direction of this velocity is not directly away from the MW. For example, IC 342 should hardly affect the GRV of KKH 98 because, as perceived by KKH 98, the MW and IC 342 are almost at right angles (angle $\sim 89.4 ^\circ$).

The perturber would also have a small effect on the motion of the MW, this being $\sim$15 km/s towards each perturber in the context of this model. For a target near a perturber, one expects them to also be nearby on the sky. Thus, the MW would be pulled towards the target to some extent, reducing its GRV. This might be why our more detailed model for tides (see below) often predicts that they would reduce the GRVs of target galaxies.

%
%
%
%
%
%

%

Our more detailed model involves two gravitating masses. We treat the MW \& M31 as a single object with mass $4 \times {10}^{12} M_\odot$ and assume $q_{_1} = 0.2$. This object represents the LG. We put the LG and the perturber along the $y$-axis and solve both objects forwards using Equation \ref{Separation_history_equation} (the relevant mass is that of the MW$+$M31$+$perturber). Their initial separation is varied so as to get a final separation equal to the observed distance between the LG barycentre and the perturber.

\begin{table}
	\begin{tabular}{lccc}
	\hline
	Galaxy & $\Delta$HRV & Effect & Effect \\
	& (km/s) & of M81 & of IC 342 \\
	\hline
	DDO 99 & $110 \pm 13$ & $-$6.5 & $-$10.3 \\
	DDO 190 & $104 \pm 5$ & $-$7.5 & $-$9.8 \\
	KKH 98 & $100 \pm 12$ & $-$3.0 & $-$9.5 \\
	DDO 125 & $95.5 \pm 4.5$ & $-$5.9 & $-$10.6 \\
	NGC 404 & $89 \pm 37$ & $-$3.6 & $-$13.2 \\
	Tucana & $48 \pm 6$ & 2.0 & 5.4 \\
	NGC 3109 & $38.1 \pm 5.1$ & $-$1.9 & 1.4 \\
	\hline
	ESO 294-G010 & $-16.8 \pm 5.8$ & 3.6 & 4.9 \\
	IC 4662 & $-24 \pm 14$ & 3.1 & 9.4 \\
	IC 5152 & $-32.3 \pm 4.0$ & 3.7 & 7.1 \\
	Leo A & $-46.8 \pm 4.4$ & $-$2.3 & $-$3.1 \\
	\hline
	\end{tabular}
	\caption{$\Delta HRV$s for the LG galaxies most discrepant with our model. Error budgets are found by adding $\sigma_{pos}$ and $\sigma_{v_h}$ in quadrature. We also give an estimate in km/s for how much M81 and IC 342 might have affected the GRV of each galaxy (method described in text).}
\label{IC342_M81_effect}
\end{table}

We then determine how a target galaxy would fit into this picture. We solve a test particle trajectory so that it ends up at the correct distance from the LG barycentre and at the correct angle to the perturber as perceived at the LG particle.\footnote{A 2D model is sufficient for this as there are three particles.} The final GRV of the test particle is determined using Equation \ref{Model_GRV}, referred to the LG particle rather than the MW.

To determine the effect of the perturber, we then (effectively) reduce the perturber mass to 0 and repeat the calculation. The final GRV of the test particle is compared between the two simulations. Some results from this procedure are shown in Table \ref{IC342_M81_effect}.

The combination of large distances from the perturbers ($\ga$ 2 Mpc) and projection effects reduce how much tides might have affected the GRVs of target galaxies. As a result, tides from IC 342 and M81 can't explain the very high HRVs of targets such as the DDO objects in the context of this model. In fact, for several galaxies like these, tides seem to reduce GRVs and thus make the discrepancy even worse. Thus, we do not believe that tides are responsible for the discrepancies, assuming we have reasonable perturber masses (Table \ref{Perturbers}) and a good method of estimating their effects.

\subsubsection{The Great Attractor}
\label{Great_Attractor}

There are additional structures in the Universe on a larger scale which might be pertinent to our analysis. In particular, the Local Group as a whole has a velocity of $\sim$630 km/s with respect to the surface of last scattering \citep{Kogut_1993}. It is thought that this is mostly due to the gravity of the Great Attractor \citep[GA,][]{Mieske_2005}. Assuming a distance of 84 Mpc, it is clear that tides raised by the GA can have a non-negligible impact on motions within the LG.

As the GA is much more distant from the MW than objects within the LG, we use the distant tide approximation. Treating the MW and other target galaxies as freely falling in the gravitational field of a distant point mass, the change in the Galactocentric radial velocity of a target galaxy is
\begin{eqnarray}
	\Delta GRV_{GA} = \left(3 \cos^2 \theta - 1 \right) \frac{v_{pec, LG} ~d}{d_{GA}} ~~~~~\text{for } d \ll d_{GA}
	\label{Distant_tide_approximation}
\end{eqnarray}

Here, $d_{GA}$ is the distance to the GA while $\theta$ is the angle on our sky between it and the target galaxy, which is at a heliocentric distance $d$. The GA is assumed to have caused the LG to gain a peculiar velocity of $v_{pec, LG} = 630$ km/s. We take the GA to be in the direction $l = 325^\circ$, $b = -7^\circ$ in Galactic co-ordinates \citep{Kraan_Korteweg_2000}.

For $\theta$ close to 0 or $180^\circ$, the GA tends to increase GRVs. However, for $\theta$ close to $90^\circ$, the GA reduces GRVs. This arises because both the MW and the target galaxy fall towards the perturber at similar rates. As their co-moving distance from the GA decreases, so also does their co-moving distance from each other.

Due to the GA, a test particle started with the same initial conditions will end up with an altered position as well as velocity. Thus, the initial position must be altered to match a fixed final position. This reduces the effect of the GA on predicted velocities of target galaxies (a similar effect is shown in Figure \ref{Impulsed_trajectories}).

Because of Hubble drag, present peculiar velocities are mostly sensitive to tides at late times. Thus, we expect Equation \ref{Distant_tide_approximation} to provide a reasonable approximation as long as we have accurate distances to the relevant objects and know their sky positions.

Although we may overestimate the magnitude of $\Delta GRV_{GA}$, it is much harder to get its sign wrong. This is because the sign is dependent on the factor of $\left( 3 \cos^2 \theta - 1 \right)$, a quantity sensitive only to the (usually well-known) sky positions of relevant objects but not to their distances. The trajectory of a distant LG galaxy is unlikely to have deviated much from a radial orbit. The GA is much more distant so it was probably always in much the same direction on the sky. As a result, we expect $\theta$ to never have been much different to its present value. Although this may not be true at very early times, Hubble drag makes the system `forget' about the forces acting at such times.

\begin{figure}
	\centering
		\includegraphics [width = 8.5cm] {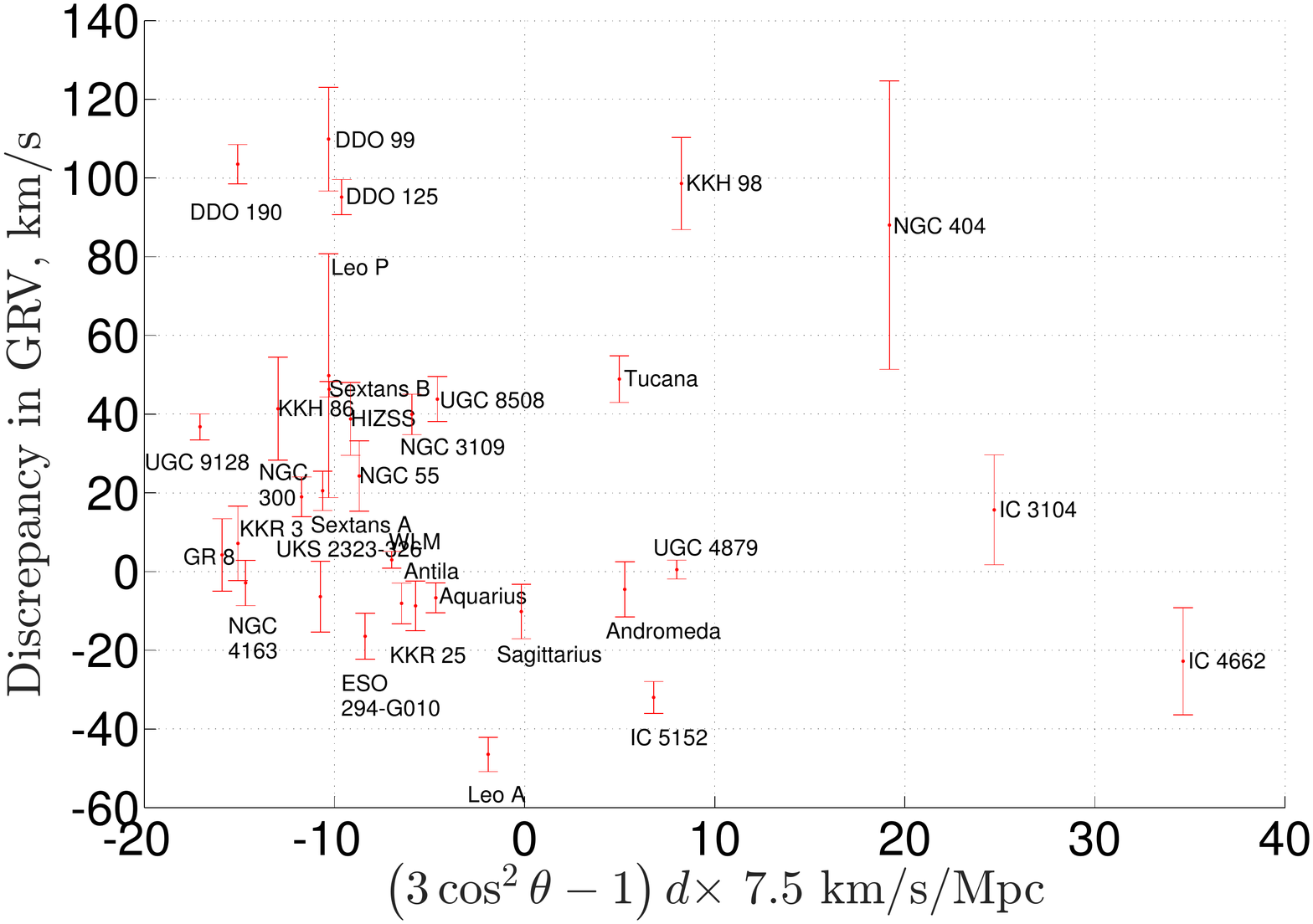}		
	\caption{$\Delta HRV$s are plotted against our estimate for how much the Great Attractor might have increased the HRV of each galaxy. The distance $d$ used here is heliocentric. Notice that the DDO objects would likely have their radial velocities reduced by tides from the Great Attractor.}
	\label{Great_Attractor_effect}
\end{figure}

In Figure \ref{Great_Attractor_effect}, we show how much tides from the GA would likely affect the HRV of each target galaxy. It is interesting to see the results for the galaxies with the highest $\Delta HRV$s, in particular the DDO objects (Table \ref{IC342_M81_effect}). Because of their positions on the sky, tides from the GA would actually \emph{reduce} their GRVs. This makes it more difficult to explain their high observed HRVs.

Perhaps more important is the lack of any apparent correlation between HRV discrepancies and the effect of tides raised by the GA. This suggests that it can't reconcile the differences between our best-fitting simulation and observations. In fact, it would probably make matters worse as it reduces HRVs for 3 out of the 4 objects which likely have $\Delta HRV > 90$ km/s (i.e. $\Delta HRV > 3 \sigma_{_H}$).

To estimate how much the GA might affect $\sigma_{extra}$, we adjusted model-predicted GRVs of all our target galaxies using Equation \ref{Distant_tide_approximation}. We then re-ran our statistical analysis. This raised $\sigma_{extra}$ by $\sim$6 km/s (or 4 km/s if the RLG is assumed empty). The actual effect of the GA is probably smaller because it affects final positions of test particles as well as their velocities. Still, it seems likely that the GA makes it harder rather than easier to explain observed HRVs.

Although the only large-scale structure we consider explicitly is the Great Attractor, it is of course possible to perform a similar analysis for an external perturber in another direction. Indeed, all possible directions can be investigated using a grid method. This was done by \citet{Jorge_2016}, who used a 1D model for the LG (see Section \ref{Comparison_with_Jorge}). Fortunately, tides raised by LSS are more important towards the edges of the LG. Here, the greater distance from the MW and M31 makes it more realistic to consider them as a single point mass. Thus, one might expect their analysis to be reasonably sensitive to tides raised by LSS. As a result, it is important to note that they `found no statistically meaningful deviation between the velocities predicted by the point-mass model and the location of galaxies on the sky.'

\subsection{Kinematic corrections due to massive satellites}
\label{Large_Magellanic_Cloud}

Massive satellite galaxies of the MW can affect our timing argument analysis because of an indirect kinematic effect. Instead of dealing with just the MW, we should really deal with the MW system ($\equiv$ MW $+$ satellite). The brightest satellite galaxy of the MW is the Large Magellanic Cloud (LMC). This is $\sim$ 50 kpc from the Sun \citep{Pietrzynski_2013}. Being much fainter than the MW, we expect it to be much less massive. As a result, it shifts the barycentre of the MW system by $\la$ 10 kpc.

Previously, we neglected errors that arise due to the heliocentric directions towards other LG galaxies not being the same as the directions from the centre of the MW. This is because the Sun is only $\sim$ 8 kpc from there \citep{McMillan_2011}. Similarly, we also neglect any errors that arise due to the LMC altering the position of the MW system's barycentre. This is because even the nearest target galaxy is $\sim$ 800 kpc away. Moreover, the directions from the Sun towards the Galactic Centre and towards the LMC are almost orthogonal, meaning that the errors due to these approximations would add in quadrature rather than linearly.

Unlike the position of the MW system's barycentre, its \emph{velocity} may be significantly altered by the LMC. Consequently, we determined $\bm v_{_{LMC}}$, its space velocity with respect to the MW. This requires knowledge of its heliocentric radial velocity \citep{McConnachie_2012} and its proper motion \citep{Kallivayalil_2013} multiplied by its distance \citep{Pietrzynski_2013}. The velocity of the Sun with respect to the MW is also required (Table \ref{Priors}).\footnote{The important quantity here is actually the velocity of the LMC with respect to the Sun, so it is essential to use the same $\bm v_\odot$ as in the rest of our analysis.} Using these references, we found that the speed of the LMC with respect to the MW is 319,845.6 m/s of which 229,401.5 m/s is directed towards the North Galactic Pole. Importantly, the component of this velocity directly away from M31 is 241,223.4 m/s.

To apply a kinematic correction for the motion of the LMC, we note that the velocity of the Sun with respect to the MW should now be altered to its velocity with respect to the barycentre of the MW system.
\begin{eqnarray}
	\label{LMC_adjustment}
	\bm v_\odot &\to& \bm v_\odot - f_{recoil}~q_{_{LMC}} \bm v_{_{LMC}} \\
	q_{_{LMC}} &\equiv& \frac{M_{LMC}}{M_{MW} + M_{LMC}}
\end{eqnarray}

Here, $\bm v_{_{LMC}}$ is the velocity of the LMC with respect to the MW disc. Note that the MW mass $M_{_{MW}}$ does not include a contribution from the LMC mass $M_{_{LMC}}$. The $-$ sign in Equation \ref{LMC_adjustment} arises because we are correcting for the recoil induced by the LMC on the MW. If the LMC were bound to the MW and the two were orbiting their common centre of mass, then a simple application of Newton's third law would show that we should set $f_{recoil} = 1$.

It is possible that the LMC is not bound to the MW but rather is on a first infall trajectory \citep[e.g.][]{Besla_2007}. Indeed, its high speed relative to the MW means it is unlikely to be a gravitationally bound satellite galaxy \citep{Wu_2008}. Although the magnitude of $\bm v_{_{LMC}}$ is now believed to be smaller than the $\sim$380 km/s assumed in this work, other considerations continue to suggest that it is unbound (see section 6.4 of \citet{Kallivayalil_2013}). Thus, most of its velocity might have been present even when it was far from the MW. In this case, only part of its velocity would have been gained due to gravity from the MW. As a result, the recoil of the MW induced by the LMC would be less than if the MW and LMC were bound.

To account for this possibility, we introduce the parameter $f_{recoil}$, the fraction of the momentum of the LMC that has been gained due to gravity from the MW. We assume that the direction of the recoil induced by the LMC on the MW is aligned with the MW$-$LMC relative velocity. The effect of the LMC on our analysis is maximized if we set $f_{recoil} = 1$ and assume the LMC is bound to the MW. The validity of this assumption remains an open question. However, to put an upper limit on the effect of the LMC, we will make this assumption.

In applying a kinematic correction for the LMC, another major uncertainty is its mass relative to the MW. Recent rotation curve measurements of the LMC based on both radial velocities and proper motions indicate a flatline level in its outer regions of $\sim$ 90 km/s \citep{Kallivayalil_2013}. Extrapolating to a tidal radius of 25 kpc as suggested by this work, we obtain an enclosed mass of $4.7 \times 10^{10} M_\odot$. The actual value is likely to be smaller because other studies indicate a slower-rotating LMC \citep{Alves_2000}.

The LMC mass can also be estimated using an abundance matching technique. This yields a pre-infall mass of $\sim 1.9 \times 10^{11} M_\odot$ \citep{LMC_Mass}. Not all of this mass can get as close as 50 kpc to the MW and exert a force on it. This is because the outer parts of the LMC's dark matter halo have likely been tidally stripped due to its close approach of the MW. The work of \citet{LMC_Mass} suggests that we should reduce the pre-infall mass of the LMC by a factor of $\sim$ 3.5 to account for this (see their Figure 11). This makes both estimates of the LMC mass agree.

However, several recent investigations suggest a much higher LMC mass, which may be possible if it has not been tidally stripped to a significant extent. This is tied to the issue of whether the LMC is on its first infall into the MW, as first suggested by \citet{Besla_2007}. Those authors conducted further investigations into this possibility \citep{Besla_2010, Besla_2012}. Recently, it was shown that a first infall of a massive LMC could induce a recoil on the MW of as much as $\sim$70 km/s \citep{Gomez_2015}, corresponding to $q_{_{LMC}} \la 0.2$. A high LMC mass is also hinted at by the discovery of stellar streams around the Magellanic Clouds \citep{Belokurov_2016} and by its high star formation rate, suggestive of a first infall \citep{Tollerud_2011}.

A very massive LMC would exert strong tides on the disc of the MW, perhaps warping it more than is observed. Assuming a bound LMC, it proved possible to reproduce important properties of the observed warp with a fairly low LMC mass of just $20 \times 10^9 M_\odot$ \citep{Weinberg_2006}. If the LMC was instead on its first infall, it would only recently have had a substantial effect on the MW. This might be compensated by a higher LMC mass. The interplay between these effects deserves further investigation.

To incorporate the LMC into our analysis, we assumed that the relevant MW mass for the purposes of the timing argument is the combined mass of the MW and the LMC. Even if the LMC was quite far from the MW in the past, it seems likely that other LG galaxies were much further still, so that the MW and LMC can be treated as a single point mass. Neglecting the small increase in MW mass due to accretion (Figure \ref{Mass_accretion}), this means that
\begin{eqnarray}
	q_{_{LMC}} ~=~ \frac{M_{LMC}}{q_{_1} M_i}
	\label{q_LMC}
\end{eqnarray}

In models with a very low total LG mass $M_i$ and a very small fraction $q_{_1}$ of this in the MW, it is possible to get $q_{_{LMC}} > 1$. To avoid this occurring, we calculated $q_{_{LMC}}$ using Equation \ref{q_LMC} and then capped its value at 0.3. This is a very generous upper limit on the ratio between the LMC and MW masses $-$ high-resolution $\Lambda$CDM simulations indicate that it is very unlikely to find a sub-halo with ${>10\%}$ as much mass as the main halo \citep{LMC_Mass}. Moreover, virial masses scale approximately as the cube of rotation velocities \citep{Evrard_2008}. Assuming the MW rotation curve in the DM-dominated regions is above 180 km/s \citep{Kafle_2012}, this suggests a LMC mass of below $\ssim \frac{1}{8}$ that of the MW.

To check if our imposed upper limit on $q_{_{LMC}}$ was affecting our analysis, we determined the best-fitting values of $M_i$ and $q_{_1}$ for each value of $M_{_{LMC}}$ that we tried. We then verified that the most likely total MW system mass ($M_i q_{_1}$) always comfortably exceeded $\frac{10}{3}$ times the LMC mass (in fact, it was never below $4.5 M_{_{LMC}} ~-$ see Figure \ref{LMC_joint_M_q}). This indicates that our analysis should not have been greatly affected by our decision to cap $q_{_{LMC}}$ at 0.3.

We investigated a wide range of LMC masses (0$-$250$\times 10^9 M_\odot$) and tried 101 linearly spaced values in this range. Each time, we recalculated the probability density function over the other 4 model parameters, meaning that we essentially added a fifth parameter using a grid method. As we are only including the kinematic effect of the LMC, it is not necessary to repeat our dynamical simulations. Only the statistical analysis had to be redone. The posterior probability distribution was then marginalized over each of the model parameters one at a time to look for trends with the LMC mass.

\begin{figure}
	\centering
		\includegraphics [width = 8.5cm] {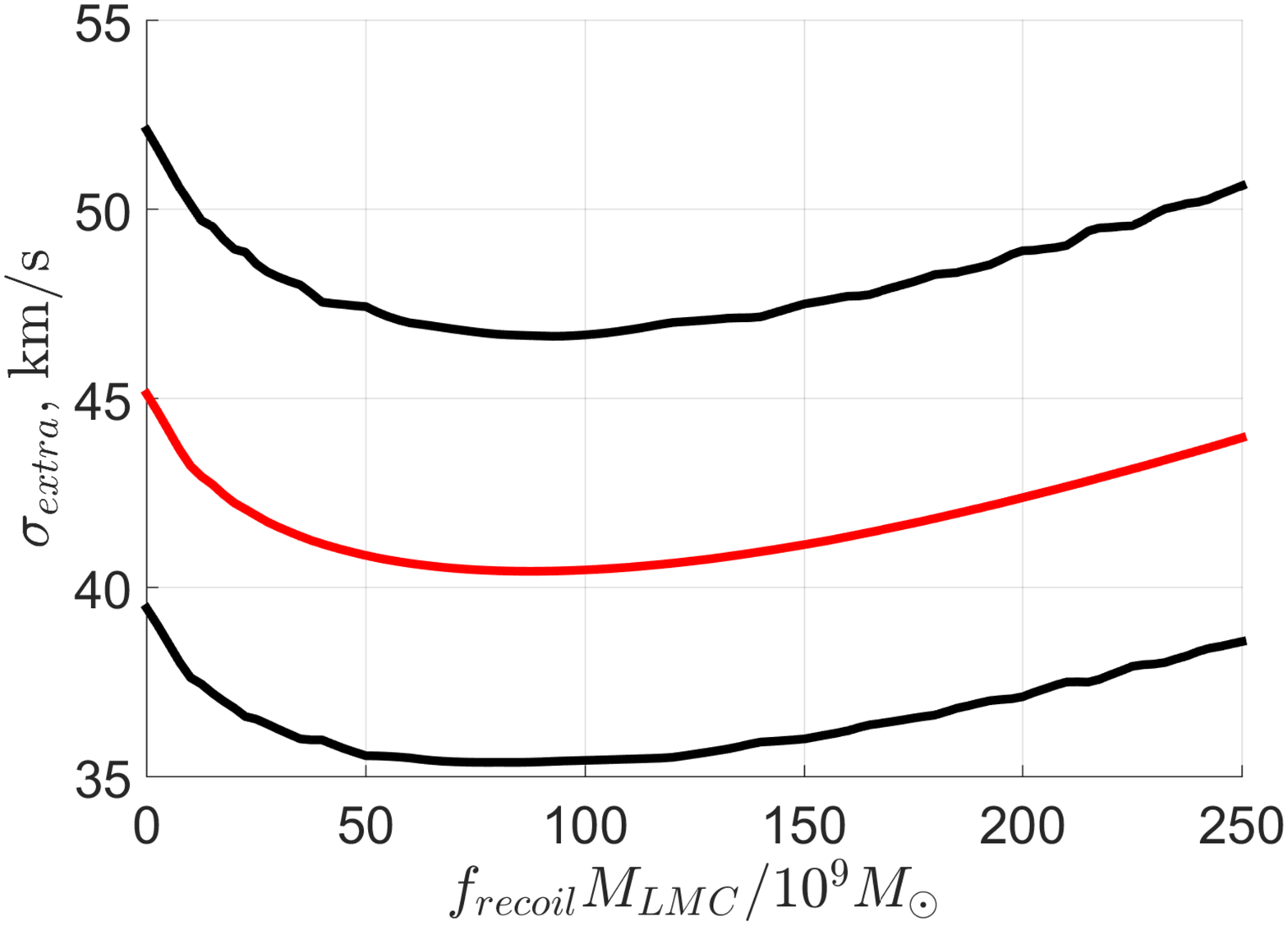}
	\caption{The most likely value of $\sigma_{extra}$ and its 1$\sigma$ uncertainty are shown here as a function of the LMC mass (assuming $f_{recoil} = 1$, see text). The LMC is included via Equation \ref{LMC_adjustment}. Notice that $\sigma_{extra}$ eventually increases with the LMC mass because very high LMC masses imply a very large kinematic correction for it.}
	\label{LMC_trend_s}
\end{figure}

\begin{figure}
	\centering
		\includegraphics [width = 8.5cm] {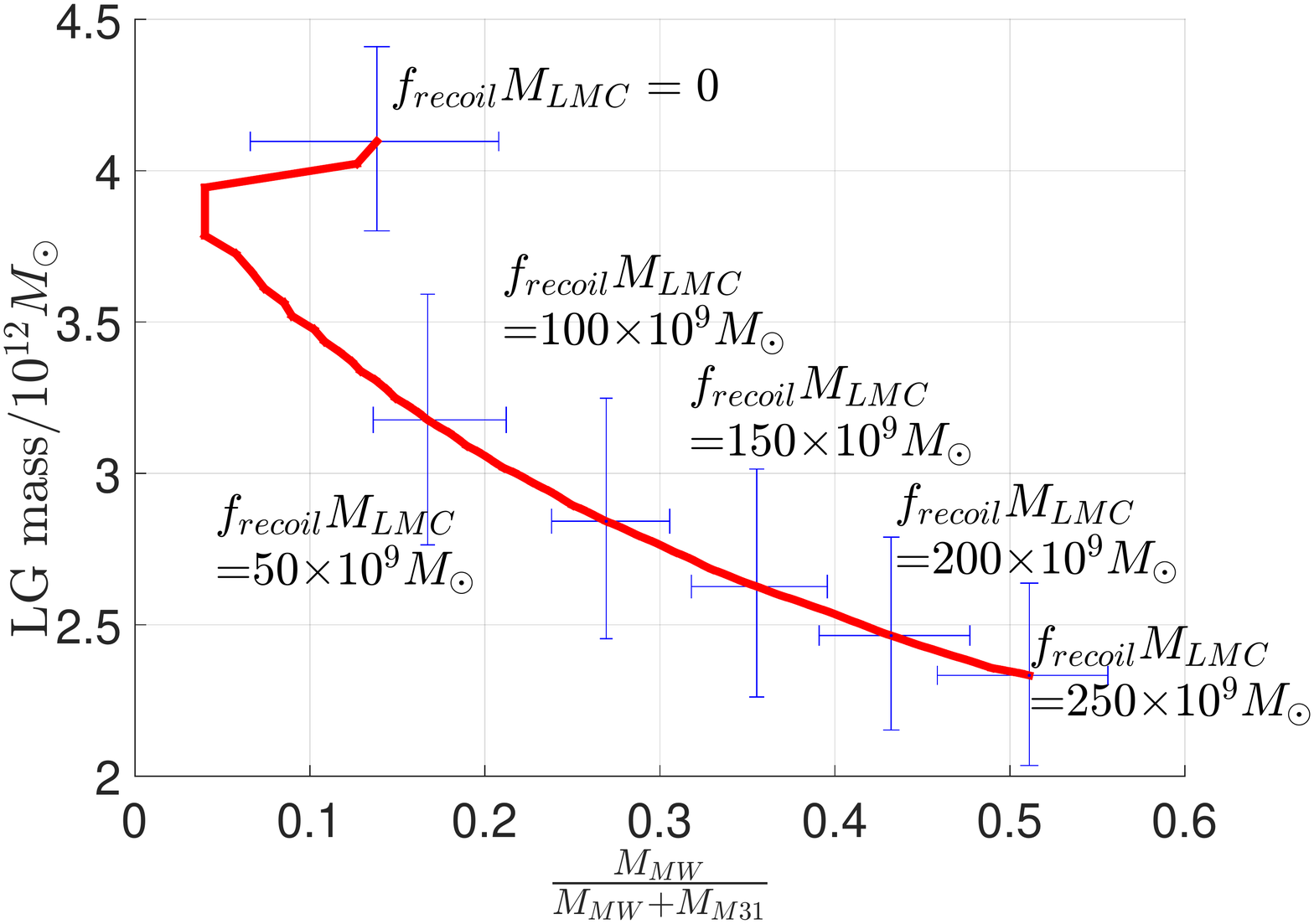}	
	\caption{The red line shows the locus of the most likely values of $M_i$ and $q_{_1}$ as a function of LMC mass. Crosses show uncertainties on each parameter for 6 different LMC masses. Figure \ref{Primary_result} suggests that uncertainties in $M_i$ and $q_{_1}$ are nearly uncorrelated.}
	\label{LMC_joint_M_q}
\end{figure}

Probably the most important result of this investigation is that the overall fit to the observations is not much improved. We quantify this using $\sigma_{extra}$, which is reduced slightly once the LMC is included (Figure \ref{LMC_trend_s}). However, for very large LMC masses, the kinematic correction it induces becomes very large, thereby worsening the fit to the data. Thus, including the LMC can't reduce $\sigma_{extra}$ by even as much as its formal uncertainty.

The correction for the LMC is implemented by altering $\bm v_\odot$ according to Equation \ref{LMC_adjustment}. At given $q_{_{LMC}}$, this adds a constant vector to the predicted velocities of all LG galaxies with respect to the Sun. If we set $q_{_{LMC}} = 0.2$, then the magnitude of this vector is 64 km/s, comfortably exceeding $\sigma_{extra}$. Nonetheless, including the LMC hardly reduces $\sigma_{extra}$. This is because the galaxies we identified as having anomalously high radial velocities (Table \ref{IC342_M81_effect}) are in several quite different sky directions. Indeed, we confirmed that assuming a large LMC mass causes some galaxies to have HRVs very substantially below the predictions of the best-fitting model.

The motion of the LMC with respect to the MW disc is mostly along the MW$-$M31 line, so one expects a strong effect on the implied total LG mass $M_i$. This is clearly borne out by our analysis (Figure \ref{LMC_joint_M_q}). The lower LG mass resulting from including the LMC is now more consistent with the works of \citet{Jorge_2014} and \citet{Diaz_2014} $-$ both give values around $2.5 \times 10^{12} M_\odot$.

Our analysis clearly prefers a non-zero value for $q_{_{LMC}}$ (Figure \ref{LMC_trend_s}). Thus, at low LMC masses, the analysis prefers low MW masses to force up $q_{_{LMC}}$ towards its preferred value. The opposite occurs at high LMC masses $-$ a rapid increase in $q_{_1}$ is required to raise the MW mass and hold down the kinematic correction due to the LMC. This is because the fit to the data is worsened if this correction is too large.


The LMC also affects the tangential velocity of M31 with respect to the MW system. Considering the M31 proper motion measurement of \citet{M31_motion}, it is likely that including the LMC slightly increases the tangential velocity of M31, though its radial velocity is increased far more. Still, M31 should remain on a nearly radial orbit with respect to the MW system if the LMC is given a reasonable mass. Any tangential motion would increase the inferred total LG mass as there would be a larger centrifugal force between the MW and M31 (which is not included in our analysis). This would tend to reduce model-predicted GRVs of LG galaxies, making it even more difficult to explain the observations.

Including the LMC hardly affects our inference on $v_{c, \odot}$ (Table \ref{Trend_LMC_v_c_Sun}). We used a prior on this parameter of 239$\pm$5 km/s \citep{McMillan_2011}. Our analysis slightly reduces its uncertainty. Based on the magnitude of this reduction, we conclude that our timing argument analysis independently constrains the speed of the LSR to within $\sim$ 15 km/s. The combination of the prior with our work yields a best-fitting LSR speed very close to that implied by the prior alone. This suggests that if we did not impose a prior constraint on $v_{c, \odot}$, then we would find a most likely value for $v_{c, \odot}$ that was within $\sim \left( \frac{15}{5} \right)^2 \times 0.5 \approx 5$ km/s of 239 km/s. Thus, there is no tension between the LSR speed preferred by our analysis and that preferred by \citet{McMillan_2011} on independent grounds, although our analysis is $\sim$ 3 times less accurate in this regard.

\begin{table}
	\begin{tabular}{cc}
	\hline
	$f_{recoil}~M_{LMC}$ & $v_{c, \odot}$ \\
	\hline
	0 &  239.53 $\pm$ 4.82 km/s \\
	125$\times 10^9 M_\odot$ &  239.44 $\pm$ 4.72 km/s \\
	250$\times 10^9 M_\odot$ &  238.55$^{+4.65}_{-4.67}$ km/s \\	
	\hline
	\end{tabular}
	\caption{The most likely value of the LSR speed $v_{c, \odot}$ with its 1$\sigma$ uncertainty is given as a function of the assumed LMC mass. The prior constraint is $239 \pm 5$ km/s \citep{McMillan_2011}. There is no tension between this value and that suggested by our analysis.}
\label{Trend_LMC_v_c_Sun}
\end{table}

It is possible for massive satellites of M31 to cause a similar kinematic correction to its adopted HRV \footnote{though see \citet{Jorge_2016} for why such corrections are likely very small, even for the brightest M31 satellites}. Our analysis places a high statistical weight on M31 by requiring our models to match its HRV fairly well. However, our previous results were almost unaffected if we used the same value of $\sigma_{extra}$ for M31 as for other LG galaxies (the inferred value of $\sigma_{extra}$ differed by $\sim$ 2 km/s). This remains true if we include the kinematic effect of the LMC in our analysis. In other words, our results are not much different if we treat M31 in exactly the same way as other LG galaxies. Consequently, even a substantial alteration to the HRV of M31 should hardly affect our analysis as it involves 31 other galaxies.

The case is different for the LMC because including a kinematic correction for it alters the predicted HRV of every galaxy in the LG rather than just the observed HRV of one galaxy. As the former does not much affect our overall conclusions, we suspect the same is true of the latter.

\subsection{Interactions With Massive Satellite Galaxies}
\label{Massive_satellites}

Models invoking gravitational slingshot encounters near the MW/M31 to fling out galaxies at high speed seem to share an important feature with the data: in Figure \ref{GRV_LCDM_Comparison}, most of the galaxies (20/32) are going outwards faster than predicted by the best-fitting model. This trend is perhaps clearer in Figure \ref{Delta_GRV_histogram_detailed}. 

We already included gravity from the MW and M31. One possibility not previously considered is that their satellites could have interacted with what are now non-satellite galaxies in the LG. For example, the Triangulum galaxy (M33) might be able to expand the region around M31 with a disturbed velocity field. This is possible via gravitational slingshot encounters with M33, using energy from its orbital motion around M31 to fling out material at high speed. 

Considering that M33 rotates at $\sim$100 km/s \citep{Corbelli_2003}, it can't have affected the motion of a passing object by much more than this without merging with it. Thus, an important issue with such a scenario is whether it can explain the fast outward motion of galaxies like DDO 125. Not only would it have to reach its present position several Mpc from M31, it would also have to possess sufficient kinetic energy to move at its present high velocity. Even if we neglect the retarding effect of gravity from the MW and M31, Hubble drag alone would mean that a peculiar velocity of 100 km/s today needed to have been 300 km/s at redshift 2, a plausible time for the interaction considering how far DDO 125 is from M31 (Figure \ref{LG_Hubble_Diagram}).  


Moreover, one expects only a small fraction of the material in the LG to have interacted with Triangulum in the narrow range of impact parameters that lead to a large impulse but avoid a merger. Some of the material that was unaffected by M33 would no doubt have interacted with the LMC or with other massive satellites. Still, we find it hard to believe that such interactions would be as likely or as strong as required to fit the observations. Achieving both simultaneously does not seem feasible.

Our models did not have particles starting too close to the MW or M31. We mapped the gravitational potential at $t = t_i$ (Equation \ref{U}) and assumed that all material below a certain level (i.e. with $U < U_{exc}$) had gone into one of these galaxies. For most parameters, this region did not split into separate regions around each galaxy but was a single region encompassing both. Test particles were not allowed to start within it.

It is possible that pockets of material within this `excluded region' did not get accreted by the MW or M31. Starting closer to one of these galaxies, this material might be more likely to interact with one of their satellites. However, it is unclear how such interactions could have been strong enough to explain the observations as the material would also have a deeper potential well to climb out of.

\subsection{Interactions with the MW and M31}
\label{MW_M31_interaction}


Other than the MW and M31, none of the objects in the LG seem heavy enough to impart a sufficiently large impulse on our target galaxies. However, our models already include gravitational slingshots caused by the MW \& M31. Such encounters provide a way of extracting energy from the motion of these galaxies and putting it into the motion of a less massive third object. 

In principle, Andromeda can exert a large impulse on a passing object $-$ perhaps up to twice Andromeda's rotation speed. Therefore, it might be able to exert an impulse of as much as $\sim$ 450 km/s on an object which approached closely enough yet avoided merging. For such an interaction to help explain the observations, the scattered object must have been fast-moving relative to M31 \emph{even when the two were far apart}. Otherwise, even fully reversing the small relative velocity `at infinity' would only lead to a small impulse.

In our simulations, the MW and M31 have never been moving very fast (Figure \ref{MW_M31_separation_history}). Their relative motion has usually been slower than at present \citep[$\sim$110 km/s,][]{M31_motion}. It is difficult to achieve an impulse much exceeding the motion of the massive body. Supposing the MW was moving at $\sim$90 km/s, a small fraction of the material in the LG received an impulse of perhaps that much.\footnote{In a logarithmic potential, an extremely eccentric orbit has an angle of $\sim$240$^\circ$ between apocentres, meaning that the deflection angle is $\sim$60$^\circ$ rather than 180$^\circ$.}

The effect of Hubble drag then reduces the peculiar velocity gained in this way. So also does the gravity of M31 (except for particles between the MW and M31, a region in which none of our target galaxies lie). Thus, the region in which the velocity field is disturbed by interactions with the MW/M31 only goes out to $\sim$1 Mpc from the LG barycentre (Figure \ref{LG_Hubble_Diagram}). At higher LG masses, this region is somewhat larger: its linear size $\appropto {M_i}^\frac{1}{3}$. It is difficult to see how this region can be made to encompass the whole LG.

Prior to the start of our simulations, the MW$-$M31 mutual gravity would not yet have had much time to retard their motion. Thus, we can assume that they were tracing the cosmic expansion, with mutual separation $d \left( t \right) \propto a\left( t \right)$. At these times, the Hubble parameter $\frac{\overset{.}{a}}{a} \sim a^{-\frac{3}{2}}$ (Equation \ref{Expansion_history}) and so we expect the velocities of the galaxies to behave as $a^{-\frac{1}{2}}$. As a result, an interaction with a passing dwarf galaxy could lead to a maximum impulse on it that depends on the encounter time as $\ssim a\left( t \right) ^{-\frac{1}{2}}$. This means that encounters of the MW/M31 with LG dwarfs at very early times may have been very powerful.

However, due to Hubble drag, the effect on the present peculiar velocity of the dwarf galaxy would be reduced by a factor of $a$ at the time of the encounter.\footnote{Figure \ref{Impulsed_trajectories} suggests that a factor of $a^{2.4}$ might be more accurate.} This means that very early encounters between the MW/M31 and LG dwarf galaxies should hardly affect our analysis, even if they were very strong. This is probably why our results changed very little when we altered the start time of our simulations to correspond to redshift 14 rather than 9 ($\sigma_{extra}$ decreased by $\sim$ 1 km/s when using the earlier start time).

It is unlikely that the MW and M31 existed at earlier times. This makes it difficult to argue that early encounters between LG dwarfs and the MW/M31 are responsible for the anomalously high HRVs of some distant LG galaxies.

Therefore, one possible solution is to suppose that the MW and M31 were moving much faster than in our model a few Gyr after the Big Bang. We mentioned in Section \ref{Introduction} that they might indeed have done so. In MOND, they would have undergone a close flyby $\sim$9 Gyr ago (unlike the $\Lambda$CDM-based trajectories used in our models, Figure \ref{MW_M31_separation_history}). The relative speed at the time of closest approach would have been $\sim$600 km/s \citep{Zhao_2013}. One could suppose that the MW was moving at 400 km/s and M31 at 200 km/s. Any passing dwarf galaxies would then have received a large impulse. Hubble drag would reduce peculiar velocities gained in this way by only a factor of $\sim$3. Thus, the MW$-$M31 relative speed in this model seems about right to explain the motions of the LG galaxies with high $\Delta GRV$s.

Of course, one does not expect \emph{all} of our target galaxies to have received an impulse quite this large. Only some of the material in the LG would have closely approached the MW or M31 at a time when their relative speed exceeded e.g. 400 km/s. This material might then get flung outwards and become an observed galaxy. It is also possible for the material to later merge with a galaxy that never strongly interacted with the MW or M31. Depending on the mass ratio of such a merger, it might leave the resulting object with a GRV only a little above that of the unperturbed galaxy before the merger.

An interesting aspect of the observations which may point towards this scenario is apparent in Figure \ref{Cylindrical_projection_GRV_map}: the galaxies with the greatest excess radial velocity relative to $\Lambda$CDM all seem to be towards the edge of the LG. This may be because those objects which were flung out at higher speeds are now further away from the MW and M31.

In this scenario, the high $\Delta GRV$ galaxies were all (roughly) at the same place at the same time: close to the LG barycentre when the MW$-$M31 flyby occurred. Thus, the magnitudes of the GRV discrepancies can be used to estimate the time of the flyby. The highest $\Delta GRV$ galaxy in our sample is DDO 99, with $\Delta GRV = 100$ km/s and a distance of $\sim$3 Mpc.\footnote{As this is the highest $\Delta GRV$ out of 32 galaxies, the true value is likely smaller than the 110 km/s obtained using nominal values.} Assuming objects at this distance would nearly follow a pure Hubble flow in $\Lambda$CDM (e.g. bottom panel of Figure \ref{LG_Hubble_Diagram}), its radial velocity should be $H_{_0} d \sim 200$ km/s with respect to the LG barycentre. Neglecting projection effects,\footnote{on the sky, DDO 99 and M31 are almost at right angles {(Figure \ref{LG_Hubble_Diagram})}, so our conclusions should not be much affected by uncertainty in the motion of the MW due to uncertainty in $q_{_1}$} the actual radial velocity is $\sim$50\% larger. This corresponds to an elapsed time since the flyby of $\ssim \frac{2}{3}$ the age of the Universe, i.e. $\sim$9 Gyr ago.

Interestingly, this is also when there appears to have been a sudden perturbation to the disc of the MW which created its thick disc \citep{Quillen_2001}. There is some circumstantial evidence that this perturbation was tidal in nature rather than a process internal to the MW \citep{Banik_2014}. Although a minor merger is an obvious possibility, it would leave accreted stars. As the accreted galaxy must have been reasonably massive compared to the MW to create its thick disc, these accreted stars should have characteristic properties. Recent attempts to find such stars have not found any \citep{Ruchti_2015}. This might be an indication that the thick disc was created by a close flyby of another massive galaxy $\sim$9 Gyr ago rather than a minor merger at that time.

In this respect, it is interesting to estimate when a MW$-$M31 interaction might have occurred if MOND were the correct description of nature. Applying this theory, it can be shown that the time from apocentre to pericentre is given by \citep[equations 15 and 29$-$30 of][]{Zhao_2010}
\begin{eqnarray}
	\label{Equation_66}
	\Delta t &\approx& \int_0^1 \left[ \frac{2}{{t_{_M}}^2} \ln{\frac{1}{x}}  - (1 - x^2) {H_{_0}}^2\Omega_{\Lambda, 0}  \right]^{-1/2} dx	~~\text{ where} \nonumber \\
	t_{_M} &=& \frac{d_{\mathrm{apo}}}{\sqrt[4]{0.61 G M a_{_0}}}
\end{eqnarray}

Note that the relevant mass $M$ is the total baryonic mass of the LG. The non-linear nature of the theory reduces the force between two particles with the same total mass if it is distributed more equally. Even assuming (conservatively) equal MW and M31 masses (leading to the factor of 0.61) and using a very low estimate for $M$ of $10^{11} M_\odot$, we would get $ \Delta t \approx 7.2$ Gyr.\footnote{We used cosmological parameters as in Table \ref{Priors} and an apocentre distance $d_{\mathrm{apo}} = 1$ Mpc, slightly less than in $\Lambda$CDM due to the stronger gravity. For $a_{_0}$, we used $1.2 \times {10}^{-10} $ m/s$^2$ \citep{McGaugh_2011}.} The MW and M31 are slightly past their apocentre now, but it is still clear that they must have had a past close flyby in this theory.

Equation \ref{Equation_66} neglects several complications which arise in MOND. Most important is the external field effect \citep{Bekenstein_Milgrom_1984, Milgrom_1986}. This arises because MOND is an acceleration-dependent theory. As a result, a constant external gravitational field acting upon a system weakens the self-gravity of objects within the system. This effect is approximately taken into account in the work of \citet{Zhao_2013}.

We are currently undertaking more accurate timing argument calculations in the context of MOND. Preliminary results indicate that the LG mass implied by the MOND timing argument is consistent with baryons only, if one requires a past close approach between the galaxies. This flyby needs to have been $\sim$8$-$9 Gyr ago. It is not feasible to construct trajectories of the MW and M31 in MOND which avoid such a close approach or have $\geq$2 such events.

In MOND, the longer range nature of gravity means that we need a more careful treatment of objects outside the LG (Table \ref{Perturbers}). It is possible that some of the anomalously high outwards velocities found in this work are due to LG galaxies $-$ especially those close to perturbers $-$ falling in towards them. Indeed, including tides from Centaurus A improved the fit to observations somewhat, even in the context of $\Lambda$CDM (Figure \ref{Sigma_extra_comparison}). This effect might be further enhanced in modified gravity scenarios.

Interestingly, the discrepancy does seem to be higher for target galaxies closer to M81 or to IC 342 (Figure \ref{Tide_correlation}). We argued that the apparent correlation could not be due to gravity from these perturbers as this would imply that they affected velocities of target galaxies more than is reasonable. However, this argument likely breaks down under a different law of gravity. In this case, tides might well be more significant than we assumed.

\subsection{Comparison with \citet{Jorge_2014}}
\label{Comparison_with_Jorge}

Our results are broadly similar to the recent study conducted by \citet{Jorge_2014}, hereafter P14. Those authors also favour a low $q_{_1}$ and a similar LG mass. We found a slightly higher value of $\sigma_{extra}$ (${45^{+7}_{-6}~\text{km/s}}$ instead of ${35^{+6}_{-4}~\text{km/s}}$), though the estimates are marginally in agreement.

However, our investigation greatly strengthens the conclusions reached by P14. We used an axisymmetric model for the Local Group rather than a spherically symmetric one. As our target galaxies are often not much further away than the MW$-$M31 separation, a spherically symmetric gravitational field may be a poor approximation. For example, gravitational slingshot encounters with the MW/M31 rely on a time-dependent non-spherical potential. Without modelling either of these effects, it would be difficult to draw reliable conclusions about whether these close encounters might have left an imprint on the present motions of target galaxies.

Even if one could be sure that such encounters were not important, a two-centred potential has other subtle consequences. A trajectory which initially went orthogonal to the MW$-$M31 line from the point halfway between them; would curve towards the heavier galaxy (almost certainly M31). This would tend to increase the HRV of the target galaxy (e.g. bottom panel of Figure \ref{GRV_Sextans_A}). Curvature of test particle trajectories seems to be important even at quite large distances from the MW and M31 (top panel of Figure \ref{LG_Hubble_Diagram}). The process is more significant if the MW and M31 masses are very unequal, which definitely seems to be the case (Figure \ref{Primary_result}).

By using the same list of target galaxies as P14, we avoid targets too close to any of the major perturbers relevant to our analysis (Table \ref{Perturbers}). However, tides from these objects must affect our results at some level. We directly include the most massive perturber (Centaurus A), exploiting its location almost along the MW$-$M31 line (Section \ref{Tides_Cen_A}). We think this greatly improves our model. We conduct a more thorough investigation of tides raised by other objects (Section \ref{Tides}) and consider the effect of the Large Magellanic Cloud in some detail (Section \ref{Large_Magellanic_Cloud}).

Our initial conditions are handled differently to P14. We use cosmological initial conditions (Equation \ref{Initial_conditions}) because of observations indicating very low peculiar velocities at early times \citep{Planck_2015}. P14 used a procedure involving non-cosmological initial conditions which does not seem entirely physical (see their section 3).


We added an extra term to our equation of motion (Equation \ref{Equation_of_motion}) to account for cosmological expansion. The idea is to recover $\bm r \propto a \left( t \right)$ in the absence of inhomogeneities. A similar approach was used by P14. However, they did not include the deceleration to the expansion rate caused by matter, leaving only the acceleration caused by dark energy. This implicitly assumes that the rest of the LG is empty. As we did not make this assumption, we suspect that the predicted HRVs in our investigation are lower. With a present dark energy fraction of $\sim$ 0.7, we expect a difference of $\ssim \left( 1 - \sqrt{0.7} \right) H_{_0} d$ for a target at distance $d$. Assuming a typical distance of 2 Mpc, this suggests a difference of $\sim$20 km/s. Because observed HRVs tend to exceed predicted ones, it is unsurprising that our estimate of $\sigma_{extra}$ is higher.

In Section \ref{Reduced_LG_Mass}, we discuss how much mass might actually be in the RLG. Here, we show that, although it is possible to have an empty RLG, this is an extreme case. It is also possible for it to have even more mass than we assumed. Our assumption is an intermediate case. Nonetheless, we performed calculations assuming an empty RLG and showed that this reduces $\sigma_{extra}$ by $\sim$7 km/s.


We argued that M31 should be treated specially in that one should use a lower value of $\sigma_{extra}$ for it than for other LG galaxies. This forces our models to match the HRV of M31 very well, restricting which models are viable and thus raising $\sigma_{extra}$. The effect is substantial for our analysis without Cen A: the most likely value of $\sigma_{extra}$ is raised from 46 to 54 km/s. However, in our more realistic analysis including Cen A, $\sigma_{extra}$ is only affected by $\sim$2 km/s. Thus, although we recommend treating M31 specially due to its much higher mass than LG dwarf galaxies, our overall conclusions are little altered if one does not do so. The parameter most affected seems to be $M_i$, which is lower if M31 is not treated specially. This is also apparent in Figure 13 of P14, especially if one imposes an independent constraint on the LSR speed \citep[e.g.][]{McMillan_2011}.


\section{Conclusions}
\label{Conclusions}

We performed a careful dynamical analysis of the Local Group (LG) to try and explain the observed positions and velocities of galaxies within it. The LG was treated as a collection of test particles and two massive ones $-$ the Milky Way (MW) and Andromeda (M31) $-$ which we put on a radial orbit. We added a third massive particle to represent Centaurus A, which is very close to the MW$-$M31 line (Table \ref{Perturbers}). All particles were started moving outwards from the centre of mass of the LG with speeds proportional to distance from there. Thus, they all started on a pure Hubble flow with no peculiar velocity (Equation \ref{Initial_conditions}).

A wide range of possible masses for the MW and M31 was investigated using a grid method (Table \ref{Priors}). Each time, we got the final MW$-$M31 and MW$-$Cen A distances to match their observed values. We also got a test particle trajectory to end at the same location as each observed LG galaxy. This gave a model-predicted velocity, whose line-of-sight component (the HRV) was compared with observations.

The best-fitting total LG mass is $4.33^{+0.37}_{-0.32} \times {10}^{12} M_\odot$, with $0.14 \pm 0.07$ of this accounted for by the MW and the rest by M31. There is almost no tension between the Local Standard of Rest rotation speed estimated by \citet{McMillan_2011} and our analysis (Table \ref{Trend_LMC_v_c_Sun}).

However, even in the best-fitting model, there was a poor match between observed and model-predicted HRVs. Thus, we tried to quantify the extra astrophysical noise $\sigma_{extra}$ that the observations imply. To do this, we added it in quadrature with the other known sources of error, which are all observational. $\sigma_{extra}$ can be constrained using
\begin{eqnarray}
	P( \text{Model} | HRV_{obs}) \propto \frac{1}{\sigma} {\rm e}^{-\frac{1}{2}\left( \frac{HRV_{obs} - HRV_{model}}{\sigma}\right)^2} 
\end{eqnarray}

Raising $\sigma_{extra}$ $-$ and thus $\sigma$ $-$ will eventually cause the probability to decrease\footnote{when $\sigma > \left| HRV_{obs} - HRV_{model} \right|$}. In this way, we found that $\sigma_{extra} = 45.1_{-5.7}^{+7.0}$ km/s (Figure \ref{Primary_result}). This rather high value is partly due to some galaxies receding from the LG even faster than a pure Hubble flow (Figure \ref{GRV_No_Gravity}). This is despite the attractive gravity of the MW and M31. 

We expect interactions between LG dwarf galaxies to have contributed somewhat to $\sigma_{extra}$, but only at the $\sim$15 km/s level. This is because they have fairly low rotation velocities/internal velocity dispersions \citep{Kirby_2014}, limiting how much they can influence each other gravitationally. If we consider just those galaxies with HRVs below the predictions of the best-fitting model (blue area in Figure \ref{Delta_GRV_histogram_detailed}), then we see that they can indeed be described quite well by a Gaussian of this width.

One might expect the same to be true for LG galaxies with HRVs that exceed the model predictions. However, this is not true (red area in Figure \ref{Delta_GRV_histogram_detailed}). Thus, there is a systematic trend for radial velocities to be higher than we expect.



We considered tides from objects outside the LG at some length (Section \ref{Tides}). The most relevant perturbers are given in Table \ref{Perturbers}. A correlation is apparent whereby the most problematic galaxies are closest to these perturbers (Figure \ref{Tide_correlation}). Thus, we constructed a simplified model to estimate how much they might have affected the GRVs of the most discrepant galaxies. Due to a combination of projection effects and large distances from the perturbers ($\ga$ 2 Mpc), we consider it unlikely that tides from IC 342 and M81 can reconcile our model with observations (Table \ref{IC342_M81_effect}). In fact, they would likely exacerbate the tension in several cases. This also seems to be true of tides raised by the Great Attractor (Figure \ref{Great_Attractor_effect}): including these raises $\sigma_{extra}$ by $\sim$5 km/s (Section \ref{Great_Attractor}).

A local under-density may help to explain the observations. We consider this possibility in Section \ref{Reduced_LG_Mass} and show that this can reduce $\sigma_{extra}$ by at most $\sim$7 km/s. Increasing the Hubble constant $H_{_0}$  has a similar effect, which we consider in Section \ref{High_H_0}. This reduces $\sigma_{extra}$ by $\sim$5 km/s (Figure \ref{Sigma_extra_comparison}). Although $H_{_0}$ is known quite accurately \citep{Planck_2015}, a higher value may be appropriate for the LG if there is a local under-density.

Satellite galaxies of the MW can affect our analysis kinematically. In this regard, we considered the Large Magellanic Cloud (LMC) in some detail (Section \ref{Massive_satellites}). Including the LMC can reduce $\sigma_{extra}$ by at most 5 km/s (Figure \ref{LMC_trend_s}). Thus, incorporating it into our models does not reconcile them with observations.

Interactions of LG dwarf galaxies with massive MW/M31 satellites (e.g. M33) would probably be too weak to explain distant non-satellite galaxies moving outwards even faster than a pure Hubble flow. As for encounters with the MW and M31 themselves, this would naturally explain why LG galaxies tend to move outwards much faster than expected. However, this process is already included in our simulations $-$ it seems to be too weak.

It might have been more powerful in reality if the MW and M31 had been moving much faster than in our models (Section \ref{MW_M31_interaction}). Although this would be very unusual in the context of $\Lambda$CDM, such fast motions arise naturally in certain modified gravity theories \citep{Zhao_2009}. For example, Modified Newtonian Dynamics \citep[MOND,][]{Milgrom_1983} leads to a past close encounter between the MW and M31 and a maximum relative speed of $\sim$600 km/s \citep{Zhao_2013}.

One aspect of our results which may point towards this scenario lies in the distances to the objects with the highest radial velocities relative to the best-fitting $\Lambda$CDM model. These distances all seem to be close to the upper limit of the distance range probed by our sample (Figure \ref{Cylindrical_projection_GRV_map}). This may be because those objects which received the strongest gravitational boost from a close encounter with the MW/M31 are currently furthest away from the location of the encounter. Considering the distances and velocities of such galaxies suggests that the encounter was $\sim$9 Gyr ago. This is around the same age as the thick disc of the MW \citep{Quillen_2001}. It is also roughly when a MOND timing argument calculation suggests that a MW$-$M31 encounter took place \citep{Zhao_2013}.



We think it likely that the observed positions and velocities of LG galaxies would be easier to explain if there was a past close MW$-$M31 flyby. This is possible only within the framework of a modified gravity theory. More work will be required to test such a scenario.

The authors wish to thank Jorge Pe{\~n}arubbia for helpful discussions and the referee for their patience while we implemented their suggestions, which greatly strengthened this contribution. IB is supported by a Science and Technology Facilities Council studentship. The algorithms were set up using \textsc{matlab}$^\text{\textregistered}$.

%
%
%
%
%
%
%
%
%

\bibliographystyle{mnras}
\bibliography{LGE_bbl}
\bsp
\label{lastpage}
\end{document}